\documentclass[a4paper,11pt]{article}
\pdfoutput=1 

\usepackage{jcappub} 
\usepackage{graphicx,subfigure}
\usepackage{physics}
\usepackage{amssymb}
\usepackage{listings}
\usepackage[table]{xcolor}
\usepackage{tikz}
\usepackage{hyperref}
\usepackage{pifont}
\newcommand{\xmark}{\ding{55}}%
\newcommand\Tstrut{\rule{0pt}{2.6ex}} 
\usepackage{adjustbox,lipsum}
\usepackage{amsmath}
\usepackage[T1]{fontenc} 

\definecolor{codegreen}{rgb}{0,0.6,0}
\definecolor{codegray}{rgb}{0.5,0.5,0.5}
\definecolor{codepurple}{rgb}{0.58,0,0.82}
\definecolor{backcolour}{rgb}{0.95,0.95,0.92}

\lstdefinestyle{mystyle}{
    backgroundcolor=\color{backcolour},   
    commentstyle=\color{codegreen},
    keywordstyle=\color{magenta},
    numberstyle=\tiny\color{codegray},
    stringstyle=\color{codepurple},
    basicstyle=\ttfamily\footnotesize,
    breakatwhitespace=false,         
    breaklines=true,                 
    captionpos=b,                    
    keepspaces=true,                 
    numbers=left,                    
    numbersep=5pt,                  
    showspaces=false,                
    showstringspaces=false,
    showtabs=false,                  
    tabsize=2
}

\usetikzlibrary{shapes.geometric,shapes.symbols,fit,positioning,shadows}
\makeatletter
\pgfkeys{/pgf/.cd,
  parallelepiped offset x/.initial=2mm,
  parallelepiped offset y/.initial=2mm
}
\pgfdeclareshape{parallelepiped}
{
  \inheritsavedanchors[from=rectangle] 
  \inheritanchorborder[from=rectangle]
  \inheritanchor[from=rectangle]{north}
  \inheritanchor[from=rectangle]{north west}
  \inheritanchor[from=rectangle]{north east}
  \inheritanchor[from=rectangle]{center}
  \inheritanchor[from=rectangle]{west}
  \inheritanchor[from=rectangle]{east}
  \inheritanchor[from=rectangle]{mid}
  \inheritanchor[from=rectangle]{mid west}
  \inheritanchor[from=rectangle]{mid east}
  \inheritanchor[from=rectangle]{base}
  \inheritanchor[from=rectangle]{base west}
  \inheritanchor[from=rectangle]{base east}
  \inheritanchor[from=rectangle]{south}
  \inheritanchor[from=rectangle]{south west}
  \inheritanchor[from=rectangle]{south east}
  \backgroundpath{
    \southwest \pgf@xa=\pgf@x \pgf@ya=\pgf@y
    \northeast \pgf@xb=\pgf@x \pgf@yb=\pgf@y
    \pgfmathsetlength\pgfutil@tempdima{\pgfkeysvalueof{/pgf/parallelepiped offset x}}
    \pgfmathsetlength\pgfutil@tempdimb{\pgfkeysvalueof{/pgf/parallelepiped offset y}}
    \def\ppd@offset{\pgfpoint{\pgfutil@tempdima}{\pgfutil@tempdimb}}
    \pgfpathmoveto{\pgfqpoint{\pgf@xa}{\pgf@ya}}
    \pgfpathlineto{\pgfqpoint{\pgf@xb}{\pgf@ya}}
    \pgfpathlineto{\pgfqpoint{\pgf@xb}{\pgf@yb}}
    \pgfpathlineto{\pgfqpoint{\pgf@xa}{\pgf@yb}}
    \pgfpathclose
    \pgfpathmoveto{\pgfqpoint{\pgf@xb}{\pgf@ya}}
    \pgfpathlineto{\pgfpointadd{\pgfpoint{\pgf@xb}{\pgf@ya}}{\ppd@offset}}
    \pgfpathlineto{\pgfpointadd{\pgfpoint{\pgf@xb}{\pgf@yb}}{\ppd@offset}}
    \pgfpathlineto{\pgfpointadd{\pgfpoint{\pgf@xa}{\pgf@yb}}{\ppd@offset}}
    \pgfpathlineto{\pgfqpoint{\pgf@xa}{\pgf@yb}}
    \pgfpathmoveto{\pgfqpoint{\pgf@xb}{\pgf@yb}}
    \pgfpathlineto{\pgfpointadd{\pgfpoint{\pgf@xb}{\pgf@yb}}{\ppd@offset}}
  }
}
\pgfdeclareshape{document}{
\inheritsavedanchors[from=rectangle] 
\inheritanchorborder[from=rectangle]
\inheritanchor[from=rectangle]{center}
\inheritanchor[from=rectangle]{north}
\inheritanchor[from=rectangle]{north east}
\inheritanchor[from=rectangle]{north west}
\inheritanchor[from=rectangle]{south}
\inheritanchor[from=rectangle]{south east}
\inheritanchor[from=rectangle]{south west}
\inheritanchor[from=rectangle]{west}
\inheritanchor[from=rectangle]{east}
\backgroundpath{%
\southwest \pgf@xa=\pgf@x \pgf@ya=\pgf@y
\northeast \pgf@xb=\pgf@x \pgf@yb=\pgf@y
\pgf@xc=\pgf@xb \advance\pgf@xc by-5pt 
\pgf@yc=\pgf@ya \advance\pgf@yc by5pt
\pgfpathmoveto{\pgfpoint{\pgf@xa}{\pgf@ya}}
\pgfpathlineto{\pgfpoint{\pgf@xa}{\pgf@yb}}
\pgfpathlineto{\pgfpoint{\pgf@xb}{\pgf@yb}}
\pgfpathlineto{\pgfpoint{\pgf@xb}{\pgf@yc}}
\pgfpathlineto{\pgfpoint{\pgf@xc}{\pgf@ya}}
\pgfpathclose
\pgfpathmoveto{\pgfpoint{\pgf@xc}{\pgf@ya}}
\pgfpathlineto{\pgfpoint{\pgf@xc}{\pgf@yc}}
\pgfpathlineto{\pgfpoint{\pgf@xb}{\pgf@yc}}
\pgfpathclose
}
}
\tikzset{
    folder/.pic={
    \draw [fill=yellow!70!blue, rounded corners] (-1.43,-1.05) -- (-1.43,1.04) -- (-0.77,1.04) -- (-0.77,0.68) -- (0.94,0.68) -- (0.94,0.0) -- cycle;
    \draw [fill=yellow!80!, rounded corners] (-1.43,-1.05) -- (-1.03,0.37) -- (1.29,.37) -- (0.90,-1.05) -- cycle;
    }
}

\makeatother
\newcommand{\paa}[1]{\left(#1\right)}

\definecolor{myred}{rgb}{0.8,0,0}

\title{cp3-bench: A tool for benchmarking symbolic regression algorithms demonstrated with cosmology} 

\author{M. E. Thing and S. M. Koksbang}
\affiliation{CP3-Origins, University of Southern Denmark, Campusvej 55, DK-5230 Odense M, Denmark}

\emailAdd{thing@cp3.sdu.dk}
\emailAdd{koksbang@cp3.sdu.dk}


\abstract{
We introduce cp3-bench, a tool for comparing/benching symbolic regression algorithms, which we make publicly available at \url{https://github.com/CP3-Origins/cp3-bench}. In its current format, cp3-bench includes 12 different symbolic regression algorithms which can be automatically installed as part of cp3-bench. The philosophy behind cp3-bench is that it should be as user-friendly as possible, available in a ready-to-use format, and allow for easy additions of new algorithms and datasets. Our hope is that users of symbolic regression algorithms can use cp3-bench to easily install and compare/bench an array of symbolic regression algorithms to better decide which algorithms to use for their specific tasks at hand.
\newline\indent
To introduce and motivate the use of cp3-bench we present a small benchmark of 12 symbolic regression algorithms applied to 28 datasets representing six different cosmological and astroparticle physics setups. Overall, we find that most of the benched algorithms do rather poorly in the benchmark and suggest possible ways to proceed with developing algorithms that will be better at identifying ground truth expressions for cosmological and astroparticle physics datasets. Our demonstration benchmark specifically studies the significance of dimensionality of the feature space and precision of datasets. We find both to be highly important for symbolic regression tasks to be successful. On the other hand, we find no indication that inter-dependence of features in datasets is particularly important, meaning that it is not in general a hindrance for symbolic regression algorithms if datasets e.g. contain both $z$ and $H(z)$ as features. Lastly, we note that we find no indication that performance of algorithms on standardized datasets are good indicators of performance on particular cosmological and astrophysical datasets. This suggests that it is not necessarily prudent to choose symbolic regression algorithms based on their performance on standardized data. Instead, a more robust approach is to consider a variety of algorithms, chosen based on the particular task at hand that one wishes to apply symbolic regression to.
}
\begin{document}
\maketitle
\flushbottom
	
\section{Introduction}
    A core aspect of physics is that we are able to use symbolic expressions to describe how the world around us works. It is therefore a significant challenge when we meet a physical system, observable, or an experimental dataset that we cannot describe through mathematical formulas. This problem arises in various places in astrophysics and cosmology where, for instance, the precise halo mass function is unknown \cite{halo_profile}, stellar density profiles are estimated based on various approximations but the true underlying relation is yet to be discovered \cite{stellar_profile}, a derived expression for the non-linear matter power spectrum does not yet exist \cite{nonlinear_power, nonlinear_power2}, and intricacies of general relativity leave useful expressions for observables such as the mean redshift drift \cite{zdrift1,zdrift2} still waiting to be found (the concept of mean redshift drift will be introduced in section \ref{sec:data}). When a ground truth expression for a quantity is desired but seems unfeasible to obtain through analytical considerations, one option is to try to use symbolic regression which is a regression method based on machine learning (ML). Symbolic regression (SR) is, at least in principle, a powerful tool for identifying unknown relations between physical quantities such as data from astrophysical observations or laboratory experiments.
    \newline\indent
    When utilizing SR algorithms in e.g. cosmology, the approach has generally been to use one or very few specific algorithms. For instance, the work in \cite{zdrift1, zdrift2} utilized the symbolic regression algorithm AI Feynman \cite{Feynman1, Feynman2} because of its user-friendliness and because it was developed by physicists with discovering physics formulae in mind, showing an improvement compared to earlier state-of-the-art algorithms when applied to a specific suite of datasets. The work presented in \cite{nonlinear_power, nonlinear_power2} instead utilized Operon \cite{operon} which was chosen due to its speed, memory efficiency and because it is based on genetic algorithms which tend to do well in SR benchmark studies and competitions (e.g. \cite{benchmark, competition}). In general, when we choose an algorithm for SR, we naturally make a decision we expect represents the state-of-the-art for the task. However, SR algorithms are usually evaluated and compared by studying their efficiency at identifying symbolic expressions corresponding to certain ensembles of datasets such as the Penn Machine Learning Benchmarks \cite{PMLB1, PMLB2}, collections of mathematical expressions (see e.g. appendix A of \cite{review}), simple physical laws as in EmpiricalBench of \cite{cranmer2023interpretable}, and some even use a dataset based on equations from the Feynman Lectures on Physics \cite{FeynmanLectures}, the so-called Feynman Symbolic Regression Database\footnote{https://space.mit.edu/home/tegmark/aifeynman.html}. The approach of evaluating SR algorithms based on standard datasets seems prudent at least for academic evaluations/comparisons in relation to developing SR algorithms. However, although these ensembles of generic datasets often do include physics datasets (not the least the Feynman Symbolic Regression Database), this does not necessarily mean that the resulting benchmark/comparison is representative of how well the individual algorithms will perform in practice on a specific real cosmology or other (astro-)physics dataset. One hindrance is that real world cosmology problems are typically based on very complex physics which means that the corresponding datasets can be difficult to obtain, understand and can end up being complicated from a physical point of view, e.g. including several dependent variables. This type of dataset simply does not seem to be represented significantly in common machine learning/symbolic regression databases. In addition, it is not clear that SR algorithms optimized with typical machine learning database problems are automatically also optimized for real world physics problems. This means that when one has a genuine SR task, one cannot expect to identify the most optimal algorithms and hyperparameters simply by looking at results from earlier benchmarks or competitions.
    \newline\indent
    To promote an easier and more diverse use of SR algorithms in especially cosmology and astro(particle) physics and to encourage researchers who wish to {\em use} but not necessarily {\em develop} SR algorithms, we here introduce cp3-bench which we release together with this paper. We have developed cp3-bench as a Python-based tool for comparing/benching SR algorithms. The current version supports 12 different SR algorithms. We provide cp3-bench in a ready-to-use format for Ubuntu machines with Python 3.10 or newer, where installation and usage is straightforward and where the user can choose to have (some or all of) the 12 SR algorithms currently supported automatically installed when installing cp3-bench without extra effort from the user. A true out-of-the-box experience can also be obtained with our Docker image which ensures that the package is ready-to-use on (nearly) any type of machine with Docker and that all dependencies are set correctly.  
    \newline\indent
    To demonstrate the use of cp3-bench and to highlight strengths and weaknesses of the 12 SR algorithms as well as obstacles related to different types of datasets, we here present a small benchmark of the 12 SR algorithms with 28 datasets. We stress that the benchmark we present should not itself be used as a guide for deciding which SR algorithm(s) to use for a specific SR task at hand. Instead, our benchmark is meant as a demonstration, and we encourage all to make their own benchmark/comparison of existing SR algorithms to supplement existing ``academic'' benchmarks such as \cite{benchmark, competition} when choosing which SR algorithms to use for practical tasks at hand. We encourage contributions to the benchmark e.g. by adding new algorithms to the suite.
    \newline\newline
    In Section \ref{sec:data} below, we introduce the datasets we used in our benchmark, including an introduction to the physical settings they represent. In Section \ref{sec:benchmark} we present the benchmark setup, our benchmark tool cp3-bench and an example wrapper Things-to-bench. Section \ref{sec:benchmark-results} presents our benchmark results. Section \ref{sec:Summary} provides a summary and discussion of future possible developments.

\section{Cosmological datasets}\label{sec:data}
    This section serves to describe the data used in our benchmark. There can be several different aims when using SR. A goal can for instance be to make a good fit of the nonlinear matter spectrum for certain $\Lambda$CDM model parameters and use the fit as a type of emulator. Another goal can be to identify new laws of nature or hitherto unknown observational relations, i.e. discovering ``ground truth'' expressions. We will focus our demonstration benchmark on the latter and our choice of benchmark data and evaluation criteria of SR results are based on having this goal in mind. Note that most of the synthetic datasets we choose do not themselves represent actual SR problems, i.e. for most of the datasets we consider, the ground truth is known. This makes it easier to evaluate the performance of SR algorithms in terms of their ability to specifically ``guess'' ground truths. The main differences here compared to standard datasets are thus that a) the data is based on real cosmology and astro(particle) physics models, making it more closely related to datasets one could imagine considered in real astropysics applications, and b) we specifically choose datasets and variations of these that highlight different qualities of datasets that may be important when seeking to use SR to identify ground truth expressions. In addition, two groups of our datasets do indeed represent datasets where the ground truth is unknown.
    \newline\indent
    Below, we introduce six overall groups of datasets. Each group of datasets represents a specific physical scenario and within each group there can be several versions of datasets representing the scenario. In the first subsection below, we introduce our first group of datasets which is based on the Hubble relation of standard cosmology. The following subsection is dedicated to the observable known as redshift drift and its Friedmann-Lemaitre-Robertson-Walker (FLRW) limit. In the subsequent two subsections we introduce a specific family of toy cosmological models called two-region models and discuss how to compute the redshift drift as well as a quantity known as cosmic backreaction in these models. We then introduce datasets based on dark matter galaxy halo profiles and lastly a group of datasets based on simplified mock gravitational wave data. The datasets are summarized in Section \ref{subsec:data_summary} which includes a list of the datasets in Table \ref{table:C}.
    \newline\indent
    Since most of the 28 datasets are cosmology-based, we will for simplicity refer to this group of datasets as the ``cosmological'' datasets to distinguish them from a group of test datasets introduced later.
    
    \begin{figure}
        \centering
        \includegraphics[scale = 0.47]{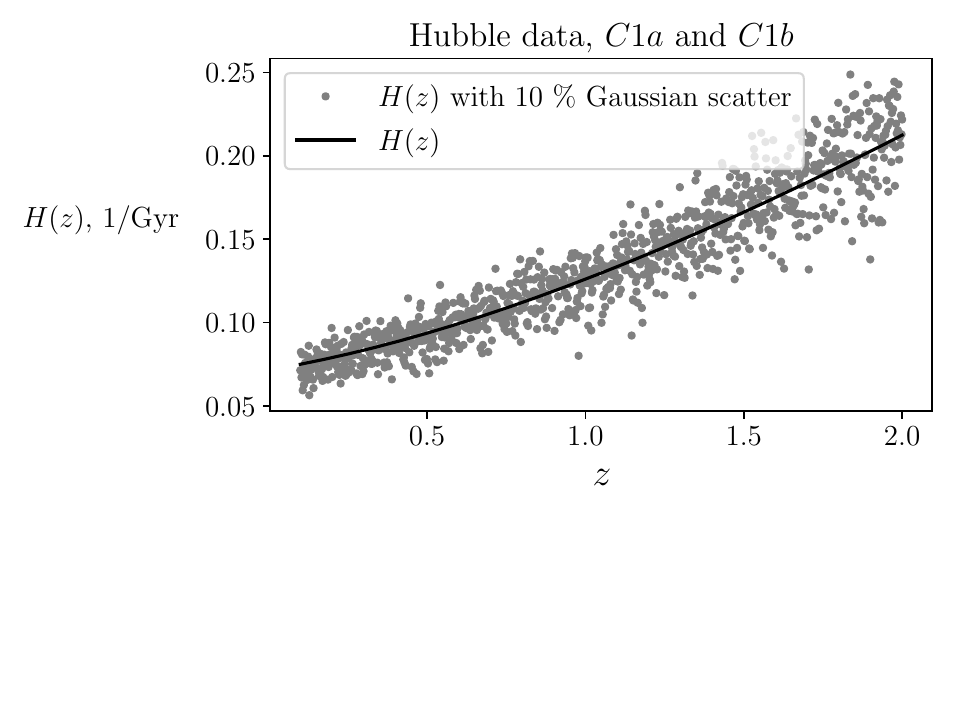}
        \includegraphics[scale = 0.47]{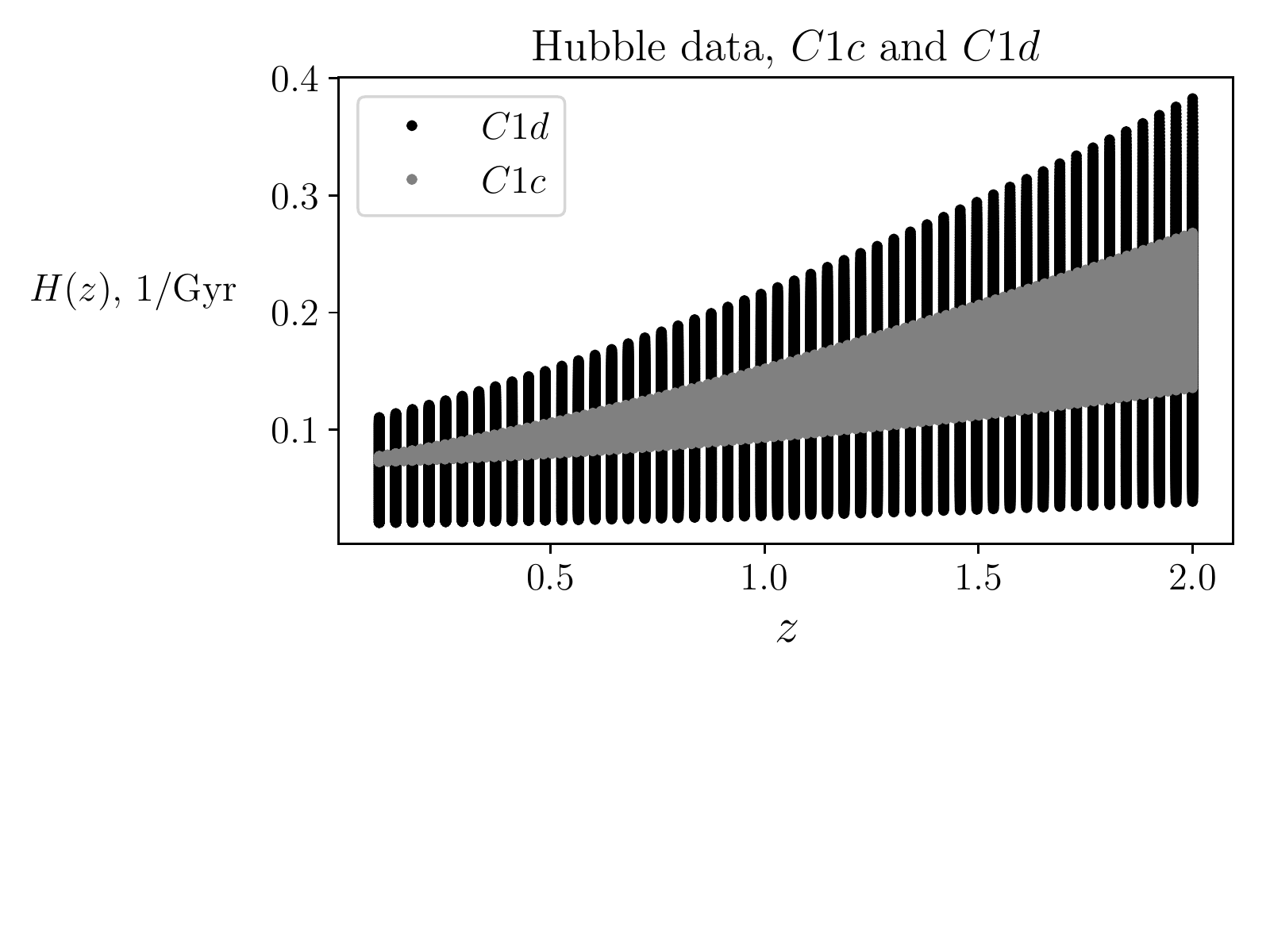}
        \caption{The left figure shows $C1a$ ($H(z)$ with no error for a single $\Lambda$CDM model) and $C1b$ ($H(z)$ with 10 \% Gaussian error for a single $\Lambda$CDM model). The figure to the right shows the datasets $C1c$ and $C1d$, i.e. $H(z)$ for $\Lambda$CDM models with varying $\Omega_m$ and both $\Omega_m$ and $H_0$, respectively. For $C1c$ and $C1d$, one can see the dependence of $\Omega_m$ (and $H_0$ in the latter) as vertical lines in the dataset at constant $z$.}
        \label{fig:C1}
    \end{figure}        	

\subsection{Hubble data}
    The Hubble parameter describes the expansion history of the Universe and has great interest in cosmology. Earlier work on symbolic regression such as \cite{physics_motivation, Hubble, language} has thus fittingly included showcasing the use of new symbolic regression algorithms for astrophysics by applying them to Hubble data either directly, or indirectly through supernova data. We will also use Hubble datasets here.
    \newline\indent
    In the $\Lambda$CDM model, the Hubble parameter is given by
    \begin{align}
    \frac{H^2(z)}{H_0^2} = \Omega_{m,0}(1+z)^3 + (1-\Omega_{m,0}).
    \end{align}
    Using this relation, we have generated three different Hubble datasets. The first two are similar to those used in \cite{physics_motivation, Hubble} since we only consider a single cosmological model, namely the (flat) $\Lambda$CDM model with $H_0 = 70$km/s/Mpc and $\Omega_{m,0}=0.3$. One dataset will consist of $(H(z), z)$ without noise while the second dataset will contain 10\% Gaussian noise on $H(z)$. By including datasets that are identical up to noise we can asses the impact that noise has on the performance of the SR algorithms. For each dataset we compute 1000 data points distributed equidistantly in the redshift interval $z\in[0.1,2]$, roughly representing the redshift interval of real datasets that directly measure the Hubble parameter (e.g. measurements based on cosmic chronometers \cite{Jimenez_2002, Renzini_2006, Moresco_2018, Moresco_2020} or on baryon acoustic oscillations, BAO \cite{BAO_1, BAO_2, BAO_3, BAO_4}). We will refer to these two datasets as $C1a$ and $C1b$. They are plotted together in Figure \ref{fig:C1}.
    \newline\indent            
    We will additionally consider a dataset that corresponds to several $\Lambda$CDM models where we let $H_0$ and $\Omega_{m,0}$ vary in the intervals $H_0\in[20,100]$ in units of km/s/Mpc and $\Omega_{m,0}\in[0.1,0.5]$. Again, we use equidistant points, now with 50 points for each feature. This dataset, as well as those discussed in the following subsections, are motivated by the ambition to decipher which symbolic regression methods are most suitable for discovering hitherto unknown observational relations. The more general an observational relation is, the more interesting it is. It is therefore desirable to reveal observational relations that are valid for not just a single $\Lambda$CDM model but rather for, say, \emph{all} FLRW models or $\Lambda$CDM models. We therefore wish to test how well symbolic regression methods can describe data representing one or more families of cosmological models such as sub-families of FLRW models rather than just a single FLRW model. Since this type of dataset will be inherently more complicated than datasets based on only a single FLRW model\footnote{Note for instance that it was in \cite{zdrift1,zdrift2} found that the symbolic regression algorithm AI Feynman \cite{Feynman1,Feynman2} was much better at finding analytical expressions for the redshift drift of a single cosmological model than for several models at once.}, we will not add noise to these datasets. We denote these two datasets $C1c$ and $C1d$, respectively, and show them in Figure \ref{fig:C1}.
    
    \begin{figure}
        \centering
        \includegraphics[scale = 0.3]{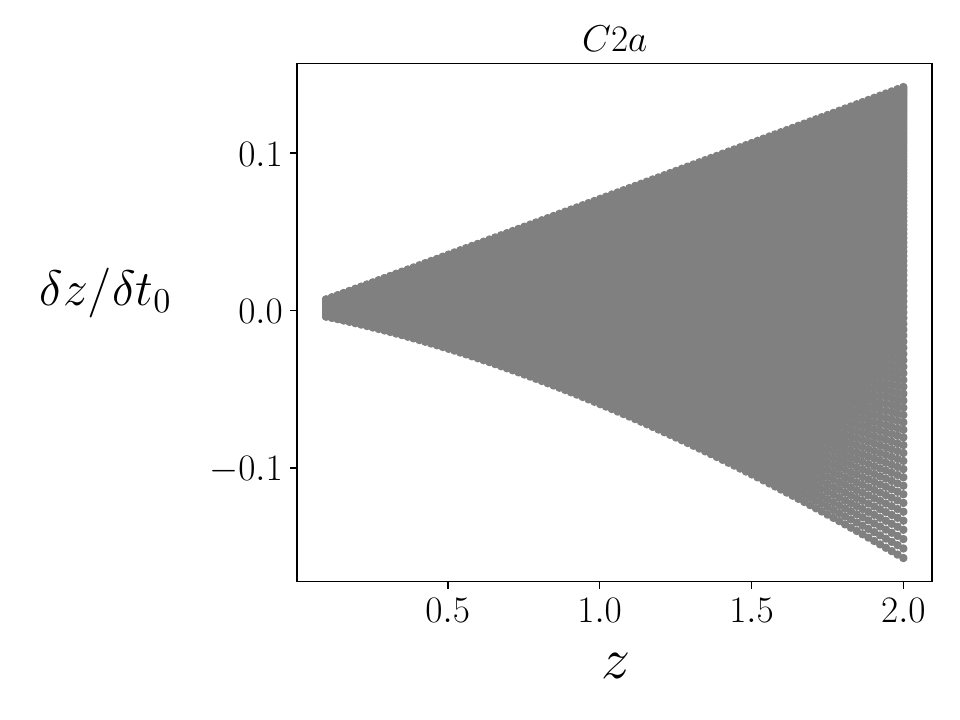}
        \includegraphics[scale = 0.3]{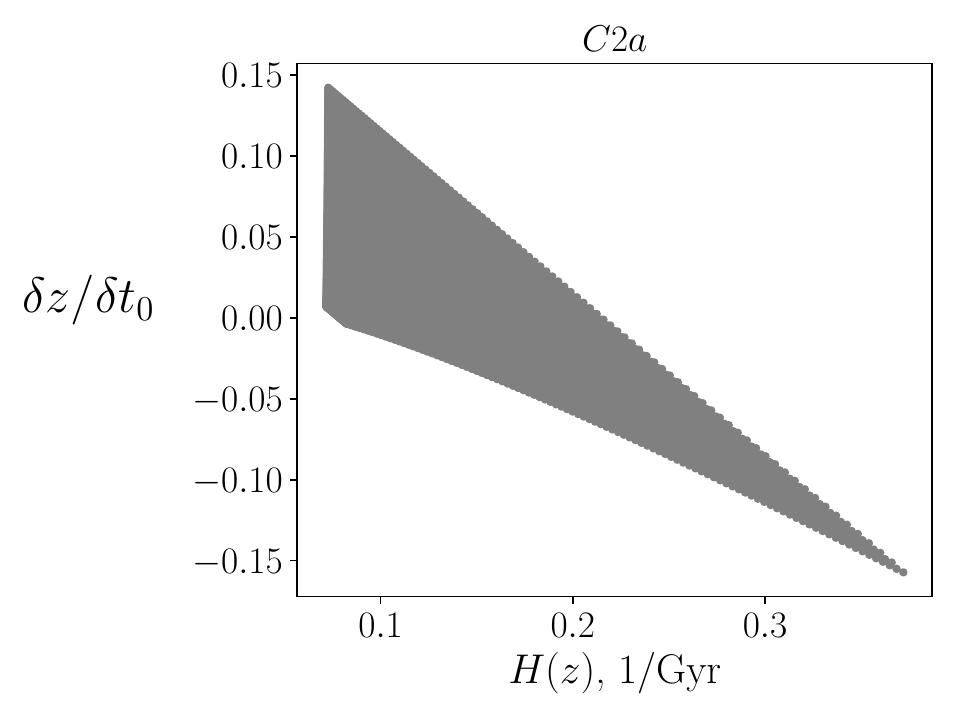}
        \includegraphics[scale = 0.3]{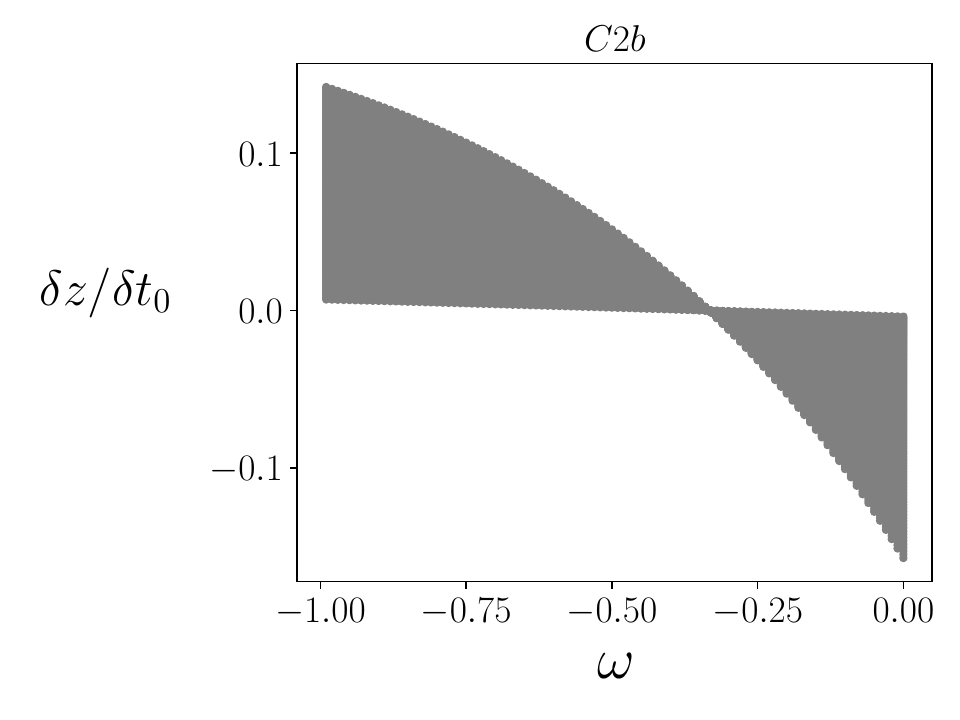}
        \caption{The redshift drift datasets $C2a$ and $C2b$ plotted against the different features used in the two datasets.}
        \label{fig:C2}
    \end{figure} 

\subsection{Redshift drift in standard cosmology}
    One of the biggest mysteries in cosmology is the physical nature of dark energy i.e. the source of accelerated expansion of the Universe \cite{DEreview}. The apparent accelerated expansion of the universe has, however, never been measured directly. The observable \emph{redshift drift} represents a possibility to remedy this. The redshift of a luminous astronomical object changes with time. This change has been dubbed the redshift drift. Redshift drift was first discussed in \cite{Sandage, Mcvittie} from 1962 but it was there assessed that the effect was too small to be measurable within a reasonable time frame. Due to technological advances, it is presently expected that the effect can be measured with e.g. the Square Kilometer Array (SKA) Observatory \cite{SKA}.
    \newline\indent
    For FLRW models, the redshift drift, $\delta z$, can be written as
    \begin{align}
        \delta z = \delta t_0[(1+z)H_0 - H(z) ],
    \end{align}
    where $H(z)$ is the Hubble parameter, $H_0$ the Hubble constant, and $\delta t_0$ is the time interval of the observation such that an object observed to have redshift $z$ at time $t_0$ will be observed to have the redshift $z+\delta z$ at time $t_0 + \delta t_0$.
    \newline\indent
    The expression for the redshift drift written above is fairly simple. Nonetheless, written as above, it represents an interesting symbolic regression problem because the right hand side of the equation contains a feature ($H(z)$) which depends on the other feature ($z$) in a non-trivial way. In addition, $H(z)$ depends on the cosmological model parameters such as $\Omega_{m,0}$ which is not included as a parameter in the dataset since $z$ and $H(z)$ are enough to obtain the expression for the redshift drift shown above. Earlier studies into machine and deep learning algorithms have included datasets where two or more variables were dependent, but these dependencies are typically linear or otherwise simple (see e.g. \cite{BIMT} for an example). However, to the authors' knowledge, there have not been conducted systematic studies of the impact of the inter-dependence of features nor studies based on datasets where features have more complicated dependencies such as in the case of the Hubble parameter and its dependence on the redshift.
    \newline\indent
    In cosmology, observable relations are often more convenient to consider in terms of several dependent features. Besides the redshift drift, this is for instance the case for redshift-distance relations which are usually written as integrals over the Hubble parameter in terms of the redshift multiplied by factors containing the redshift itself as well. In our benchmark dataset we will therefore include the redshift drift as written above, i.e. with the target being $\delta z/\delta t_0$ and the features being $z$ and $H(z)$. However, we will consider only flat, single-component FLRW models when generating the datasets. In this case, a simple rewriting of the above expression yields
    \begin{align}
        \delta z = \delta t_0 H_0[(1+z) - (1+z)^{3(1+\omega)/2}],
    \end{align}
    where $\omega$ is the equation-of-state parameter of the single component of the content of the model universe\footnote{We could equally well have written out the Hubble parameter in terms of $\Omega_{m,0}$, $z$, and $H_0$ as when we considered the Hubble parameter but since the difference between the targets in the datasets with the Hubble parameter versus the redshift drift would very small, we expect the symbolic regression algorithms would perform similarly to as on the Hubble datasets.}.
    \newline\indent
    Using the formalism discussed above, we have generated two redshift drift datasets, $(\delta z/\delta t_0, z, H(z, \omega))$ and $(\delta z/\delta t_0, z, \omega)$. The two models contain the same number of features, has the same target and contains the same number of data points generated using 50 equidistant points in each of the two intervals $z\in[0.1,1]$ and $\omega\in[-0.99,1/3]$, resulting in a total of $50^2$ data points per dataset. When choosing the redshift interval, we deliberately avoid the very lowest redshifts to avoid a target very close to zero. This is sensible since some of the SR algorithms we bench (e.g. AI Feynman) remove data with low values of the target which we wish to avoid. We choose the upper limit $z = 1$ since this is the upper limit expected for main redshift drift measurements from SKA \cite{SKA}. For $\omega$ we choose the upper limit of $1/3$ which corresponds to the equation-of-state parameter for radiation while we choose a lower limit close to that of a cosmological constant, although neglecting the value $-1$ to keep the overall form of the Hubble parameter the same for all models\footnote{Remember that for flat, single component universes, $H(z) = \frac{2}{3(1+\omega)t}$ except for the case $\omega = -1$ where the Hubble parameter is constant.}. We denote the redshift drift datasets $C1a$ and $C1b$ and plot the redshift drift against $z, H(z)$ and $\omega$ in Figure \ref{fig:C2}. 
    
    \begin{figure}
        \centering
        \includegraphics[scale = 0.3]{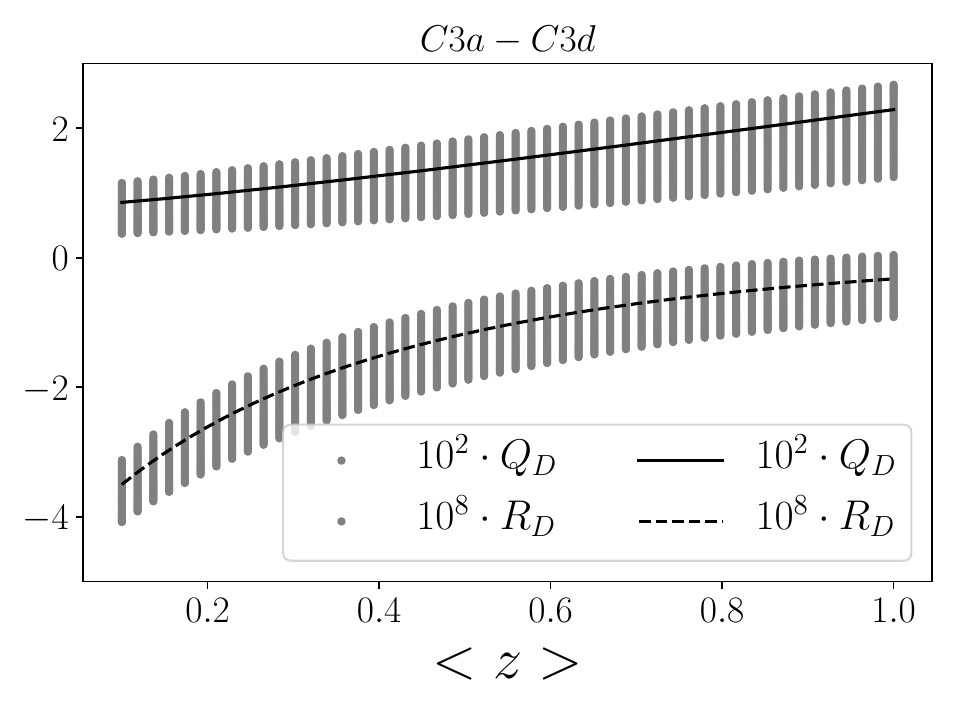}
        \includegraphics[scale = 0.3]{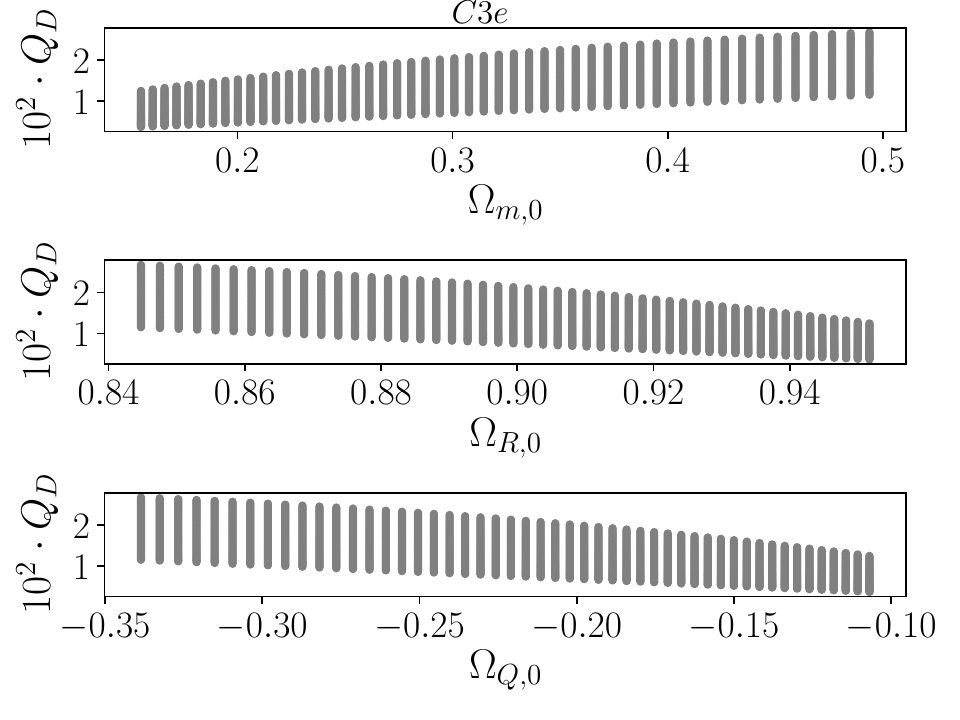}
        \includegraphics[scale = 0.3]{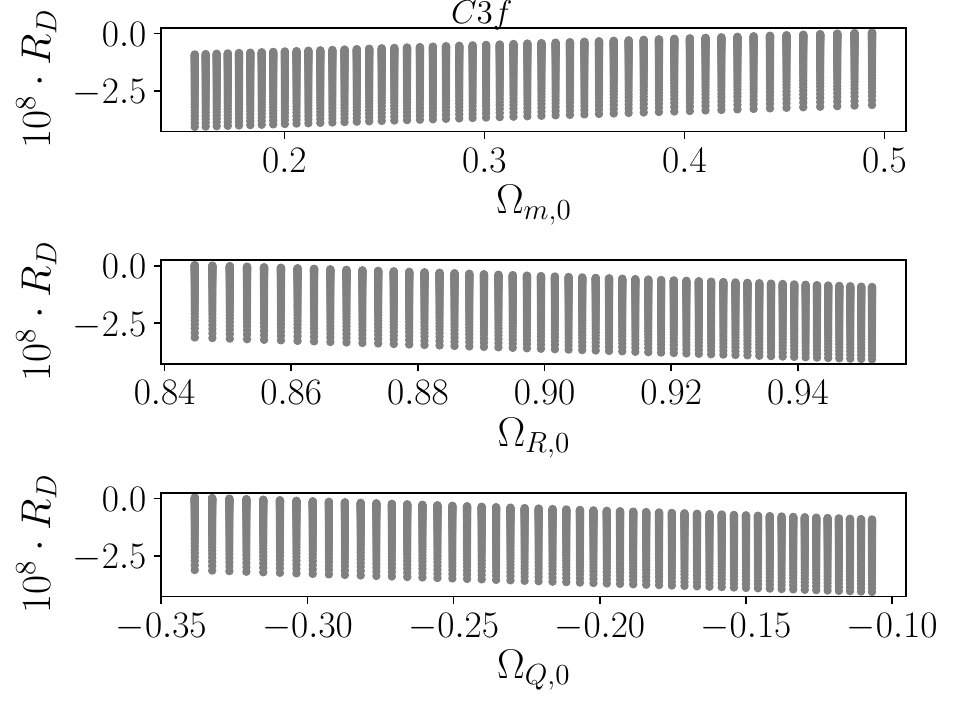}
        \caption{Left: $Q_D$ and $R_D$ plotted against $z$, corresponding to the datasets $C3a-C3d$. Middle: $Q_D$ plotted against different features (dataset $C3e$). Right: $R_D$ plotted against different features (dataset $C3f$).}
        \label{fig:C3abcd}
    \end{figure} 

\subsection{Cosmic backreaction and two-region models}
    The expression for the redshift drift given in the previous section is valid only for FLRW spacetimes, i.e. in spacetimes which have no structures on any scales. Standard cosmology is based on the assumption that the large-scale dynamics of the Universe can be approximated well by the FLRW models but it has also been discussed whether the structures on smaller scales can affect the large-scale/averaged dynamics. This possibility is known as cosmic backreaction \cite{bc_review1, bc_review2, bc_review3, bc_review4} and it has even been suggested that the seemingly accelerated expansion of the Universe is an artifact due to this phenomenon (the co-called backreaction conjecture -- see e.g. \cite{conjecture1, conjecture3, conjecture4}). Backreaction and the backreaction conjecture are most commonly considered through the Buchert equations presented in \cite{dust, perfect_fluid, general} where dynamical equations describing the average (large-scale) evolution of the Universe are obtained by averaging Raychaudhuri's equation and the Hamiltonian constraint using the averaging definition
    \begin{align}
        s_D:=\frac{\int_D sdV}{\int_D dV},
    \end{align}
    where $s$ is some scalar and $dV$ is the proper infinitesimal volume element on the spatial domain $D$ assumed to be larger than the homogeneity scale. This definition can be used when considering spacetimes foliated with spatial hypersurfaces orthogonal to the fluid flow (possible for irrotational fluids such as dust) and where the line element is written as
    \begin{align}
        ds^2 = -dt^2 + g_{ij}dx^idx^j.
    \end{align}
    The resulting averaged Hamiltonian constraint and Raychaudhuri equation can be written as (setting $c=1$)
    \begin{align}\label{eq:Buchert_av}
    \begin{split}
        3H_D^2 &= 8\pi G\rho_D - \frac{1}{2}R_D - \frac{1}{2}Q_D\\
        3\frac{\ddot a_D}{a_D} &= -4\pi G\rho_D + Q_D.
    \end{split}
    \end{align}
    $H_D:=\dot a_D/a_D$ denotes the average Hubble parameter and is related to the local fluid expansion scalar $\theta$ by $H_D=\theta_D/3$ (dots denote derivative with respect to time). The volume/averaged scale factor $a_D$ is, unlike the scale factor in FLRW spacetimes, not related to the metric (which is not considered in the Buchert averaging scheme) but is instead defined as $a_D:=(V_D/V_{D_0})^{1/3}$, where $V_D$ is the proper volume of the spatial domain $D$ and subscripts zero indicate evaluation at present time.
    \newline\indent
    The dynamical equations for the average universe shown above are reminiscent of the Friedmann equation and acceleration equation but contain an extra term, namely $Q_D:=2/3\left[  (\theta^2)_D - (\theta_D ^2)\right]  - \left( \sigma_{\mu\nu}\sigma^{\mu\nu}\right) _D$, where $\sigma_{\mu\nu}$ is the shear tensor of the fluid. In addition, the curvature term $R_D$ can deviate from the FLRW behavior where the curvature must be proportional to the inverse of the squared scale factor. $Q_D$ is sometimes referred to as the kinematical backreaction while the deviation of $R_D$ from having FLRW evolution is referred to as intrinsic backreaction. Combined, these two differences with respect to the FLRW models mean that the large-scale dynamics of an inhomogeneous universe does not necessarily follow FLRW dynamics even when averaging above the homogeneity scale.
    \newline\newline
    The two backreaction terms are coupled through the integrability condition
    \begin{align}
        a_D^{-6}\partial_t(a_D^6Q_D) + a_D^{-2}\partial_t(a_D^2R_D) = 0.
    \end{align}
    This equation ensures that the two Buchert equations are consistent with each other. However, the two Buchert equations and the integrability condition do not form a closed set so cannot be used to predict backreaction, not even qualitatively. Therefore, it is unknown how backreaction is realistically parameterized in terms of the scale factor (or equivalently, the mean redshift $\expval{z}+1\approx 1/a_D$ \cite{light1, light2}, where triangular brackets indicate taking the mean over many lines of sight to remove statistical fluctuations). As discussed in e.g. \cite{zdrift1,zdrift2}, this means that backreaction cannot be sensibly constrained with current cosmological data. This motivated an attempt (presented in \cite{zdrift1,zdrift2}) to obtain parameterizations for backreaction using toy cosmological models where backreaction can be computed, combined with symbolic regression to identify symbolic expressions for backreaction. Building on that study, our benchmark dataset includes several datasets based on the so-called two-region models introduced in \cite{2region1, 2region2}.
    \newline\newline
    Two-region models are toy cosmological models consisting of an ensemble of two different FLRW regions. Since the ensemble is disjoint, the model is not an exact solution to Einstein's equation. The model is nonetheless convenient to use for studying backreaction because the model is numerically fast and simple to work with. Following the procedure in \cite{2region1, 2region2} we will consider a two-region model where one model is the empty (Milne) solution while the other region is an overdense (density larger than the critical density) matter+curvature model. In this case, the scale factors, $a_o$ and $a_u$, of the two regions can be related by a common time coordinate, $t$, by using a parameter, $\phi$ (sometimes denoted the development angle), according to
    \begin{align}
    \begin{split}
        t &= t_0\frac{\phi - \sin(\phi)}{\phi_0 - \sin(\phi_0)}\\
        a_u & = \frac{f_u^{1/3}}{\pi}(\phi - \sin(\phi))\\
        a_o & = \frac{f_o^{1/3}}{\pi}(1-\cos(\phi) ).
    \end{split}
    \end{align}
    The parameters $f_u$ and $f_o$ denote the present-time relative volume fractions of the two regions in the ensemble and are related by $f_u = 1-f_o$. As in \cite{2region1,2region2}, present time is defined to correspond to $\phi_0 = 3/2\pi$. As noted in e.g. \cite{zdrift2}, once $f_u$ or $f_o$ has been fixed, there is still one parameter which needs to be set in order to specify a two-region model uniquely. This is seen by noting that $H_D = H_u(1-v + vh)$, where $v$ is the time-dependent volume fraction of the overdense region and $h:=H_o/H_u$. Since $H_u=1/t$ we therefore have $t_0=(1-v_0+v_0h_0)/H_{D_0}$ and thus need to fix either $t_0$ or $H_{D_0}$. We set $H_{D_0} = 70$km/s/Mpc.
    \newline\newline
    Since both $Q_D$ and $R_D$ have unknown parameterizations in terms of $a_D$, we have used two types of two-region datasets for the benchmark, namely with $Q_D$ and $R_D$ as target, respectively. We first consider only a single two-region model specified by $f_o = 0.2$, yielding the datasets $(Q_D, \expval{z})$ and $(R_D,\expval{z})$. We denote these two datasets $C3a$ and $C3b$, respectively. We then vary $f_o$ as well as $z$, leading to the datasets $C3c = (Q_D, \expval{z}, f_o)$ and $C3d = (R_D, \expval{z}, f_o)$. Since $f_o$ cannot easily be related to FLRW-quantities and hence quantities we often seek to constrain through observations, we also consider the datasets $C3e = (Q_D, \expval{z}, \Omega_{m,0}, \Omega_{R,0}, \Omega_{Q,0}, H_{D_0})$ and $C3d = (R_D, \expval{z}, \Omega_{m,0}, \Omega_{R,0}, \Omega_{Q,0}, H_{D_0})$, where we have defined
    \begin{align}
    \begin{split}
        \Omega_{m,0} & :=8\pi/(3H_{D_0}^2)\rho_{D,0}\\
        \Omega_{R,0} & :=-R_{D,0}/(6H_{D_0}^2)\\
        \Omega_{Q,0} & :=-Q_{D,0}/(6H_{D_0}^2).
    \end{split}
    \end{align}
    Based on the findings in \cite{zdrift1,zdrift2}, the data points were generated using equidistant points in the intervals $\expval{z}\in[0.1,1]$ and $f_o\in[0.1,0.25]$ with $50$ points in both dimensions (giving $50^2$ points in total per dataset). We note that $H_{D_0}$ is included as a feature in these datasets even though it takes the same value in all data points. Ideally, SR algorithms should not include this feature in their results. The datasets are shown in Figure \ref{fig:C3abcd}.
    \newline\newline
    Since the ground truths of the symbolic expressions for $Q_D$ and $R_D$ in terms of the features in these datasets are unknown, we asses the success of the symbolic regression algorithms by 1) evaluating their precision on the training subspace using the mean squared error (MSE), and 2) evaluating their precision on data generated outside the parameter subspace used for training. Specifically, if a symbolic expression reproduces the data well within the training region, we also consider its performance on a test dataset corresponding to models in the larger interval $z\in[0,5]$ and $f_o\in[0.05, 0.3]$. The reasoning here is that if a symbolic expression represents the ground truth, it must also be able to reproduce data outside the region it was trained on. Evaluating the performance of SR algorithms based on the ability of the resulting symbolic expressions to extrapolate well outside the training region is only sensible because we are specifically interested in the ability of SR algorithms to identify ground truth expression which must naturally extrapolate well outside the training region. If we were merely interested in good fits/emulators within a specific region, a more prudent approach would be to make training data within the entire desired region and only evaluate performance based on this region.
    \newline\indent
    
    \begin{figure}
        \centering
        \includegraphics[scale = 0.45]{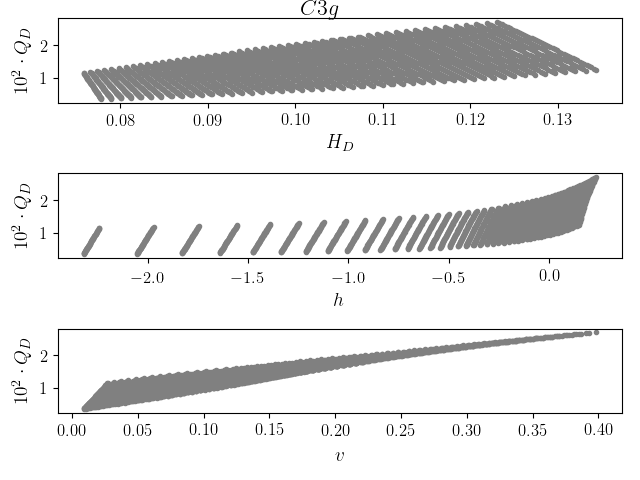}
        \includegraphics[scale = 0.45]{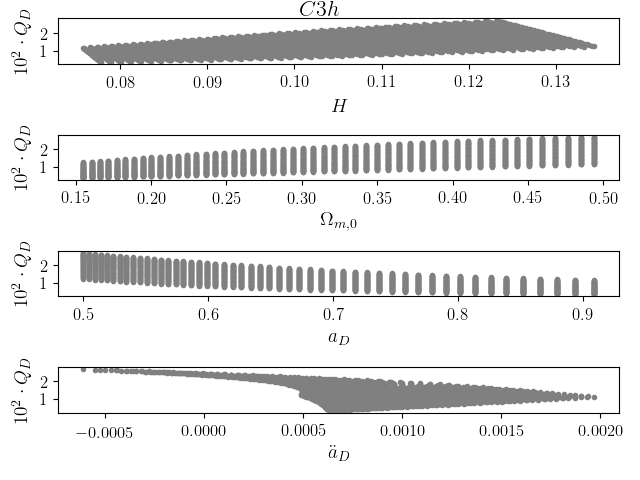}
        \caption{Left: $C3g$ shown as $Q_D$ plotted against the different features of the dataset. Right: $C3h$ shown as $Q_D$ plotted against the different features of the dataset.}
        \label{fig:Cgh}
    \end{figure} 
    
    The datasets $C3e$ and $C3f$ are plotted in Figure \ref{fig:C3abcd}. We note that the plots of $C3e$ and $C3f$ show $Q_D$ and $R_D$ to be close to linear in $\Omega_{m,0}, \Omega_{R,0}$ and $\Omega_{Q,0}$ which would indicate it reasonable to expect the SR algorithms to identify good fits within the training region (which might or might not represent ground truths).
    \newline\newline
    As detailed in \cite{2region1,2region2}, the averaged deceleration parameter of two-region models, $q_D$, can be written as
    \begin{align}
    \begin{split}
         q_D := -\frac{1}{H_D^2}\frac{\ddot a_D}{a_D}& = q_u\frac{1-v}{(1-v+hv)^2} + q_o\frac{vh^2}{(1-v+hv)^2} - 2\frac{v(1-v)(1-h)^2}{(1-v+hv)^2}\\
        & = q_o\frac{vh^2}{(1-v+hv)^2} - 2\frac{v(1-v)(1-h)^2}{(1-v+hv)^2},
    \end{split}
    \end{align}
    where the second line follows from the empty region being coasting (hence $q_u = 0$). The last term is the backreaction term due to $Q_D$. By comparing with the Equations \ref{eq:Buchert_av} we see that we can write $Q_D$ as e.g. 
    \begin{align}
    \begin{split}
    Q_D & = 6H_D^2\frac{v(1-v)(1-h)^2}{(1-v+hv)^2}\\
    &= -6\left( \frac{\ddot a_D}{a_D} + \frac{1}{2a_D^3}H_{D_0}^2\Omega_{m,0}\right). 
    \end{split}
    \end{align}
    The two above expressions for $Q_D$ are fairly simple analytical expressions. However, both expressions contain several dependent variables since $v, h, H_D, \ddot a_D$ and $a_D$ all depend on time (differently) which may make the regression task more complicated. We will therefore include $\left(Q_D, H_D, v, h \right) $ and $\left( Q_D, H_{D_0}, \Omega_{m,0}, a_D, \ddot a_D \right) $ as datasets used for our benchmark to complement the earlier datasets for the target $Q_D$ where there is no known ground truth expression. The two datasets (denoted $C3g$ and $C3h$) are shown in Figure \ref{fig:Cgh}.
    
    \begin{figure}
        \centering
        \includegraphics[scale = 0.45]{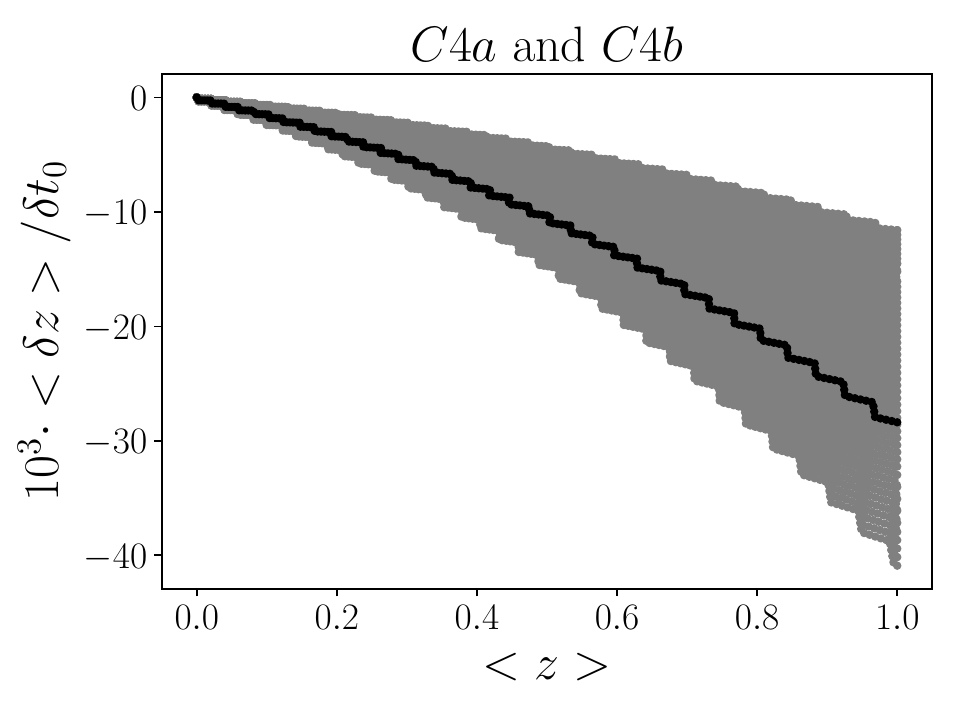}
        \includegraphics[scale=0.45]{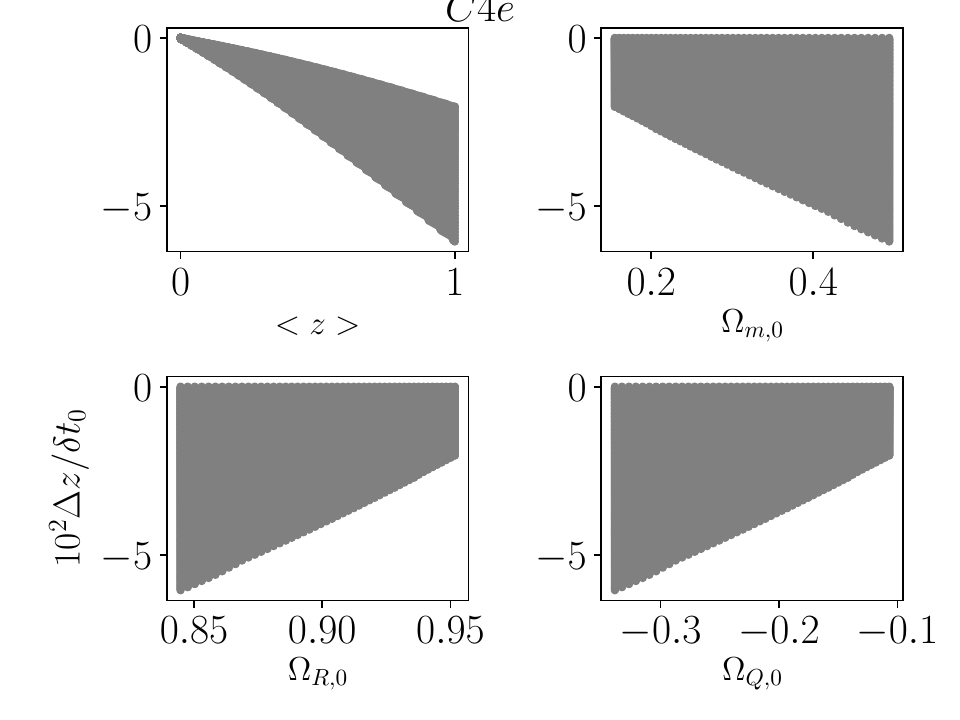}
        \includegraphics[scale = 0.5]{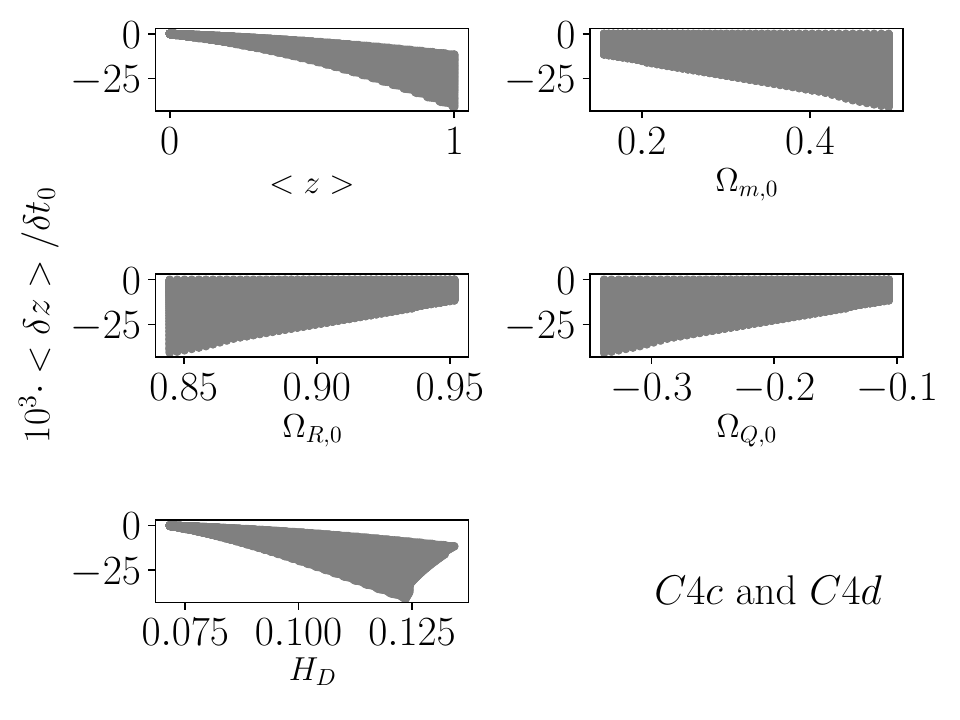}
        \caption{Top left: $\expval{\delta z}/\delta t_0$ plotted against $\expval{z}$, corresponding to $C4a$ and $C4b$. Top right: Scaled $\Delta z:=(\expval{\delta z}- \delta \expval{z})$ plotted against the features in dataset $C4e$. Bottom: $\expval{\delta z}/\delta t_0$ plotted against features in datasets $C4c$ and $C4d$.}
        \label{fig:C4cde}
    \end{figure} 
    
\subsubsection{Redshift drift in 2-region models}\label{subsubsec:zdrift_2region}
    The redshift drift is a particularly interesting observable in relation to backreaction and the backreaction conjecture because it can be used as a smoking gun signal for distinguishing between accelerated expansion due to dark energy versus apparent acceleration due to backreaction \cite{smokinggun1, smokinggun2, smokinggun3, smokinggun4}. The foundation for the smoking gun signal is that the mean redshift drift is not in general equal to the drift of the mean redshift in an inhomogeneous spacetime even if that spacetime is statistically homogeneous and isotropic (where ``mean'' observations here means observations after taking the mean over many lines of sight in order to remove statistical fluctuations). Mathematically, we can write this statement as
    \begin{align}
        \expval{\delta z}\neq \delta \expval{z} = \delta t_0[(1+\expval{z})H_{D,0}-H_D(\expval{z})],
    \end{align}
    where we set $\expval{z} +1  =1/a_D$. As discussed in \cite{zdrift1,zdrift2}, an expression for $\expval{\delta z}$ in inhomogeneous cosmological models exhibiting backreaction is unknown. The redshift drift can nonetheless be computed in concrete cosmological models exhibiting backreaction as was done in \cite{smokinggun1,smokinggun2}, with the former utilizing two-region models.
            
    The procedure for computing the redshift drift in two-region models is to arrange copies of the two FLRW regions consecutively after one another along a light ray. The regions must be arranged so that their proper distances appear in the same fraction as the proper volume fractions used for computing the volume averages (see e.g. section III a of \cite{Bull_2012} for details). As discussed in \cite{smokinggun1}, each region must be small enough for them not to evolve much during the time it takes a light ray to traverse the homogeneity scale (set by the size of two consecutive regions). The regions were here chosen to have present-time size of order $10$Mpc. Small regions of proper length $\sim 10$Mpc will lead to small, rapid fluctuations around the mean. Larger regions would lead to large deviations from the mean which would imply that a single light ray could not capture the mean behavior of the redshift drift.
    \newline\indent
    The light ray is propagated through the consecutive regions according to the equations
    \begin{align}
        \frac{dt}{dr} &= -a\\
        \frac{dz}{dr}& = (1+z)\dot a\\
        \frac{d\delta z}{dr} &= \dot a\delta z + (1+z)\ddot a\delta t\\
        \frac{d\delta t}{dr} &=-\dot a	\delta t,
    \end{align}
    where the scale factor is evaluated according to which local FLRW model the light ray is in at any given point along its path. The redshift $z$ is the local redshift along a light ray which approximates $\expval{z}$ as discussed in \cite{smokinggun1}. The resulting $z$ and $\delta z$ data only approximate their mean counterparts since they contain statistical fluctuations as discussed above. The statistical fluctuations mean that these datasets have a statistical error applied to them. This can be seen in the left plot in Figure \ref{fig:C4cde} which shows $C4a$ and $C4b$ plotted together. For a single value of $f_o$, the smoothed data that we wish to identify the ground truth for would correspond to a smooth/``best fit'' line to the black squiggly data corresponding to $C4a$. It would be possible to provide smoothed data by computing the redshift and redshift drift along many ($\sim 100-1000)$ light rays and taking the mean. However, since real data contains similar fluctuations (as well as errors from imprecise measurements), we here choose to use the dataset reflecting the statistical fluctuations.
    \newline\newline
    By solving the above equations, we generate datasets for a single two-region model with $f_o = 0.2$. From this, we generate $C4a = (\expval{\delta z}/\delta t_0, \expval{z})$. We then vary $f_o$ to obtain datasets of the form:
    \begin{itemize}
        \item $C4b = (\expval{\delta z}/\delta t_0, \expval{z}, f_o)$
        \item $C4c = (\expval{\delta z}/\delta t_0, \expval{z}, \Omega_{m,0}, \Omega_{R,0}, \Omega_{Q,0})$
        \item $C4d = (\expval{\delta z}/\delta t_0, \expval{z}, H_D(\expval{z}),\Omega_{m,0}, \Omega_{R,0}, \Omega_{Q,0})$
        \item $C4e = ((\expval{\delta z} - \delta \expval{z})/\delta t_0, \expval{z}, \Omega_{m,0}, \Omega_{R,0}, \Omega_{Q,0})$,
    \end{itemize}
    where the two latter datasets are inspired by the fact that we know that the symbolic expression for the FLRW redshift drift requires either an equation of state parameter or Hubble parameter. The datasets were generated using 50 equidistant points of $f_o\in[0.1,0.25]$ and approximately 200 values of the redshift in the interval $z\in[0,1]$\footnote{The exact number of redshift values in the dataset is difficult to control for these computations where we propagate light rays through an inhomogeneous universe and for this reason we chose, for convenience, to use equidistant points in the affine parameter rather than the redshift. This has the risk of leading to severe under-sampling in redshift space at high redshifts compared to lower redshifts. However, as long as we do not go further than $\sim z = 1$ in the training sample, the distribution in redshift space stays fairly uniform.}. Note in particular that $C4c$ and $C4e$ have the same features and that the difference is thus in the target where we in $C4e$ have subtracted the naive generalisation of the FLRW redshift drift. The datasets are plotted in Figure \ref{fig:C4cde}. It is not apparent from these plots that we gain much from subtracting $\delta\expval{z}$ from the target in $C4e$, although the plots for $C4e$ may be a bit less curved than those for $C4c$. Nonetheless, there is good physical reason to expect that a significant part of the redshift drift signal is given by the naive counterpart, and we therefore retain $C4e$ as part of our benchmark dataset.
    \newline\newline
    As with the earlier two-region model datasets, we will assess the quality of the symbolic expressions obtained by the regression algorithms by seeing how well they generalized outside the parameter subspace that the algorithms were trained on. Specifically, we make test datasets with the feature intervals $z\in[0,5]$ and $f_o\in[0.01,0.3]$.  
    
    \begin{figure}
        \centering
        \includegraphics[scale = 0.3]{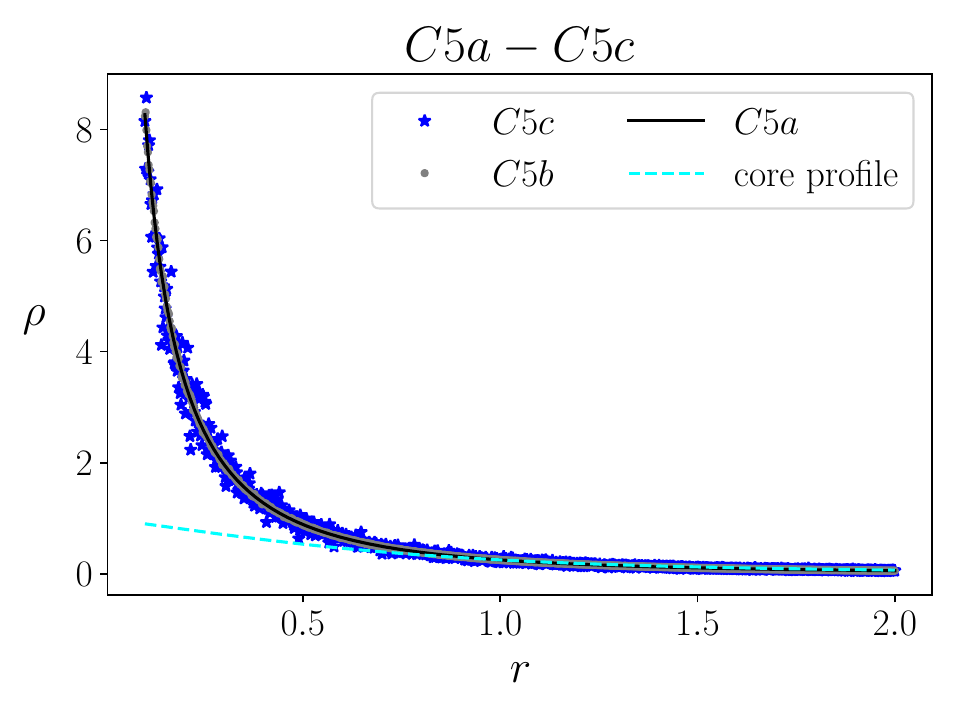}
        \includegraphics[scale = 0.3]{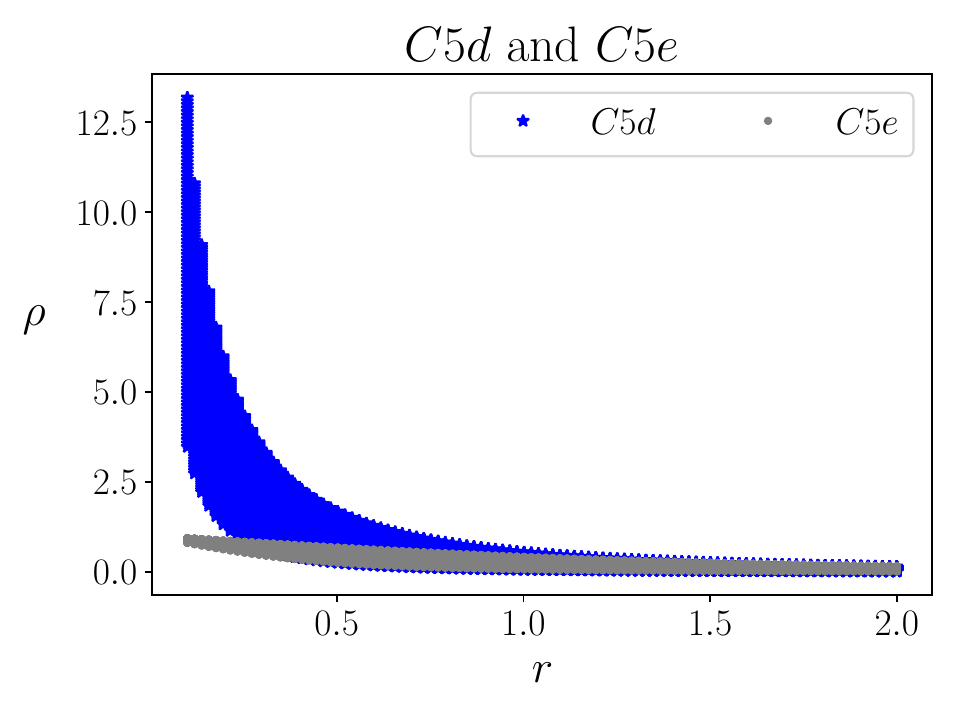}
        \includegraphics[scale = 0.3]{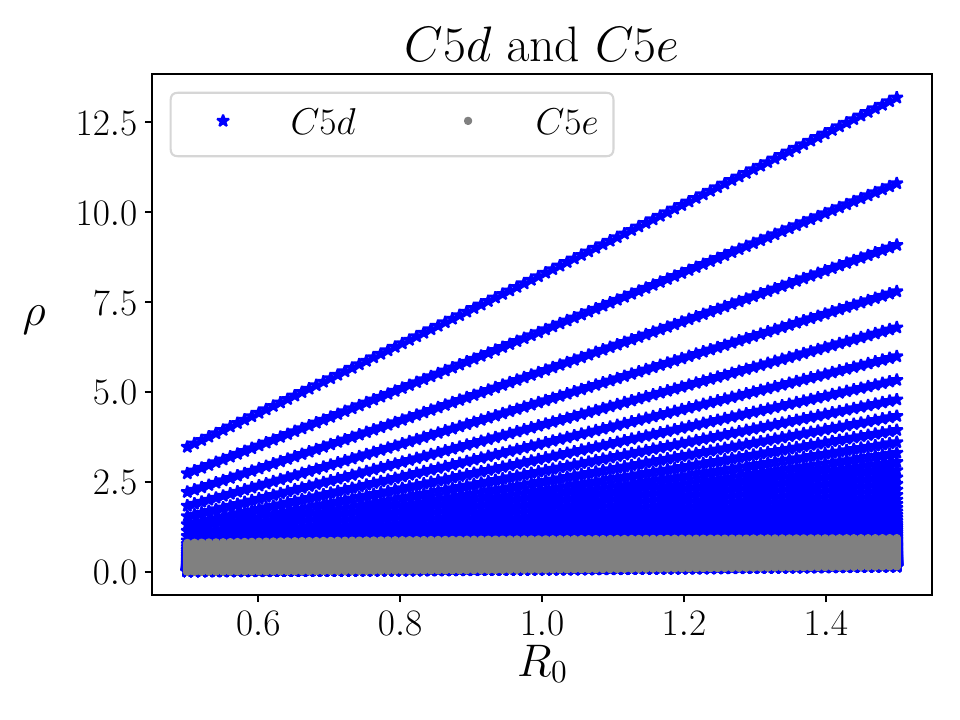}
        \caption{Datasets $C5a-e$. Note that the dataset $C5f$ would be a combination of the plots in the middle and right figure.}
        \label{fig:C5}
    \end{figure} 

\subsection{Dark matter halo profiles}
    Next after dark energy, the most dominant component in the Universe is dark matter (see e.g. \cite{Lisanti_2016} for an introduction).  A pressing task in astroparticle physics is determining the distribution of dark matter in large objects such as galaxies. Simulations based on the $\Lambda$CDM model suggest a cuspy matter density profile with an increasing density towards the core of the galaxy. Although galaxies with a cusp profile indeed have been observed, other observations of objects such as dwarf galaxies show a core type density distribution with a constant density near the core (for more details see e.g. \cite{de_Blok_2009}). The density profiles are sensitive to the specific qualities of the dark matter particles such as whether or not they interact through other forces than gravity. The study \cite{Nguyen_2020} for instance indicates that self-interactions of dark matter particles can lead to more core-like profiles within the $\Lambda$CDM model. Such interactions could thus be necessary for reconciling observations with model predictions of the dark matter density profiles. Overall, the physical origin of dark matter is still one of the biggest mysteries in fundamental physics and it is even not clear that dark matter represents new particle types and not e.g. a need for a modification of our theory of gravity -- or a combination, as discussed (and challenged) in e.g. \cite{Eriksen_2021, frandsen_2018}. Learning more about the dark matter halo profiles of galaxies is one possible step towards a better understanding of dark matter phenomenology. The observational appearance of different dark matter halo density profiles (cusp versus core) for instance prompt the question of whether it is possible to identify another underlying more general density profile that can reconcile the cuspy and core profiles discussed today. To assess the utility of symbolic regression for identifying such unknown general density profiles, we include synthetic dark matter halo density profiles in our benchmark.
\newline\indent
    We consider two dark matter halo profiles. One represents a core profile while the other represents the cusp profile. The Navarro-Frenk-White (NFW) density profile  (Equation 3 of \cite{Navarro_1996}) representing a cusp profile is given as
    \begin{align}
        \rho_{\rm NFW} \propto \frac{1}{\frac{r}{R_0}(1+r/R_0)^2}
    \end{align}
    and the density profile for a core profile is given as \cite{Burkert_1995}
    \begin{align}
        \rho_{\rm core} \propto \frac{R_0^3}{(r+R_0)(r^2+ R_0^2)},
    \end{align}
    where $R_0$ is a characteristic radius. We will set the proportionality constant to $1$ in the above relations and construct six datasets from the resulting halo profiles. The first three datasets will be with constant $R_0 = 1$ and use the NFW profile where we sample the density on a uniform grid of $N = 1000$ points with $r\in[0.1,2]$. The three datasets will differ by the first having no error and the others having 1\% and 10\% Gaussian noise, respectively. We will additionally add a classification parameter $x$ as an extra feature which we set to $1$ for the NFW profile and $-1$ for the core profile. The datasets thus take the form $C5a = (\rho_{\rm NFW}, r, x)$ and similarly for $C5b$ and $C5c$. For the remaining three datasets we will vary $R_0\in[0.5,1.5]$ and $r\in[0.1,2]$ on uniform grids with $N = 100$ in each feature dimension. We include no errors in these datasets. We make one dataset for each of the two density profiles, yielding $C5d = (\rho_{\rm NFW}, r, R_0, x = 1)$ and $C5e = (\rho_{\rm core}, r, R_0, x = -1)$. Since the classification feature has no impact on the target value in the five datasets, the SR algorithms should ideally be able to ignore this feature. However, we also consider a sixth dataset which is made by combining $C5d$ and $C5e$ to see if the SR algorithms can produce an analytical expression that faithfully represents both profiles. The resulting dataset is $C5f = (\rho_{\rm NFW/core}, r, R_0, x)$, where
    \begin{align}
        \rho_{\rm NFW/core} = \frac{1}{2}\left(\rho_{\rm NFW} + \rho_{\rm core}\right) + \frac{x}{2}\left(\rho_{\rm NFW} - \rho_{\rm core}\right).
    \end{align}
    For this dataset, the value of $x$ is crucial for the algorithms to be able to distinguish between the two profiles. Such an expression extrapolates between two different dark matter density profiles, although using a fabricated classification feature. In reality, the desire is to use other features in real dark matter halo catalogs such as the SPARC\footnote{http://astroweb.cwru.edu/SPARC/} catalog to identify a single universal expression for dark matter halo profiles. Since we are here interested in a demonstration benchmark we prefer to consider synthetic datasets with known ground truths and will therefore not consider real dark matter density profile data.
    \newline\indent
    The datasets $C5a-C5f$ are shown in Figure \ref{fig:C5}. In the figure to the left we show the datasets $C5a, C5b$ and $C5c$. The figure also shows $\rho_{\rm core}$ for the same value of $R_0$ for comparison. The two other figures show the datasets $C5d$ and $C5e$, and then $C5f$ is a combination of these two.
    
    \begin{figure}
        \centering
        \includegraphics[scale = 0.3]{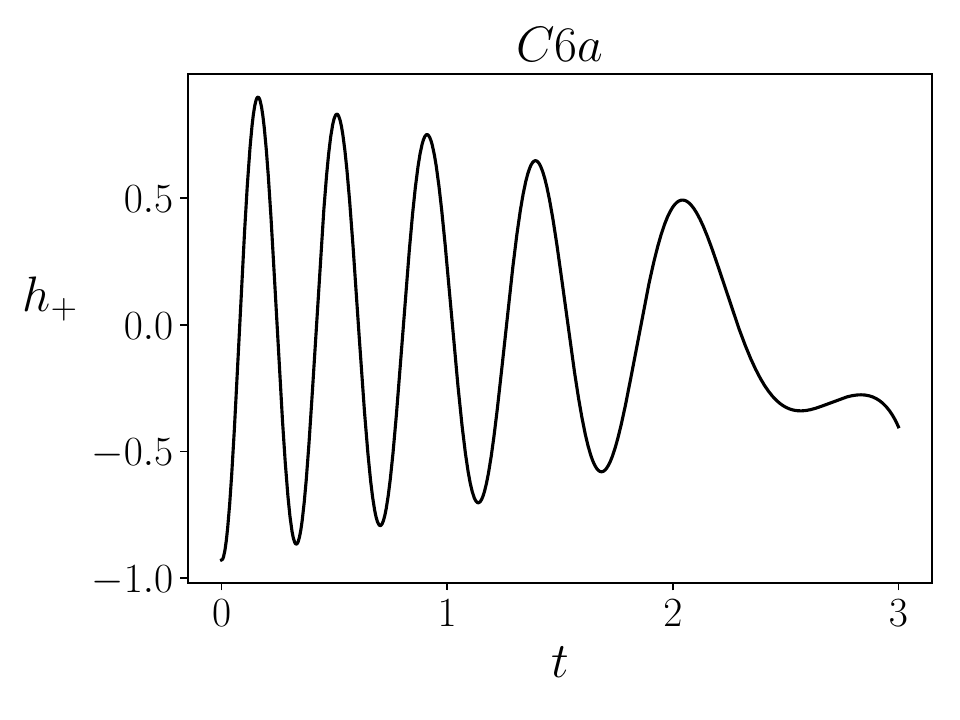}
        \includegraphics[scale = 0.3]{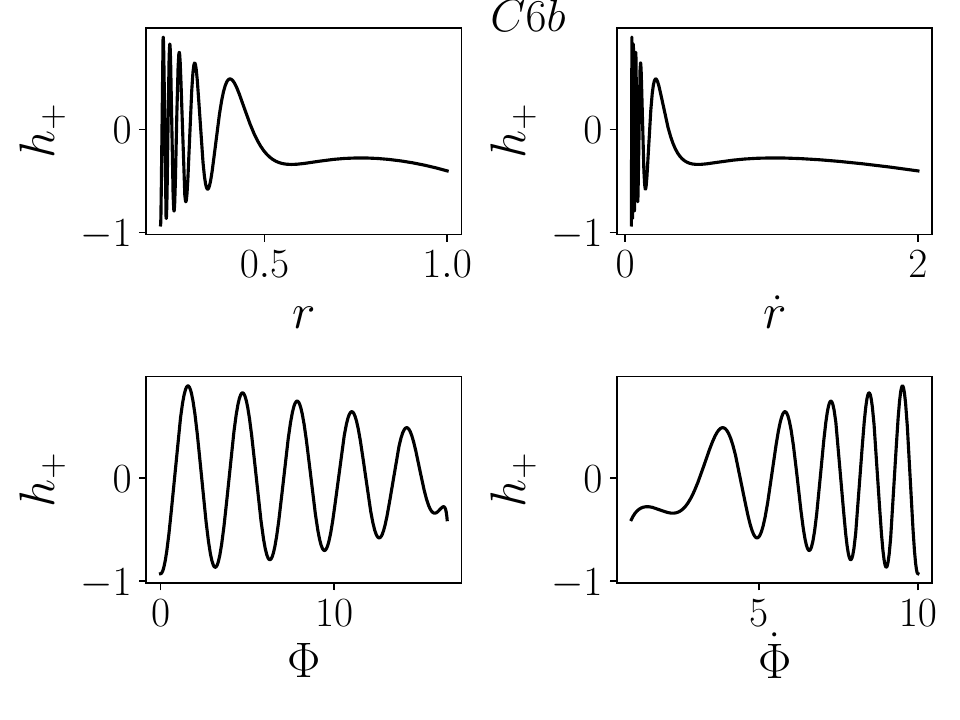}
        \includegraphics[scale = 0.3]{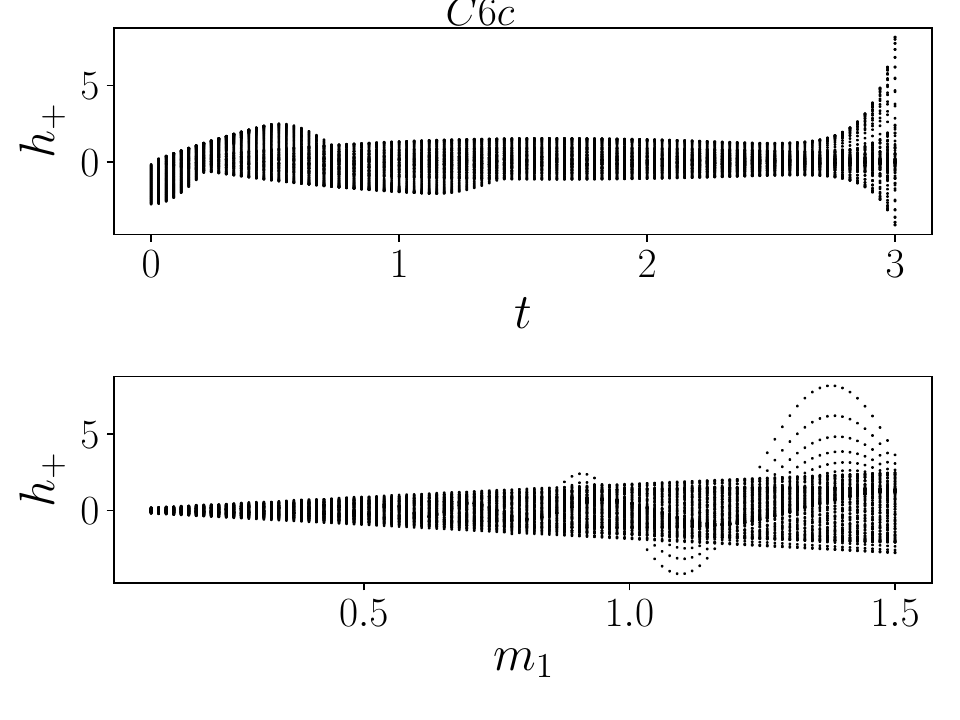}
        \caption{Datasets $C6a$, $C6b$ and $C6c$. The subfigure to the left shows the dataset $C6a$ which contains $t$ as its single feature. The middle subfigure shows $C6b$ where the target ($h_+$) is plotted against all four features. The figure to the right shows $C6c$ plotted against its two parameters $t$ and $m_1$.}
        \label{fig:C6}
    \end{figure}
    
\subsection{Gravitational wave-inspired oscillating dataset}
    Several cosmological datasets represent oscillating behavior, with prime examples being matter power spectra and the power spectrum of the CMB (cosmic microwave background) temperature fluctuations. Another well-known example of oscillating data in astrophysics is that from gravitational waves. We would therefore like our demonstration benchmark to include a dataset exhibiting oscillating behavior. However, for this demonstration we prefer datasets with known ground truths since this makes it easier to evaluate the performance of the SR algorithms. We will therefore not consider genuine CMB power spectra, matter power spectrum data or gravitational wave data as these do not represent known analytical expressions we can easily write up. Instead, we will introduce data based on an analytical expression that approximates a (very) simplified gravitational wave signal emitted from the in-spiraling of a two-body system. We construct the model by simplifying the inspiral model presented in \cite{Buskirk_2019}. In particular, we consider the plus polarization mode (setting $c = G = 1$)
    \begin{align}
        h_{+} = -\frac{2M\eta}{R}\left[ \left(-\dot r^2 + r^2\dot \Phi^2 + \frac{M}{r}\right)\cos(2\Phi) +2r\dot r\dot \Phi\sin(2\Phi)  \right],
    \end{align}
    where $R$ is the distance from the binary to the detector, $M= m_1 + m_2$ is the total mass of the binary, $\eta = m_1m_2/M^2$, $r$ is the orbital separation between the two bodies, $\Phi$ is the phase of the gravitational wave signal, and dots denote time derivatives. This equation is supplemented by the auxiliary equations 
    \begin{align}
        M\frac{d\Phi}{dt} = v^3
    \end{align}
    and
    \begin{align}
        \frac{dv}{dt} = -\frac{M}{r^2},
    \end{align}
    where $v$ is the orbital velocity. In \cite{Buskirk_2019}, the equation for $dv/dt$ is provided to several post-Newtonian orders, leading to numerical solutions for the gravitational waves. In order to best assess if some SR algorithms perform significantly better/worse on oscillating datasets than monotonic datasets, we here wish to construct datasets with known analytical ground truths. To obtain this we will construct mock gravitational wave data  by making the (non-physical) ansatz
    \begin{align}
        v(t) = \left(v_0-3t\right)^{1/3},
    \end{align}
    where we ignore units as we will eventually scale the target to take on values close to 1 anyway. Using this ansatz, we find
    \begin{align}
        \begin{split}
            M\dot \Phi = -3t+v_0\implies M\Phi=-\frac{3}{2}t^2+v_0t + \Phi_0.
        \end{split}
    \end{align}
    Furthermore, from Kepler's third law we have (see \cite{Buskirk_2019}) $r = M/v^2$ from which we obtain
    \begin{align}
        \begin{split}
     r &= M\left( -3t + v_0 \right)^{-2/3}\\ 
     \dot r &= 2M\left(-3t + v_0\right)^{-5/3}.
        \end{split}
    \end{align}
    Using the above with $\Phi_0=0$, we can write
    \begin{align}
    \begin{split}
        h_+ &= -\frac{2M\eta}{R}\Big[ \left(-4M^2\left(-3t + v_0\right)^{-10/3} + 2\left( -3t+v_0 \right)^{2/3}\right)\cos\left( -3t^2/M + 2v_0t/M \right) \\
        &+ 4M\left( (-3t + v_0) \right)^{-4/3}\sin\left( -3t^2/M + v_0t/M \right)\Big].
    \end{split}
    \end{align}
    We do {\bf \em not} claim that the resulting expression for $h_+$ is particularly realistic\footnote{In fact, our ansatz for $v(t)$ is inconsistent with $r = M/v^2$. However, the ansatz leads to oscillating behavior of the type we are interested in here and since we are merely interested in mock data, we will not concern ourselves with the non-consistency of the solution.}, but it nonetheless provides us with an oscillating analytical expression {\em inspired} by genuine cosmological/astrophysical data. We specifically make 3 datasets based on this data. The first dataset will be for a single binary system where we fix $m_1=0.5 = m_2$, $R_0 = 5$ and $v_0 = 10$. These values were chosen somewhat adhoc, keeping in mind that we prefer obtaining target values of the order close to $1$. We then make a dataset by choosing a uniform grid for $t\in[0,3]$ and $N = 1000$ (we neglect units since we do not consider this a realistic model and all parameter values are chosen out of convenience rather than physical meaningfulness). We will consider both $C6a = (h_+, t)$ and $C6b = (h_+, r, \dot r, \Phi, \dot\Phi)$. While the latter expression is analytically simpler, the former expression corresponds more closely to a real signal, since time is one of the observables we can directly measure. The last dataset, $C6c$, is similar to $C6a$ but we now vary the extra model parameter $m_1$ so that the dataset becomes $C6c = \left(h_+, t, m_1\right)$. For generating this dataset we choose $t\in[0,3]$ and $m_1\in[0.1,1.5]$ and $N = 100$ in both dimensions.
    \newline\indent
    The datasets in group $C6$ are shown in figure \ref{fig:C6}.
    
\subsection{Summary of datasets}\label{subsec:data_summary}
    The cosmological datasets used for our benchmark are summarized in Table \ref{table:C}. Note that some targets are scaled to be of the order 1-10 in order to improve the symbolic regression. The feature intervals used for the benchmark are those indicated in the table. For datasets with unknown ground truth we assess the output from the symbolic regressions algorithms by comparing with a test dataset where the feature intervals are broadened. If the obtained symbolic expressions signify (the hitherto unknown) ground truths, they must perform well not only within the training interval but also in the test data representing the broadened parameter spaces. Otherwise, a symbolic expression with small mean square error (MSE) on the training data can simply represent a good (over-)fit without any (clear) physically relevant information. For the test data without known ground truths we use $z\in[0,5]$ and $f_o\in[0.01,0.3]$ to test the symbolic expressions outside the training regions.
    \newline\indent
    Several datasets correspond to the same setup and to the same data but with different features provided in the datasets. We therefore split our data into several groups, with each group corresponding to a specific physical setup. For instance, the group $C1$ contains the datasets $C1a, C1b, C1c$ and $C1d$. Below, we list the dataset groups and summarize what questions we wish to answer/qualities we wish to demonstrate by using each group.
    \begin{itemize}
        \item Group $C1$ consisting of datasets $C1a, C1b, C1c, C1d$: Dataset $C1a$ represents what should be a fairly simple regression task, where the target variable depends on a single feature with the ground truth being a simple rational function. We would thus expect most SR algorithms to be able to identify this expression. In $C1b$ we have added noise and by comparing the SR algorithms' performances on $C1a$ versus $C1b$ we wish to demonstrate how much the added error affects this performance. The final datasets, $C1c$ and $C1d$, are the same as $C1a$ but augmented to include more cosmological models and hence represents datasets with more features. In $C1c$ we vary one model parameter while we vary two model parameters in $C1d$. By comparing the performance of SR algorithms on $C1a$ versus $C1c$ and $C1d$ we wish to demonstrate how much the dimension of the feature/parameter space affects the performance of the SR algorithms. Note that the extra parameter in $C1d$ compared to $C1c$ is $H_0$ which simply gives an overall scaling of $H(z)$ which means that the overall complexity of the final expression for the two datasets are very similar. This means that we can be fairly certain that any difference in performance on dataset $C1c$ versus $C1d$ is specifically related to the dimension of the feature space and not complexity of the ground truth.
        \item Group $C2$ consisting of $C2a, C2b$: These two represent the redshift drift for a family of FLRW models. The two datasets represent the same data but in $C2a$ the target is parameterized by two features which are inter-dependent while in $C2b$ the features are independent. With these datasets we wish to demonstrate whether or not the interdependence of features is important for the performance of SR algorithms.
        \item Group $C3$ consisting of $C3a-C3h$: These datasets all represent backreaction for the two-region toy-models. For $C3a-C3f$, the ground truth of the target ($Q_D$ or $R_D$) is unknown. We include these datasets in our demonstration benchmark partially to showcase datasets where SR could potentially be utilized to reveal a new physical relation. We also include this dataset to highlight the difficulties one faces when seeking to use SR to reveal ground truth relations. The first two datasets in this group, $C3a$ and $C3b$, represent the simpler situation where only a single two-region model is considered and hence where the target depends only on a single parameter/feature. The two datasets differ by their target since we do not have prior knowledge regarding whether it it easier to obtain a symbolic expressions for one or the other. The datasets $C3c-C3f$ correspond to data where we also vary the model parameter $f_o$ so that the datasets $C3c$ and $C3d$ both have two features and where, again, we make two datasets to have both $Q_D$ and $R_D$ as targets in case it turns out to be easier for the SR algorithms to obtain the ground truth expression for one over the other. Datasets $C3e$ and $C3f$ represent the same datasets (i.e. same model parameters and same redshift) but where we supply other features than $f_o$. Specifically, we introduce features similar to those used for FLRW spacetimes. Based on the dynamical equations in Equation \ref{eq:Buchert_av} we expect that relations between the target and provided features exist. However, we note that one of the features in these two datasets, namely $H_{D_0}$ takes the same value at each data point in these two datasets. It will therefore be interesting to see if this parameter is included in the expressions reported by the SR algorithms since optimally this should not be the case. Finally, datasets $C3g$ and $C3h$ are included because the targets are here chosen such that we actually know the ground truth expression of the target in terms of the provided features. However, the features are inter-dependent which we suspect may compromise the SR algorithms' ability to identify the ground truth expressions.
        \item Group $C4$ consisting of $C4a-C4e$: As $C3$, this group of datasets represent unknown ground truths, this time of a direct observable (the redshift drift). We chose to introduce two groups of datasets representing ground truths to give the SR algorithms more opportunities to identify unknown ground truths. Dataset $C4a$ represents data for a single two-region model and hence contains only a single feature. The dataset $C4b$ represents data for a family of two-region models, parameterized by $f_o$. Datasets $C4c$ and $C4d$ represent the same data parameterized by different features. Based on the analytical considerations of e.g. \cite{smokinggun3}, we expect that is may be possible to express the target through some (or all) of the supplied features. The last dataset in this group, $C4e$, represents the same dataset as $C4d$, but where the target has been rewritten in terms of the FLRW limit of the general symbolic expression/ground truth we are seeking. All datasets in this group have a statistical error. This error comes from the method used for generating data, where we consider a single light ray in an inhomogeneous spacetime rather than smoothing over $\sim 100-1000$ light rays to remove statistical fluctuations. Because of this statistical error, we expect that the SR algorithms are more likely to identify ground truths for the dataset $C3$ (which has no error) than for $C4$.
        \item Group $C5$ consisting of $C5a-C5e$: This group represents data with known ground truths. As demonstrated in $C3$ and $C4$, when we do not know the ground truth, we also do not know what parameters/features to add to our datasets. It is therefore likely that we will add features that are not relevant for the target. With this dataset, we wish to see to what extent the extra features affect the SR algorithms' performance by adding the superfluous classification feature $x$ to the datasets $C5a-C5d$. To highlight the importance of data precision we also consider versions of $C5a$ with 1\% and 10\% Gaussian noise. We lastly add $C5e$ which is simply the combination of $C5a$ and $C5d$. The corresponding analytical expression of the target extrapolates between the two density profiles. Such an extrapolation is of interest for dark matter halo research and the task thus resembles a possible genuine SR task within this field. 
        \item Group $C6$ consisting of $C6a-C6c$. This dataset represents a simplified gravitational wave dataset which is first of all considered in order to assess how well the considered SR algorithms perform on datasets that exhibit oscillation compared to the more monotonous behavior of the other datasets we consider. $C6a$ contains a single feature, $t$, and represents a fairly complicated analytical expression. $C6b$ represents the same data points, but rather than having $t$ as feature, we introduced four features that are all functions of $t$ but which can be combined into a simpler analytical expression. This dataset is thus similar to $C2a$ where the two features are interdependent but combine to form the target through a relatively simple analytical expression. Lastly, $C6c$ is similar to the dataset $C6a$ except that we now vary an extra model parameter ($m_1$) to see how much the extra feature affects the performance of the SR algorithms.
        
    \end{itemize}
        
\begin{table}[!htb]
    \centering
    \begin{adjustbox}{max width=\textwidth}
    \begin{tabular}{c c c c}
    \hline\hline
    Dataset name & Dataset & Data interval &  Error\\
    \hline
    \Tstrut
    $C1a$ & $\left(H|\,\,\,\, z\right)$ & $z\in[0.1,2]$, $N = 1000$ & \xmark \\
    $C1b$& $\left(H|\,\,\,\, z\right)$ &$z\in[0.1,2]$, $N = 1000$ & $10\%$\\
    $C1c$& $\left( H|\,\,\,\, z, \Omega_{m,0} \right)$&$z\in[0.1,2], \Omega_{m,0}\in[0.1,0.5]$, $N = 100$ & \xmark  \\
    $C1d$& $\left( H|\,\,\,\, z, H_0, \Omega_{m,0} \right)$ & $z\in[0.1,2], \Omega_{m,0}\in[0.1,0.5], h\in[20,100]$, $N = 50$ &  \xmark  \\\\
    $C2a$& $\left( \delta z/\delta t_0|\,\,\,\, z, H(z,\omega) \right)$& $z\in[0.1,2], \omega\in[-0.99,1/3]$, $N = 100$ & \xmark\\
    $C2b$& $\left( \delta z/\delta t_0|\,\,\,\, z, \omega\right)$&$z\in[0.1,2], \omega\in[-0.99,1/3]$, $N = 100$  & \xmark\\\\
    $C3a$ &$\left( 10^2\cdot Q_D|\,\,\,\, \left<z\right>\right)$ &$z\in[0.1,1], N = 1000$& \xmark\\
    $C3b$ &$\left( 10^8\cdot R_D|\,\,\,\, \left<z\right>\right)$ &$z\in[0.1,1], N = 1000$& \xmark\\
    $C3c$&$\left( 10^2\cdot Q_D|\,\,\,\, \left<z\right>, f_o\right)$ &$z\in[0.1,1], f_o\in[0.1,0.25], N = 50$& \xmark\\
    $C3d $&$\left( 10^8\cdot R_D|\,\,\,\, \left<z\right>, f_o\right)$ &$z\in[0.1,1], f_o\in[0.1,0.25], N = 50$& \xmark\\
    $C3e$ & $\left(10^2\cdot Q_D|\,\,\,\, \left<z\right>, \Omega_{m,0}, \Omega_{R,0}, \Omega_{Q,0}, H_{D_0} \right)$ &$z\in[0.1,1], f_o\in[0.1,0.25], N = 50$& \xmark\\
    $C3f$ &$\left( 10^8\cdot R_D|\,\,\,\, \left<z\right>, \Omega_{m,0}, \Omega_{R,0}, \Omega_{Q,0}, H_{D_0} \right)$&$z\in[0.1,1], f_o\in[0.1,0.25], N = 50$ &  \xmark\\
    $C3g$ & $ \left( 10^2\cdot Q_D|\,\,\,\, H_D, v, h \right)$ &$z\in[0.1,1], f_o\in[0.1,0.25], N = 50$& \xmark\\
    $C3h$& $\left(10^2\cdot Q_D|\,\,\,\, H_{D_0}, \Omega_{m,0}, a_D, \ddot a_D \right)$ &$z\in[0.1,1], f_o\in[0.1,0.25], N = 50$& \xmark\\\\
    $C4a$ &$\left( 10^3\cdot \left<\delta z \right>/\delta t_0|\,\,\,\, \left<z\right>\right)$ &$z\in[0,1], f_o\in[0.1,0.25], N = 50$& (\xmark)\\
    $C4b$&$\left( 10^3\cdot\left<\delta z \right>/\delta t_0|\,\,\,\, \left<z\right>, f_o\right)$ &$z\in[0,1], f_o\in[0.1,0.25], N = 50$& (\xmark)\\
    $C4c$& $\left( 10^3\cdot\left<\delta z \right>/\delta t_0|\,\,\,\, \left<z\right>, \Omega_{m,0}, \Omega_{R,0}, \Omega_{Q,0}\right)$ &$z\in[0,1], f_o\in[0.1,0.25], N = 50$ &(\xmark)\\
    $C4d$& $\left( 10^3\cdot\left<\delta z\right>/\delta t_0|\,\,\,\, \left<z\right>, H_D, \Omega_{m,0}, \Omega_{R,0}, \Omega_{Q,0}  \right)$ &$z\in[0,1], f_o\in[0.1,0.25], N = 50$& (\xmark)\\
    $C4e$ & $\left( 10^2\cdot\left(\left<\delta z\right> - \delta\left<z\right>\right)/\delta t_0|\,\,\,\, \left<z\right>, \Omega_{m,0}, \Omega_{R,0}, \Omega_{Q,0}  \right)$ &$z\in[0,1], f_o\in[0.1,0.25], N = 50$& (\xmark)\\\\
    $C5a$ & $\left( \rho_{\rm NFW}|\,\,\,\, r, x \right)$ & $r\in[0.1,2], N = 100$, $x = 1$ & \xmark\\
    $C5b$ & $\left( \rho_{\rm NFW}|\,\,\,\, r, x \right)$ &  $r\in[0.1,2], N = 100$, $x = 1$ & 1\%\\
    $C5c$ & $\left( \rho_{\rm NFW}|\,\,\,\, r, x \right)$ &  $r\in[0.1,2], N = 100$, $x = 1$ & 10\%\\
    $C5d$ & $\left( \rho_{\rm NFW}|\,\,\,\, r, R_0, x \right)$ &$R_0\in[0.5,1.5]$, $r\in[0.1,2], N = 100$, $x = -1$& \xmark\\
    $C5e$ & $\left( \rho_{\rm core}|\,\,\,\, r, R_0, x \right)$ &$R_0\in[0.5,1.5]$, $r\in[0.1,2], N = 100$, $x = -1$& \xmark\\
    $C5f$ & $\left( \rho_{\rm NFW/core} |\,\,\,\, r, R_0, x \right)$ &$R_0\in[0.5,1.5]$, $r\in[0.1,2], N = 100, x = \pm1$& \xmark\\\\
    $C6a$ & $\left( h_+ |\,\,\,\, t \right)$ &$t\in[0,3], N = 1000$& \xmark\\
    $C6b$ & $\left( h_+ |\,\,\,\, r,\dot r, \Phi, \dot\Phi \right)$ &  $t\in[0,3], N = 1000$& \xmark\\
    $C6c$ & $\left( h_+ |\,\,\,\, t, m_1 \right)$ &$t\in[0,3], m_1\in[0.1,1.5], N = 100$& \xmark\\
    \hline
    \end{tabular}
    \end{adjustbox}
    \caption{Summary of cosmological datasets used in the demonstration benchmark. In the rightmost column we indicate whether the given dataset has been attributed a Gaussian distributed 10 \% error, 1\% error or no error (indicated by an \xmark). As discussed in Section \ref{subsubsec:zdrift_2region}, datasets based on redshift drift in two-region models have an un-quantified error which we indicate by (\xmark). $N$ indicates the number of equidistant points along each feature dimension (so for a $d$-dimensional dataset with $N$ points in each dimension, the total number of points in the dataset would be $N^d$). For the two-region redshift drift data, the data points are not exactly equidistant along the $z-$dimension. In addition, $N$ for the $z-$interval in these datasets are approximately 200 while the $N$ indicated in the table is for $f_o$ only. Datasets are indicated as $(\rm target| features)$. The range is always given for the parameter that was varied even though the selected features may be different. For instance, $C6b$ was obtained by varying $t$ but $t$ was not itself used as a feature in this dataset where instead we use the features $r,\dot r, \Phi, \dot\Phi$.}
\label{table:C}
\end{table}
    
\section{Benchmarking with cosmological datasets}\label{sec:benchmark}
    In this section we provide the details of cp3-bench and our demonstration benchmark. We begin by giving an overview of the 12 algorithms included in the benchmark and discussing their hyperparameter tuning.\newline\newline\noindent
    \begin{table}
        \centering
        \begin{tabular}{ |c|c|c|c|c| }
            \hline
            \multicolumn{5}{|c|}{Algorithm List} \\
            \hline
            Algorithm Name & Ref & Method & Multi-core & Year \\
            \hline
            AI-Feynman & \cite{AIFeynman1, AIFeynman2}  & PI,NN & No & 2020\\
            Deep Symbolic Optimization (DSO) & \cite{DSO1, DSO2}  & RNN,RL & Yes & 2021\\
            Deep Symbolic Regression (DSR) &  \cite{DSR} & RNN,RL,GP & Yes & 2021\\
            Fast Function Extraction (FFX)  & \cite{FFX}  & GP & Yes & 2011\\
            Genetic Engine (GE) & \cite{GeneticEngine} & GP &  No & 2022\\
            GPG & \cite{gpg} & GP &  Yes & 2022 \\
            GPZGD & \cite{GPZGD} & GP &  Yes & 2020 \\
            ITEA & \cite{ITEA} & GP & Yes & 2019\\
            Operon & \cite{operon} & GP &  Yes & 2020 \\
            PySR & \cite{cranmer2023interpretable}   & GP,NN & Yes & 2023\\
            QLattice Clinical Omics (QLattice) & \cite{QLattice} & GP & Yes & 2022\\
            Unified Deep Symbolic Regression (uDSR) & \cite{DSO1, uDSR}  & RNN,RL & Yes & 2022\\
            \hline
        \end{tabular}
        \caption{Algorithms considered in this paper including the ML methods they use, if they support multi-core in the current implementation and the year of their release. We denote Physics Inspired as PI, Neural Network as NN, Recurrent Neural Network as RNN, Reinforcement Learning as RL, and Genetric Programming as GP. AI-Feynman predominantly runs single-core. DSO and uDSR are not using GP methods due to a limitation of the current implementation of cp3-bench (cp3-bench uses the scikit-learn Python wrapper but as stated in the documentation for DSO and uDSR, GP is not supported when using this wrapper). }
        \label{table:alg}
    \end{table}
    We bench the symbolic regression algorithms/models listed in Table \ref{table:alg}. We assembled this collection of algorithms based on the availability of algorithms and a wish to consider algorithms based on a variety of methods. Each model (except FFX) has several hyperparameters and other features which can be tuned. As described below, we have tuned the hyperparameters using a set of seven standard datasets with one to three features each. These datasets were chosen from \cite{makke2023interpretable} among standard datasets used for tuning hyperparameters of symbolic regression algorithms and are listed in Appendix \ref{app:StandardDataset}. Tuning hyperparameters on a generic test dataset rather than the cosmological datasets described above is a simple way of avoiding biasing for/against any of the algorithms or datasets. We simply tune using standard datasets used in the symbolic regression community and which have no relation to cosmological datasets (see Section \ref{sec:procedure} for more details on the hyperparameter tuning). The tuned hyperparameters can be found in the publicly available code where all the algorithms are in the \texttt{methods} folder and hyperparameters are set in \texttt{procedure.py} (see Figure \ref{fig:architecture} for details). Note that the algorithms DSO and uDSR in principle support Genetic Programming (GP) methods but the interface we use currently does not support this feature.

    Regarding the algorithms we note that DSR and uDSR are further developments of DSO. From the test dataset we find that these algorithms are in some cases indeed able to find the correct symbolic expressions corresponding to the datasets. FFX is fast, has no hyperparameters and would only be able to provide the correct symbolic equation in very limited cases where only multiplication, polynomials, and division is used. GPZGD and Genetic Engine (GE) were hard to tune and the former did not recover any of our test equations. GPG is reasonably fast but struggles with rather simple equations. AI-Feynman is rather slow, but was able to find one correct symbolic expression for the test data. Of the purely genetic algorithms Operon, QLattice, and ITEA perform the best on the test data. PySR is also one of the more successful algorithms according to the training done on the test datasets.
    \newline\newline
    In the following subsections we detail the procedure of benchmarking with cp3-bench and Things-to-bench. We first introduce the framework for cp3-bench and installation options before moving on to discussing the specific procedure (rules and criteria) we applied in our demonstration benchmark.

    We are aware of the existence of the benchmark tool SRBench \cite{lacava2021contemporary} and have used it as inspiration. The main guide for designing our tool has been to make it as user-friendly as possible both in terms of implementation and adding extensions. Since SRBench has come to represent the standard with which SR algorithms are benched by developers, we compare cp3-bench and SRBench at the end of the section.
    
    \subsection{Benchmarking method}
        In this subsection we discuss the benchmark tools developed for this paper and the procedure we have used to achieve the results presented in Section \ref{sec:benchmark-results}.

        \subsubsection{Framework}
            As part of this paper we release the code cp3-bench and Things-to-bench used for the benchmarking presented in this paper. The release contains two components. Firstly, we are releasing the benchmark engine which we call {\bf\em c}osmology and {\bf\em p}article {\bf\em p}hysics {\bf\em p}henomenology benchmark (cp3-bench). Secondly, we release our Things-to-bench wrapper repository showcasing how one can leverage the benchmark tool in an easy way, and create a simple script to loop over datasets of interest. A user can also use cp3-bench as project folder and put in the datasets and utility scripts straight into the cp3-bench folder and build a Docker image on that. The disadvantage of this is that changing versions of cp3-bench is less convenient as cp3-bench would not be separated from the datasets and other scripts in this case. The latter repository contains the datasets used for this work, the code used to manage the benchmark of them, as well as the raw results including Mathematica notebooks used to simplify and prepare the data for presentation in Appendix \ref{app:full-results}. The code can be found at:
            \begin{itemize}
                \item cp3-bench (v1.1.1): \url{https://github.com/CP3-Origins/cp3-bench}
                \item Things-to-bench: \url{https://github.com/CP3-Origins/Things-to-bench}
            \end{itemize}

            \begin{figure}
                \centering
                \tikzset{doc/.style={document,fill=blue!10,draw,thin,minimum
                height=1.2cm,align=center}}
                \begin{tikzpicture}[font=\sffamily,every label/.append
                    style={font=\small\sffamily,align=center}]
                    \node[tape, draw,thin, tape bend top=none,fill=purple,
                    text=white,minimum width=2.2cm,double copy shadow,minimum height=1.5cm]
                    (Datasets) {Datasets};
                    \node[above=1cm of Datasets, doc] (main) {main.py};
                    \node[draw,dashed,rounded corners,fit=(Datasets) (main),inner
                    xsep=10pt,inner ysep=30pt,label={above:{Things-to-bench}}](fit1){};
                    \node[right=4.5cm of main, doc] (evaluate) {evaluate\_dataset.py};
                    \node[right=1cm of evaluate, doc, fill=green!30] (result) {results/result.csv};
                    \node[below=1.5cm of evaluate, parallelepiped,draw=red,fill=red!80,
                      minimum width=2cm,minimum height=1.5cm,align=center,text=white]
                      (Algorithms) {methods};
                    \node[left=1.9cm of Algorithms, doc] (install) {install.py};
                    \node[draw,dashed,rounded corners,fit=(Algorithms) (evaluate) (result)  (install),inner
                    xsep=10pt,inner ysep=30pt,label={above:{cp3-bench}}](fit2){}; 
                    \node[below=2cm of fit2, doc] (procedure) {procedure.py};
                    \node[right=0.5cm of procedure, doc] (init) {\_\_init\_\_.py};
                    \node[left=0.5cm of procedure, doc, fill=black!15] (requirements) {requirements.txt};
                    \node[left=0.5cm of requirements, doc, fill=purple!25] (config) {config.json};
                    \pic[right=1.1cm of init,local bounding box=folder,scale=0.5] (folder) {folder};
                    \node[below=1mm of folder,font=\small\sffamily,align=center]{tests};
                    \node[draw,dashed,rounded corners,fit=(config) (procedure) (folder),inner
                    xsep=10pt,inner ysep=30pt,label={below:{Algorithm folder}}](fit3){};
                    \draw[-latex] (Datasets) -- (main);
                    \draw[-latex] (main) -- (evaluate) node[midway,below,font=\small\sffamily]{/bench};
                    \draw[-latex] (install) -- (Algorithms) node[midway,below,font=\small\sffamily]{CLI};
                    \draw[-latex] (evaluate) -- (Algorithms) node[midway,left,font=\small\sffamily]{CLI};
                    \draw[-latex] (Algorithms) -- (evaluate);
                    \draw[-latex] (evaluate) -- (result);
                    \draw(fit3) -- (Algorithms.south) node[pos=.22,left,font=\small\sffamily]{pyenv virtualenv};
                \end{tikzpicture}
                \caption{Architecture diagram showcasing the interaction between Things-to-bench and cp3-bench. The diagram contains an exploded view of the algorithm folder structure running in a \texttt{pyenv} virtual environment.}
                \label{fig:architecture}
            \end{figure}
            
            Things-to-bench represents an example of interacting with the benchmark engine, cp3-bench, and it is used for inputting datasets of interest into the benchmark engine as shown in Figure \ref{fig:architecture}. In this example, the interaction consists of importing the \texttt{evaluate\_dataset.py} module found in the \texttt{bench} folder of cp3-bench. This is the main module for benchmarking.

            We release cp3-bench with 12 algorithms implemented and readily available for use. Adding new algorithms is straightforward by creating a new folder in the \texttt{methods} folder with the structure described in Figure \ref{fig:architecture}. The program uses the folders in \texttt{methods} to recognize which methods are implemented. The default is to install all models when running the module \texttt{install.py}. Both modules can be used by other scripts or run with CLI (command-line interface) and have the optional flag \texttt{---methods} which allows the user to select specific algorithms to install or benchmark. The install module also allows for the optional flag \texttt{---reinstall} which by default is false, but can be chosen to be true to reinstall a misbehaving algorithm. The module \texttt{evaluate\_dataset.py} requires an argument specifying the path to a \texttt{.csv} dataset to evaluate. This flag is called \texttt{---data\_path}.
            
            Each algorithm runs in a \texttt{pyenv} virtual environment allowing each algorithms to run a specific version of Python with packages and dependencies installed specifically to the needs of the given algorithm.

            The \texttt{tests} folder contains the configuration of universal tests, which each algorithm should pass. During installation these are used to verify the functioning of the algorithms. A summery of the installation is logged in a file called \texttt{STATUS.md} in the \texttt{bench} folder which indicate if there were any failures during the installation. You can verify that everything is installed correctly by checking this file. There are four columns to consider:
            \begin{itemize}
                \item \textbf{Config:} Checks the configuration is valid
                \item \textbf{Environment:} Checks the environment is correctly setup
                \item \textbf{Installation:} Checks for successful installation of installed components
                \item \textbf{Tests:} Checks that all tests pass successfully
            \end{itemize}
            If a column is marked \texttt{[x]} then this step passed while \texttt{[-]} means failure. Consider the reinstall flag to try to reinstall the failed installation attempt for a given algorithm.

            In the \texttt{requirements.txt} file, Python dependencies of the package are specified. The \texttt{config.json} file contains the name of the algorithm, the key, which must match the folder name of the algorithm, the python version to use for creating the virtual environment with \texttt{pyenv}, and install commands which specifies any additional steps to perform in order to install the algorithm. 
            
            The most important file is the \texttt{procedure.py} script. Here the algorithm is imported and defined according to the framework. It is also where the hyperparameters are set. The script defines a class which initializes the algorithm and defines a procedure for training this algorithm. Note that some algorithms have a \texttt{format\_output} function to reformat the output equation. Some implementations may also drop coefficients that are small or set small numbers to zero and round numbers to fewer significant digits to produce a shorter equation that is easier to read. In some cases, this formatting turns the symbolic expression found by the algorithm into the ground truth. In such a case, the MSE reported will not be zero, even though the output is exactly correct since the MSE is based on the unformatted result which contained small (errornous) additional terms or deviations to coefficients. Some of the methods cp3-bench imports have similar features built into the algorithm, one example being uDSR which can drop coefficients smaller than a specified value.
            
            The algorithm classes inherit elements from a global superclass called \texttt{MethodEvaluator}. Most importantly, this class manages some logging levels and the evaluate method that is run during benchmarks. Specifically, it is run using Command Line Interface (CLI) that is used by cp3-bench to interact between the algorithms and the \texttt{evaluate\_dataset.py} module.
            
            The files and folders marked in the ``Algorithm folder'' part in Figure \ref{fig:architecture} are required for all algorithms. The individual algorithm may contain more files or folders beyond this minimum content. Anything that is created during the installation is put into a \texttt{deps} folder which is also created during installation.

            The results of running the benchmark is saved as \texttt{.csv} files in a folder called \texttt{results} which is created when running the \texttt{evaluate\_dataset.py} module the first time.

        \subsubsection{Getting started with a manual installation}
            This section aims at getting the reader started on using cp3-bench. To use this package you need the following:
            \begin{itemize}
                \item Ubuntu 22.04 or newer
                \item Python 3.10 or newer
                \item \texttt{pyenv}
            \end{itemize}
            Since this package relies on \texttt{pyenv} to manage virtual environments, it is key that this dependency is correctly installed and initialized. Otherwise you can face a wide variety of issues as mentioned in the \texttt{README.md} file on the GibHub repository page. Start by cloning the repository:
            \begin{lstlisting}[language=Python]
git clone https://github.com/CP3-Origins/cp3-bench.git
\end{lstlisting}
            With the package downloaded from the root folder, and with \texttt{pyenv} correctly setup you can install Python dependencies:
            \begin{lstlisting}[language=Python]
pip install -r requirements.txt
\end{lstlisting}
            Next, you can install the packages you want. \textbf{Remember} to correctly initialize \texttt{pyenv} otherwise almost nothing will work (see the documentation and trouble shooting section on the Github page of cp3-bench). The default is to install all algorithms:
            \begin{lstlisting}[language=Python]
python bench/install.py
\end{lstlisting}
            but you can optionally pass \texttt{---methods} to specify algorithms such as DSR, FFX, and ITEA as in the example here:
            \begin{lstlisting}[language=Python]
python bench/install.py --methods dsr,ffx,itea
\end{lstlisting}

            To use the package you need to parse a \texttt{.csv} file with the output value column named ``target''. You can find a short example of what this looks like in the \texttt{tests/utils} folder by looking at the \texttt{test\_dataset.csv} file.

            In the setup described in Figure \ref{fig:architecture} we import \texttt{evaluate\_dataset.py} as a module through:
            \begin{lstlisting}[language=Python]
from cp3bench.bench.evaluate_dataset import evaluate_dataset
evaluate_dataset(data_path="<path_to_dataset>")
\end{lstlisting}
            and then parse the input to it as a regular python module in our script in Things-to bench. Alternatively you can use the CLI to evaluate a dataset using the following command from the root folder of cp3-bench:
            \begin{lstlisting}[language=Python]
python bench/evaluate_dataset.py --data_path <path_to_dataset>
\end{lstlisting}
            Another option is to specify which algorithms to use by setting the appropriate flag, as with the installation command. The output is in the \texttt{results} folder. For research projects we recommend using cp3-bench as an imported submodule to another repository managing the benchmark project as we have done in Things-to-bench.

        \subsubsection{Getting started with an automated installation}
            To give the user an out-of-the-box experience we also offer a Dockerfile to support an automated installation with Docker. One disadvantage with using a Dockerfile is that you need to compile a new docker image to update your project, and making the same integration as we have done with Things-to-bench will require some reconfiguration by the user. Users who are simply interested in leveraging cp3-bench for benchmarking can simply clone the project, add a new folder with the datasets of interest (and potentially a script to manage them) before creating the Docker image. The added datasets will then be available to benchmark in the Docker image. To install the benchmark in an automated fashion you need:
            \begin{itemize}
                \item Docker
            \end{itemize}
            Start by cloning the repository:
            \begin{lstlisting}[language=Python]
git clone https://github.com/CP3-Origins/cp3-bench.git
\end{lstlisting}
            For compatibility reasons we cannot used Docker BuildKit, which is enabled per default on the latest version. Ensure BuildKit is disabled by setting the following environmental variable: \texttt{DOCKER\_BUILDKIT=0}. With this done and any dataset added to your cp3-bench folder you can build the package using:
            \begin{lstlisting}[language=Python]
docker build . -t cp3-bench
\end{lstlisting}
            As an alternative to setting \texttt{DOCKER\_BUILDKIT=0} as an environmental variable, you can also specify it during the build using the command:
            \begin{lstlisting}[language=Python]
DOCKER_BUILDKIT=0 docker build . -t cp3-bench
\end{lstlisting}
            If you are using MacOS with ARM chips you need to use x86 emulation. This can be done by adding the following parameter \texttt{--platform=linux/amd64}. In this case the command including disabling BuildKit becomes:
            \begin{lstlisting}[language=Python]
DOCKER_BUILDKIT=0 docker build . --platform=linux/amd64 -t cp3-bench
\end{lstlisting}
            Like with other commands of this tool, the default is to install all algorithms but you can specify only a subset of algorithms to install. For the Docker build you can use the following syntax to specify the subset with a comma-separated list:
            \begin{lstlisting}[language=Python]
docker build . -t cp3-bench --build-arg METHODS=ffx,gpg
\end{lstlisting}
            Here \texttt{METHODS} is the argument which stores what algorithms to install, the default is \texttt{all}, but in this example two algorithms are choosing namely \texttt{ffx} and \texttt{gpg}.

            Depending on how many algorithms you install and the hardware you use, the installation may take above 30 minutes. Upon completion you can setup and enter a Docker container with your fully functional benchmark tool using the command:
            \begin{lstlisting}[language=Python]
docker run -it cp3-bench
\end{lstlisting}
            This will mount your terminal to that of the Docker container and you are now in the isolated environment of the cp3-bench package with any additional files you have added available to bench. For Mac on ARM chips this becomes:
            \begin{lstlisting}[language=Python]
docker run --platform=linux/amd64 -it cp3-bench
\end{lstlisting}
            Early chips like the M1 does not support DSO, DSR, Operon, PySR and uDSR. We expect the remaining methods would work on these chips.
        
        \subsubsection{Benchmark procedure}\label{sec:procedure}
            The primary goal of this paper is to evaluate the performance of symbolic regression models on cosmological datasets. Most of the models we include in the benchmark were originally tested by their creators on artificial standard datasets without direct application to cosmology or astro(particle) physics. It is not clear that performance on standard datasets can in general be expected to be representative of performance on real datasets e.g. representing cosmological data. We will therefore also compare the performance of the SR algorithms on the standard test datasets that we use for hyperparameter tuning with their performance on the cosmological datasets. 

            We optimize the hyperparamters to perform well across the standard test datasets listed in Appendix \ref{app:StandardDataset}. The metric for successful tuning has been to increase the number of successfully identified symbolic expressions. Since the algorithms use different ML methods, it is not straightforward to compare the models one-to-one. For instance, certain models may be slower than others, while some models do not benefit from extra time, as they have reached a minimum. It is therefore not generally meaningful to aim for a similar runtime. Runtime is nonetheless one quantitative measure of the usefulness of the model and we will therefore include it when comparing the algorithms below.

            The models are all-in-all evaluated using three parameters: The resultant symbolic expression, the mean squared error (MSE), and runtime of the model. We define the MSE as
            \begin{gather}\label{eq:MSE}
                MSE = \frac{1}{n} \sum_{i=1}^n \paa{\hat{y_i} - y_i}^2,
            \end{gather}
            where $n$ is the number of values in the target vector $y_i$, and $\hat{y_i}$ represents the corresponding SR algorithm prediction.
            
            We can thereby recognize that an algorithm may be somewhat slow compared to others, but if the symbolic expression is correct, then this is generally more important than having the model completing faster. On the other hand, if two models are equally good or poor at obtaining the correct symbolic expression, then a comparison of run times becomes relevant for differentiating the algorithms. The MSE is simply a measure of the global fit of the symbolic expressions. If an algorithm obtains the correct symbolic equation and the data has no error, the MSE will vanish. For all other situations, the MSE can be considered a measure of how far the resultant symbolic expression is from producing a meaningful result and possible ground truth (for data without error). One should nonetheless remember that a low MSE is not necessarily equal to being close to finding the ground truth since an inherently incorrect expression may be fine-tuned to fit the data well within the training region. The MSE can additionally become misleading for datasets with errors since a ground truth will clearly not in general be achieved by having the lowest possible MSE. We also emphasize that the symbolic expressions identified by the algorithms are generally {\bf\em not} obtained through only considering the MSE of the expressions. Several of the algorithms for instance also take the complexity of the expressions into account, preferring simpler expressions over more complex ones. We do not implement any selection strategy into cp3-bench but simply pull out the symbolic expression each algorithm has ranked as ``the best'', using their individual built-in decision rules. We are thus simply including the MSE here as a simple way of comparing the obtained expressions.

            We categorize the correctness of the result in the following three ways:
            \begin{itemize}
                \item \checkmark: A check-mark is used when the algorithm produced the correct equation up to a few conditions:
                \begin{enumerate}
                    \item The resultant equation is equal to the expected, possibly written in a different way. Consider the example where the expected equation is
                    \begin{gather}\label{eq:test-function}
                        f(x) = x^2 + 1. 
                    \end{gather}
                    Assume the algorithm resulted in
                    \begin{gather}
                        f(x) = \frac{x^3 + x}{x}.
                    \end{gather}
                    This expression is considered correct.
                    \item The coefficients of the result are correct to the first significant digits. Assuming the equation from above, we would accept 
                    \begin{gather}
                        f(x) = \frac{1.1x^3 + x}{x}
                    \end{gather}
                    as correct.
                    \item Numbers smaller than or of the order $\mathcal{O}(10^{-10})$ are set to 0. For example,
                    \begin{gather}
                        f(x) = x^2 - 10^{-10}x + 1 \rightarrow f(x) =  x^2 + 1
                    \end{gather}
                    is considered correct. In cases where the ground truth is unknown, a \checkmark will not be given unless the MSE is negligible both inside the training region and on any reasonable parameter region beyond the training region. The reasonable parameter region must be determined for each relevant dataset, based on what parameter regions are physically sensible. Only in such a case could we claim to believe we have identified a new physical relation.
                \end{enumerate}
                \item  \checkmark (MSE): We give a check-mark with the MSE in parenthesis to indicate that the algorithms derived the correct equation at least structurally, but there are larger deviations than for the correct case defined above, such as:
                \begin{enumerate}
                    \item The coefficients are not correct to the first significant digit. In this category, the result must still contain all the correct terms but there may be an additional term with a significantly smaller coefficient in front. Continuing with Equation \ref{eq:test-function} as the correct function, we would in this category allow larger deviations of the coefficient such as
                    \begin{gather}
                        f(x) = 2x^2 + 0.3.
                    \end{gather}
                    \item Residual terms that have smaller prefactors than the main terms of the result are accepted. For instance,
                    \begin{gather}
                        f(x) = x^2 - 0.0001x + 1
                    \end{gather}
                    is considered correct. The expression
                    \begin{gather}
                        f(x) = x^2 - x + 1
                    \end{gather}
                    is, however, not accepted in this category since the coefficient of the extra term is of the same order as the correct terms and the extra term is no longer considered a residue. In cases where we do not know the ground truth, a ``\checkmark (MSE)'' is given if the MSE is sub-percent both inside the training region and on a parameter region that extends to the main part of the physically reasonable parameter region. In this case we expect that the expression found may be close to representing a genuine physical relation, while it being clear that an exactly correct expression has not been identified.
                \end{enumerate}
                \item MSE: Any expression deviating from the ground truth in ways not described above are considered incorrect and we indicate the deviation in terms of the MSE. For cases where the ground truth is not known, we present the MSE when the expression cannot be categorized in either of the above categories.
            \end{itemize}
            A scheme for categorizing expressions will inevitably contain some arbitrariness unless we very strictly consider an expression correct only if it is {\em exactly} so and otherwise consider the expression incorrect. However, such a very strict scheme seems unfair and prone to misleading categorizations and can easily become biased because some of the SR algorithms themselves contain e.g. round offs. Although there is thus some arbitrariness to our choices such as ignoring numbers smaller than $10^{-10}$, our opinion is that such a scheme is more appropriate than a more strict one. If one is trying to discover new equations with SR methods, the deviations in our ``almost correct'' category would presumably be detected manually if the general structure of the symbolic expression is correct. Note also that we show the symbolic expressions identified by each SR algorithm for each dataset in Appendix \ref{app:full-results} and the reader may consult that appendix to get a fuller impression of the applicability of each expression identified by the 12 algorithms for the 28 datasets.

            In the results section we also include a check-mark in parenthesis, (\checkmark), which is particularly relevant for the datasets in group $C5$. In this case the parenthesis indicates that the result contains a constant pseudo-parameter, that ideally shouldn't have been in the expression, but evaluating this constant will yield the correct result. E.g. consider the superficial constant parameter $a=1$. In the case that the true expression is $f(x)= x^2 + 1$, but the output expression is $f(x) = x^2 + a$, then setting $a\rightarrow 1$ clearly yields the correct equation, but we show that the algorithm failed to identify this pseudo-parameter as something to be ignored and replaced with a constant by marking it with (\checkmark).
            \newline\newline
            After tuning the models on the standard dataset (the result can be seen in Table \ref{fig:test-results}), we run the cosmological datasets on the models and evaluate the results. Note specifically that our hyperparameter fine-tuning is blind to the cosmological datasets and that we have only seen the results from benchmarking the cosmological data after we have locked the hyperparameters. This way one can test the meaningfulness of optimizing for standard datasets and how well the models perform on cosmological datasets. The hardware used is described in Table \ref{table:specs}.

            \begin{table}
                \centering
                \begin{tabular}{ |c|c| }
                    \hline
                    \multicolumn{2}{|c|}{Compute hardware} \\
                    \hline
                    Type & Specification \\
                    \hline
                    OS & Ubuntu 24.04 \\
                    CPU & 2x Intel(R) Xeon(R) Gold 6130 @ 2.1GHz \\
                    Total core count & 32 Cores\\     
                    Total thread count & 64 Threads\\  
                    GPU & Integrated Matrox G200eW3 Graphics \\
                    Memory & 384 GB \\\hline
                \end{tabular}
                \caption{Specification of the hardware used for benchmarking.}
                \label{table:specs}
            \end{table}

            Out-of-the-box, the method \texttt{evaluate\_dataset.py} will run multiple algorithms in parallel, which might lead to some fight over resources if CPU is overloaded, but overall it will speed up the procedure and yield results faster. For the testing we kept the code as it is out-of-the-box, \textbf{but} for the benchmarking of the cosmological dataset we specifically run only one algorithm per dataset at a time. This can be done by modifying the code, or by simply selecting to only use one algorithm at a time, instead of the default which is to use all algorithms.

            We note that initially cp3-bench (v1.0.0) only came with ten algorithms. Due to feedback from an anonymous referee and a code developer, we added two additional algorithms, Operon and GPG, as part of a revised version of this work. These two algorithms where implemented in the same fashion as the original ten, where we first optimize on the test datasets and then, after fine-tuning, apply them on the cosmological datasets. However, when benchmarking the first ten algorithms we gained experience of how models optimized on the training set behave on the cosmological datasets. For instance, we became aware of how strong polynomial performance is important for performing well on the cosmological datasets. This knowledge may subconsiously have affected the hyperparameter tuning of the two newly added algorithms Operon and GPG. Since we have also added nine additional datasets to the benchmark, we expect that at least for these datasets this potential bias has a vanishing effect.
            
\subsubsection{Comparison with SRBench}
     As stated on the homepage of SRBench\footnote{https://cavalab.org/srbench/} \cite{lacava2021contemporary}, the main motivation for that tool is to make a unified platform and framework for SR and genetic programming. This includes having a single platform with a large dataset to perform uniform benchmarking on the considered algorithms. SRBench therefore largely serves the needs of SR developers and thus what we consider more ``academic'' studies into SR algorithms. Our tool, cp3-bench, aims at easing the practical use of multiple SR algorithms simultaneously and especially in making many algorithms easily accessible to researchers not actively developing SR algorithms but interested in their application to their given research tasks. We believe one big obstacle in choosing the most optimal SR algorithm for this group of researchers is installing the algorithms. Most of the algorithms we considered (but not all) are in themselves fairly difficult to install. This in itself is a hindrance for researchers to consider a broad range of SR algorithms. The hindrance is significantly exacerbated by various algorithms requiring different versions of e.g. Python and TensorFlow. For instance some algorithm require older Tensorflow 1.x versions which are incompatible with Tensorflow 2.x versions. To improve user experience, we handle the setup of individual environments for each algorithms for the user.
    
    While SRBench provides environments for the individual algorithms to handle dependencies, there may still be some difficulties in archiving a successful installation, and there is no easy way to install all algorithms in one go and benchmark a dataset on several methods using SRBench. In addition, older algorithms such as FFX currently do not work on SRBench due to a deprecation issue with the FFX package. With cp3-bench we have ways to patch such issues and can thus present a working version of the FFX algorithm. 
    
    To complement the SRBench setup and make SR algorithms more easily available to potential users of SR algorithms, we provide cp3-bench in a format where the user simply installs \texttt{pyenv} \footnote{https://github.com/pyenv/pyenv} and afterwards the installation of cp3-bench automatically includes the installation of all algorithms unless the user actively chooses only to install some (or none) of the algorithms. Thus, a main difference between SRBench and cp3-bench is that the former is aimed at supplying a benchmark while the latter is aimed at being a tool that makes using and comparing SR algorithms easy for non-developers. If the user has Docker it is possible to install everything with all dependencies robustly set with a single command to provide a true out-of-the-box user experience. Since we are not specifically aiming cp3-bench at the SR developer community, we decided to use a data format we expect the cosmology and astroparticle physics community will be well-familiar with and find easy and intuitive to use. This is also why we make use of simple csv files as input and output, but there is no reason why one could not develop PMLB\footnote{PMLB (Penn Machine Learning Benchmark, https://epistasislab.github.io/pmlb/, is a large ML database with a different file format than that used here.} support if one wishes to leverage this.
    
\subsection{Benchmarking results: Cosmological datasets}\label{sec:benchmark-results}
    In this subsection we present the results from benchmarking the algorithms on our 28 cosmological datasets.
    \newline\newline
    Table \ref{fig:cosmo-results} gives a summary of the regression task for each algorithm in terms of the resulting MSEs. More details can be found in Appendix \ref{app:full-results} where we show the symbolic expressions found by each algorithm for each dataset with the raw data being available in the Things-to-bench repository. Table \ref{table:best_results} summarizes the datasets, comparing the ground truths with the most accurate expressions found by the algorithms. 
    \newline\indent
    Before discussing the results, we provide some comments regarding the presented MSE values and their reported equations: The algorithms considered are created by other authors and can come with unique features, and thus the algorithms behave differently from each other. For instance, some algorithms allow you to define thresholds for dropping certain residue terms, and some algorithms even do these simplifications out-of-the-box without user input. This means that the symbolic expression outputs can deviate from the ``un-simplified''  expressions that the algorithms use for computing MSEs. 
    
    When we compute the MSEs presented in the forthcoming tables, we use the full symbolic expression found by the algorithm, before the algorithm simplifies the expression by rounding to a certain number of digits etc. We will refer to this at \emph{the reported MSE}. The full symbolic expression is not always identical to the reported expression (the expression printed out by the algorithm). This is because the reported expression may, as noted above, have residue terms dropped and presented with less significant digits in order to make the output equation more presentable. As part of the result-processing we have extracted the reported equations into Mathematica notebooks where we have recomputed the MSE based on the reported equation. These notebooks can be found in the Things-to-bench repository. We find that due to the mentioned simplifications, the reported MSE does not always match the MSE computed from the reported equation. We emphasize that the MSE values reported in this paper are those coming from the full symbolic expression identified by the given algorithm. This is also the MSE which is provided by cp3-bench.
    
    For most algorithms we do not find significant deviations between the MSE computed using the full expression and the reported expression. However, there are some exceptions worth mentioning. We first note that GPZGD consistently showed a significantly lower reported MSE compared to the MSE of the reported expression. This was also the case for FFX for the more complicated equations corresponding to the datasets $C1d$, $C2b$, $C3a$, $C3f$, $C3g$, $C4a$, $C4c$, $C5c$, $C5d$, $C5e$, $C5f$, $C6a$ and $C6b$. We attribute this to the fact that FFX only outputs 3 significant digits in the output equation, but the underlying equation identified by the algorithm can have more significant digits. The resulting differences between the reported MSE and the MSE obtained from the reported expression thus deviate more for more complicated and long expressions. We further note that Genetic Engine and AI-Feynman do not protect against producing imaginary numbers. The results obtained from Genetic Engine on dataset $C4e$ and from AI Feynman on dataset $C2b$ and $C6c$ for instance contained imaginary parts. In these cases, only the real value is reported by the algorithm. Such situations can also lead to differences in the reported MSE and the MSE of the reported algorithm. We also note that the reported equation of Genetic Engine would yield division by zero errors for one or more parameters considered for the datasets $C2b$, $C3b$, $C4a$, $C4b$, $C4c$ and $C6b$. This division by zero would presumably not appear in the exact expressions obtained before rounding etc. In general, the reported MSE and the MSE based on the reported equation were for ITEA very similar. The only exception is for $C3e$ where the reported equation produces an MSE that is orders of magnitude worse than the reported MSE based on the exact expression. Therefore, we have put this value in red in Tables \ref{fig:cosmo-results} and \ref{table:best_results}.

    \begin{table}
        \centering
        \begin{adjustbox}{max width=\textwidth}
        \begin{tabular}{ |c||c|c|c|c|c|c|c|c|c|c|c|c| }
        \hline
        \multicolumn{13}{|c|}{Cosmological dataset results} \\
        \hline
        \hline
        &\multicolumn{12}{c|}{Algorithm} \\
        \hline
         Dataset & AI-Feynman & DSO  & DSR & FFX & GE & GPG & GPZGD & ITEA & Operon & PySR & QLattice & uDSR \\
        \hline
        \hyperref[tab:C1a]{C1a}& \cellcolor{red!20}$10^{-8}$ & \cellcolor{red!30}$10^{-5}$ & \cellcolor{red!30}$10^{-5}$ & \cellcolor{red!20}$10^{-6}$ & \cellcolor{red!20}$10^{-6}$ & \cellcolor{red!30}$10^{-2}$ & \cellcolor{red!10}$10^{-13}$ & \cellcolor{red!10}$10^{-14}$ & \cellcolor{red!30}$10^{-5}$ & \cellcolor{red!20}$10^{-9}$ & \cellcolor{red!20}$10^{-10}$ & \cellcolor{green!25}\checkmark\\
        \hline
        \hyperref[tab:C1b]{C1b}& \cellcolor{red!10}$10^{-14}$ & \cellcolor{red!10}$10^{-14}$ & \cellcolor{red!10}$10^{-14}$ & \cellcolor{red!10}$10^{-14}$ &  \cellcolor{red!10}$10^{-14}$ & \cellcolor{red!30}$10^{-3}$ & \cellcolor{red!10}$10^{-14}$ & \cellcolor{red!10}$10^{-14}$ & \cellcolor{red!30}$10^{-5}$ & \cellcolor{red!10}$10^{-14}$ & \cellcolor{red!10}$10^{-14}$ & \cellcolor{red!10}$10^{-14}$\\
        \hline
        \hyperref[tab:C1c]{C1c}& \cellcolor{red!20}$10^{-6}$ & \cellcolor{red!30}$10^{-4}$ & \cellcolor{red!30}$10^{-4}$ & \cellcolor{red!20}$10^{-6}$ & \cellcolor{red!30}$10^{-5}$ & \cellcolor{red!20}$10^{-8}$ & \cellcolor{red!20}$10^{-6}$ & \cellcolor{red!20}$10^{-9}$ & \cellcolor{red!30}$10^{-4}$ & \cellcolor{red!20}$10^{-6}$ & \cellcolor{red!20}$10^{-9}$ & \cellcolor{green!25}\checkmark\\
        \hline
        \hyperref[tab:C1d]{C1d}& \cellcolor{red!30}$10^{-5}$ & \cellcolor{red!20}$10^{-6}$ & \cellcolor{red!30}$10^{-5}$ & \cellcolor{red!30}$10^{-5}$ & \cellcolor{red!20}$10^{-6}$ & \cellcolor{red!20}$10^{-6}$ & \cellcolor{red!30}$10^{-5}$ & \cellcolor{red!20}$10^{-8}$ & \cellcolor{red!20}$10^{-4}$ & \cellcolor{red!20}$10^{-6}$ & \cellcolor{red!20}$10^{-7}$ & \cellcolor{green!25}\checkmark\\
        \hline
        \hyperref[tab:C2a]{C2a}& \cellcolor{green!25}\checkmark & \cellcolor{red!30}$10^{-4}$ & \cellcolor{red!30}$10^{-4}$ & \cellcolor{green!25}\checkmark & \cellcolor{red!30}$10^{-5}$ & \cellcolor{green!25}\checkmark & \cellcolor{orange!25}\checkmark ($10^{-31}$) & \cellcolor{green!25}\checkmark & \cellcolor{red!30}$10^{-3}$ & \cellcolor{green!25}\checkmark & \cellcolor{green!25}\checkmark & \cellcolor{green!25}\checkmark\\
        \hline
        \hyperref[tab:C2b]{C2b}& \cellcolor{red!30}$10^{-3}$ & \cellcolor{red!30}$10^{-4}$ & \cellcolor{red!30}$10^{-4}$ & \cellcolor{red!20}$10^{-6}$ & \cellcolor{red!30}$10^{-5}$ & \cellcolor{red!20}$10^{-7}$ & \cellcolor{red!20}$10^{-7}$ & \cellcolor{red!20}$10^{-7}$ & \cellcolor{red!30}$10^{-3}$ & \cellcolor{red!30}$10^{-5}$ & \cellcolor{green!25}\checkmark & \cellcolor{red!20}$10^{-8}$\\
        \hline
        \hyperref[tab:C3a]{C3a}& \cellcolor{red!20}$10^{-6}$ & \cellcolor{red!30}$10^{-4}$ & \cellcolor{red!30}$10^{-5}$ & \cellcolor{red!30}$10^{-4}$ & \cellcolor{red!20}$10^{-6}$ & \cellcolor{red!30}$10^{-1}$ & \cellcolor{red!20}$10^{-10}$ & \cellcolor{red!10}$10^{-13}$ & \cellcolor{red!30}$10^{-4}$ & \cellcolor{red!20}$10^{-7}$ & \cellcolor{red!10}$10^{-11}$ & \cellcolor{red!10}$10^{-13}$\\
        \hline
        \hyperref[tab:C3b]{C3b}& \cellcolor{red!30}$10^{-4}$ & \cellcolor{red!30}$10^{-3}$ & \cellcolor{red!30}$10^{-4}$ & \cellcolor{red!30}$10^{-4}$ & \cellcolor{red!30}$10^{-4}$ & \cellcolor{red!40}$10^{0}$ & \cellcolor{red!20}$10^{-7}$ & \cellcolor{red!20}$10^{-10}$ & \cellcolor{red!30}$10^{-4}$ & \cellcolor{red!20}$10^{-6}$ & \cellcolor{red!20}$10^{-9}$ & \cellcolor{red!10}$10^{-11}$\\
        \hline
        \hyperref[tab:C3c]{C3c}& \cellcolor{red!30}$10^{-3}$ & \cellcolor{red!30}$10^{-3}$ & \cellcolor{red!30}$10^{-3}$ & \cellcolor{red!30}$10^{-4}$ & \cellcolor{red!30}$10^{-4}$ & \cellcolor{red!30}$10^{-5}$ & \cellcolor{red!30}$10^{-5}$ & \cellcolor{red!20}$10^{-6}$ & \cellcolor{red!30}$10^{-3}$ & \cellcolor{red!30}$10^{-3}$ & \cellcolor{red!20}$10^{-6}$ & \cellcolor{red!20}$10^{-7}$\\
        \hline
        \hyperref[tab:C3d]{C3d}& \cellcolor{red!30}$10^{-3}$ & \cellcolor{red!30}$10^{-2}$ & \cellcolor{red!30}$10^{-2}$ & \cellcolor{red!30}$10^{-3}$ & \cellcolor{red!30}$10^{-3}$ & \cellcolor{red!30}$10^{-5}$ & \cellcolor{red!30}$10^{-3}$ & \cellcolor{red!30}$10^{-5}$ & \cellcolor{red!30}$10^{-3}$ & \cellcolor{red!30}$10^{-2}$ & \cellcolor{red!30}$10^{-5}$ & \cellcolor{red!20}$10^{-6}$\\
        \hline
        \hyperref[tab:C3e]{C3e}& \cellcolor{red!30}$10^{-3}$ & \cellcolor{red!30}$10^{-3}$ & \cellcolor{red!30}$10^{-3}$ & \cellcolor{red!30}$10^{-4}$ & \cellcolor{red!30}$10^{-3}$ & \cellcolor{red!30}$10^{-5}$ & \cellcolor{red!30}$10^{-4}$ & \cellcolor{red!20}\color{red}$10^{-7}$ & \cellcolor{red!30}$10^{-3}$ & \cellcolor{red!30}$10^{-3}$ & \cellcolor{red!20}$10^{-6}$ & \cellcolor{red!30}$10^{-3}$\\
        \hline
        \hyperref[tab:C3f]{C3f}& \cellcolor{red!30}$10^{-4}$ & \cellcolor{red!30}$10^{-2}$ & \cellcolor{red!30}$10^{-3}$ & \cellcolor{red!30}$10^{-3}$ & \cellcolor{red!30}$10^{-2}$ & \cellcolor{red!30}$10^{-5}$ & \cellcolor{red!30}$10^{-1}$ & \cellcolor{red!20}$10^{-6}$ & \cellcolor{red!30}$10^{-3}$ & \cellcolor{red!30}$10^{-3}$ & \cellcolor{red!20}$10^{-6}$ & \cellcolor{red!30}$10^{-2}$\\
        \hline
        \hyperref[tab:C3g]{C3g}& \cellcolor{red!30}$10^{-3}$ & \cellcolor{red!30}$10^{-2}$ & \cellcolor{red!30}$10^{-3}$ & \cellcolor{red!30}$10^{-4}$ & \cellcolor{red!30}$10^{-4}$ & \cellcolor{red!30}$10^{-4}$ & \cellcolor{red!30}$10^{-3}$ & \cellcolor{red!30}$10^{-5}$ & \cellcolor{red!30}$10^{-3}$ & \cellcolor{red!30}$10^{-4}$ & \cellcolor{red!20}$10^{-6}$ & \cellcolor{red!30}$10^{-2}$\\
        \hline
        \hyperref[tab:C3h]{C3h}& \cellcolor{red!30}$10^{-4}$ & \cellcolor{red!30}$10^{-3}$ & \cellcolor{red!30}$10^{-3}$ & \cellcolor{red!30}$10^{-4}$ & \cellcolor{red!30}$10^{-4}$ & \cellcolor{red!30}$10^{-5}$ & \cellcolor{red!30}$10^{-5}$ & \cellcolor{red!30}$10^{-2}$ & \cellcolor{red!30}$10^{-3}$ & \cellcolor{red!30}$10^{-3}$ & \cellcolor{red!20}$10^{-6}$ & \cellcolor{red!30}$10^{-3}$\\
        \hline
        \hyperref[tab:C4a]{C4a}& \cellcolor{red!30}$10^{-2}$ & \cellcolor{red!30}$10^{-1}$ & \cellcolor{red!30}$10^{-1}$ & \cellcolor{red!30}$10^{-1}$ & \cellcolor{red!30}$10^{-2}$ & \cellcolor{red!40}$10^{2}$ & \cellcolor{red!30}$10^{-1}$ & \cellcolor{red!30}$10^{-2}$ & \cellcolor{red!30}$10^{-2}$ & \cellcolor{red!30}$10^{-1}$ & \cellcolor{red!30}$10^{-2}$ & \cellcolor{red!30}$10^{-2}$\\
        \hline
        \hyperref[tab:C4b]{C4b}& \cellcolor{red!40}$10^{0}$ & \cellcolor{red!30}$10^{-1}$ & \cellcolor{red!30}$10^{-1}$ & \cellcolor{red!30}$10^{-1}$ & \cellcolor{red!30}$10^{-2}$ & \cellcolor{red!30}$10^{-2}$ & \cellcolor{red!30}$10^{-2}$ & \cellcolor{red!30}$10^{-2}$ & \cellcolor{red!30}$10^{-2}$ & \cellcolor{red!30}$10^{-2}$ & \cellcolor{red!30}$10^{-2}$ & \cellcolor{red!30}$10^{-2}$\\
        \hline
        \hyperref[tab:C4c]{C4c}& \cellcolor{red!30}$10^{-1}$ & \cellcolor{red!40}$10^{0}$ & \cellcolor{red!30}$10^{-1}$ & \cellcolor{red!30}$10^{-1}$ & \cellcolor{red!30}$10^{-2}$ & \cellcolor{red!30}$10^{-2}$ & \cellcolor{red!30}$10^{-2}$ & \cellcolor{red!30}$10^{-2}$ & \cellcolor{red!30}$10^{-2}$ & \cellcolor{red!30}$10^{-2}$ & \cellcolor{red!30}$10^{-2}$ & \cellcolor{red!40}$10^{0}$\\
        \hline
        \hyperref[tab:C4d]{C4d}& \cellcolor{red!40}$10^{0}$ & \cellcolor{red!30}$10^{-1}$ & \cellcolor{red!30}$10^{-1}$ & \cellcolor{red!30}$10^{-2}$ & \cellcolor{red!30}$10^{-2}$ & \cellcolor{red!30}$10^{-2}$ & \cellcolor{red!30}$10^{-3}$ & \cellcolor{red!30}$10^{-3}$ & \cellcolor{red!30}$10^{-2}$ & \cellcolor{red!30}$10^{-2}$ & \cellcolor{red!30}$10^{-3}$ & \cellcolor{red!40}$10^{0}$\\
        \hline
        \hyperref[tab:C4e]{C4e}& \cellcolor{red!30}$10^{-3}$ & \cellcolor{red!30}$10^{-4}$ & \cellcolor{red!30}$10^{-3}$ & \cellcolor{red!30}$10^{-3}$ & \cellcolor{red!30}$10^{-4}$ & \cellcolor{red!30}$10^{-4}$ & \cellcolor{red!30}$10^{-4}$ & \cellcolor{red!30}$10^{-4}$ & \cellcolor{red!30}$10^{-5}$ & \cellcolor{red!30}$10^{-4}$ & \cellcolor{red!30}$10^{-4}$ & \cellcolor{red!30}$10^{-2}$\\
        \hline
        \hyperref[tab:C5a]{C5a}& \cellcolor{green!25}\checkmark & \cellcolor{green!15}(\checkmark) & \cellcolor{green!15}(\checkmark) & \cellcolor{red!30}$10^{-2}$ & \cellcolor{green!15}(\checkmark) & \cellcolor{orange!25}(\checkmark) ($10^{-13}$) & \cellcolor{red!30}$10^{-1}$ & \cellcolor{red!20}$10^{-9}$ & \cellcolor{red!30}$10^{-4}$ & \cellcolor{green!15}(\checkmark) & \cellcolor{orange!25}\checkmark ($10^{-16}$) & \cellcolor{red!30}$10^{-4}$\\
        \hline
        \hyperref[tab:C5b]{C5b}& \cellcolor{green!25}\checkmark & \cellcolor{green!15}(\checkmark) & \cellcolor{red!30}$10^{-4}$ & \cellcolor{red!30}$10^{-2}$ & \cellcolor{red!30}$10^{-4}$ & \cellcolor{red!30}$10^{-4}$ & \cellcolor{red!30}$10^{-1}$ & \cellcolor{red!30}$10^{-4}$ & \cellcolor{red!30}$10^{-4}$ & \cellcolor{green!15}(\checkmark) & \cellcolor{orange!25}\checkmark ($10^{-4}$) & \cellcolor{red!30}$10^{-3}$\\
        \hline
        \hyperref[tab:C5c]{C5c}& \cellcolor{red!30}$10^{-2}$ & \cellcolor{red!30}$10^{-2}$ & \cellcolor{red!30}$10^{-2}$ & \cellcolor{red!30}$10^{-2}$ & \cellcolor{red!30}$10^{-2}$ & \cellcolor{red!30}$10^{-1}$ & \cellcolor{red!30}$10^{-1}$ & \cellcolor{red!30}$10^{-2}$ & \cellcolor{red!30}$10^{-2}$ & \cellcolor{green!15}(\checkmark) & \cellcolor{red!30}$10^{-2}$ & \cellcolor{red!30}$10^{-2}$\\
        \hline
        \hyperref[tab:C5d]{C5d}& \cellcolor{red!40}$10^{0}$ & \cellcolor{red!30}$10^{-3}$ & \cellcolor{red!30}$10^{-3}$ & \cellcolor{red!30}$10^{-2}$ & \cellcolor{red!30}$10^{-3}$ & \cellcolor{red!20}$10^{-6}$ & \cellcolor{red!30}$10^{-1}$ & \cellcolor{red!20}$10^{-7}$ & \cellcolor{red!30}$10^{-3}$ & \cellcolor{green!25}\checkmark & \cellcolor{red!30}$10^{-4}$ & \cellcolor{red!30}$10^{-2}$\\
        \hline
        \hyperref[tab:C5e]{C5e}& \cellcolor{red!30}$10^{-1}$ & \cellcolor{red!30}$10^{-4}$ & \cellcolor{red!30}$10^{-4}$ & \cellcolor{red!30}$10^{-4}$ & \cellcolor{red!30}$10^{-4}$ & \cellcolor{red!20}$10^{-7}$ & \cellcolor{red!30}$10^{-4}$ & \cellcolor{red!20}$10^{-7}$ & \cellcolor{red!30}$10^{-5}$ & \cellcolor{green!25}\checkmark & \cellcolor{red!20}$10^{-8}$ & \cellcolor{red!30}$10^{-3}$\\
        \hline
        \hyperref[tab:C5f]{C5f}& \cellcolor{red!40}$10^{0}$ & \cellcolor{red!30}$10^{-2}$ & \cellcolor{red!30}$10^{-2}$ & \cellcolor{red!30}$10^{-1}$ & \cellcolor{red!30}$10^{-2}$ & \cellcolor{red!30}$10^{-5}$ & \cellcolor{red!30}$10^{-1}$ & \cellcolor{red!30}$10^{-3}$ & \cellcolor{red!30}$10^{-3}$ & \cellcolor{red!30}$10^{-2}$ & \cellcolor{red!30}$10^{-3}$ & \cellcolor{red!30}$10^{-1}$\\
        \hline
        \hyperref[tab:C6a]{C6a}& \cellcolor{red!30}$10^{-1}$ & \cellcolor{red!30}$10^{-1}$ & \cellcolor{red!30}$10^{-1}$ & \cellcolor{red!30}$10^{-1}$ & \cellcolor{red!30}$10^{-1}$ & \cellcolor{red!30}$10^{-1}$ & \cellcolor{red!30}$10^{-2}$ & \cellcolor{red!30}$10^{-1}$ & \cellcolor{red!30}$10^{-2}$ & \cellcolor{red!30}$10^{-1}$ & \cellcolor{red!30}$10^{-1}$ & \cellcolor{red!30}$10^{-1}$\\
        \hline
        \hyperref[tab:C6b]{C6b}& \cellcolor{red!30}$10^{-3}$ & \cellcolor{red!30}$10^{-1}$ & \cellcolor{red!30}$10^{-1}$ & \cellcolor{red!30}$10^{-1}$ & \cellcolor{red!30}$10^{-1}$ & \cellcolor{red!30}$10^{-5}$ & \cellcolor{red!30}$10^{-5}$ & \cellcolor{red!30}$10^{-3}$ & \cellcolor{red!30}$10^{-2}$ & \cellcolor{red!30}$10^{-3}$ & \cellcolor{red!30}$10^{-2}$ & \cellcolor{red!30}$10^{-1}$\\
        \hline
        \hyperref[tab:C6c]{C6c}& \cellcolor{red!40}$10^{0}$ & \cellcolor{red!40}$10^{0}$ & \cellcolor{red!40}$10^{0}$ & \cellcolor{red!40}$10^{0}$ & \cellcolor{red!40}$10^{0}$ & \cellcolor{red!30}$10^{-1}$ & \cellcolor{red!40}$10^{0}$ & \cellcolor{red!40}$10^{0}$ & \cellcolor{red!40}$10^{0}$ & \cellcolor{red!40}$10^{0}$ & \cellcolor{red!40}$10^{0}$ & \cellcolor{red!40}$10^{0}$\\
        \hline
        \end{tabular}
        \end{adjustbox}
        \caption{Result of evaluating the cosmological datasets across all algorithms. We highlight a single MSE in red because the reported MSE (shown in the table) deviates significantly from the MSE of the reported equation. The green background colors indicate a correct symbolic expression, orange is an almost correct expression, and the different red colors indicate the size of the MSE.}
        \label{fig:cosmo-results}
    \end{table}

    \begin{table}
        \centering
        \begin{adjustbox}{max width=\textwidth}
        \begin{tabular}{ |p{1.30cm}|p{7.55cm}|p{11cm}|p{0.9cm}|p{2.15cm}|  }
            \hline\hline
            Dataset & Ground truth & Best result &  MSE & Algorithm\\
            \hline
            \Tstrut
            C1a & $H = 0.0716/\rm{Gyr} \sqrt{0.3 (1+z)^3 + 0.7}$ & $H = 0.0716\cdot \sqrt{0.3\cdot (1+z)^3 + 0.7}$ & $10^{-30}$ &uDSR \\
            \hline
            C1b & $H = 0.0716/\rm{Gyr} \sqrt{0.3 (1+z)^3 + 0.7}+\rm{(10\% \ error)}$ &$0.932 + z(0.279 + z (-1.31 + z (2.61 + (-2.08 + 0.576 z) z))) +  \sin\paa{e^{z + \cos\paa{e^{\sin(z)}}}}$ &$10^{-4}$ & uDSR \\
            \hline
            C1c &$H = 0.0716/\rm Gyr\sqrt{\Omega_{m,0}(1+z)^3 + (1-\Omega_{m,0})}$ &$ H=\sqrt{0.0051 +  z (0.015 + (0.015 + 0.0051 z) z) \Omega_{m,0}}$ &$10^{-30}$ &uDSR \\
            \hline
            C1d &$H = H_0\sqrt{\Omega_{m,0}(1+z)^3 + (1-\Omega_{m,0})}$ & $H=1.732 H_0 \sqrt{0.33 + z (1 + z + 0.33 z^2) \Omega_{m,0}}$ & $10^{-29}$ & uDSR \\
            \hline
            C2a & $\delta z/\delta t_0 = 0.0716/\rm{Gyr}[(1+z)-H(z)]$ & $\delta z/\delta t_0 = 0.0716 - H + 0.0716 z$ & $10^{-32}$ & ITEA \\
            \hline
            C2b & $\delta z/\delta t_0 =0.0716/\rm{Gyr}[ (1+z) - (1+z)^{3(1+\omega)/2}]$ & $\delta z/\delta t_0 = 0.0716 + 0.0716 z - 0.0716 (1 + z)^{1.5 + 1.5\omega}$ &  $10^{-16}$& Qlattice \\
            \hline
            C3a & unknown &$10^2Q_D =\Big(\big(0.51 + z (2.2 + z (11 + z (9.9 + (32 - 4.6 z) z)))\big) \cos\paa{\sin\paa{z + \cos\paa{z}}}\Big)^{1/4}$ & $10^{-13}$  & uDSR\\
            \hline
            C3b & unknown & $10^8R_D =\frac{7.24 e^{-z} \cos\paa{z - \cos\paa{z}}}{-0.81 + z \paa{-3.2 + z \paa{-3.6 + z (-2.1 + z (1.5 + z))}}}$& $10^{-11}$  & uDSR\\
            \hline
            C3c & unknown &   $10^2Q_D =f_o (-f_o - \sqrt{z} + (18 z)/(f_o^4 (-0.00080 - 0.0025 z) - 0.00029 f_o^2 z^2 + f_o^3 (-0.00032 + z (-0.0040 + 8.4 z)) + f_o z (-7.5 + z (17 + z (-13 + 3.9 z))) + z (5.9 + z (-10 + z (9.4 + z (-4.7 + z))))))$ &$10^{-7}$  & uDSR\\
            \hline
            C3d & unknown &  $10^8R_D=\frac{1}{f_o - \sqrt{\cos\paa{\cos\paa{z^2}}}} \Big(4.0 +  f_o^4 (-0.033 - 0.0048 z) +  f_o^3 (-0.011 + (-0.0078 - 0.00045 z) z) +  f_o^2 (-0.000026 +  z (-0.0011 + (-0.00058 - 0.000042 z) z)) + f_o (-7.30 + z^2 (13 + z (-18 + 7.3 z))) + z (-7.9 + z (11 + z (-11 + (7.2 - 2.3 z) z)))\Big)$& $10^{-6}$ & uDSR\\
            \hline
            C3e & unknown & \small $10^2Q_D = -244 +  0.0275 e^{\frac{H_0 \Omega_{Q,0}^3}{ z^2 \Omega_{m,0}^5 \Omega_{R,0}}} + 250 e^{\frac{H_0 z^2 \Omega_{Q,0}^4 \Omega_{R,0}^5}{\Omega_{m,0}^3}} - \frac{0.6278 \abs{\Omega_{Q,0}} \abs{\Omega_{R,0}}}{\abs{H_0^{5/2} \Omega_{m,0}}} + 370 \sqrt{\abs{\frac{H_0^3 z^2 \Omega_{Q,0}^4 \Omega_{R,0}^3}{\Omega_{m,0}^2}}} - 1.15 \sin\paa{\frac{z^3 \Omega_{m,0}^4 \Omega_{R,0}}{\Omega_{Q,0}^2}}$  & $10^{-7}$ & \color{red}ITEA\\
            \hline
            C3f & unknown & $10^8 R_D = 64 - 4.7 e^{\frac{H_0 \Omega_{Q,0} \Omega_{R,0}^3}{z \Omega_{m,0}}} - 0.0048 \sqrt{\abs{\frac{z^5 \Omega_{Q,0}^3}{H_0^5 \Omega_{m,0}^2}}} - 11\cdot10^{4} \sqrt{\abs{\frac{H_0^5 \Omega_{Q,0}^5}{\Omega_{m,0}^5}}} - 1.1\ln\paa{\frac{\Omega_{Q,0}^4 \Omega_{R,0}^5}{z^2 \Omega_{m,0}^4}} + 1.2 \sin\paa{ \frac{z^4 \Omega_{m,0}^3 \Omega_{R,0}^5}{\Omega_{Q,0}^2}}$ & $10^{-6}$ & ITEA\\
            \hline
            C3g & $Q_D = 6H_D^2\frac{v(1-v)(1-h)^2}{(1-v+hv)^2}$  & $-0.015 + (9.3 - 4.3 v) v + (0.0037 - 4.6 v) \ln \paa{e^{(1.7 - 0.76 h) h + (4.5 - 50 H) H}}$ & $10^{-6}$  & QLattice\\
            \hline
            C3h & $Q_D= -6\left( \frac{\ddot a_D}{a_D} + \frac{1}{2a_D^3}H_{D_0}^2\Omega_{m,0}\right)$ & $10^2 Q_D =\exp\Big((-0.00054 e^{((2.8 - 3.7 a_D) a_D)}\times (e^{27 H})^{e^{-30 H^2}} + (-0.47 - 0.24 \Omega_{m,0}) \Omega_{m,0} + e^{(2.8 - 3.7 a_D) a_D + (27- 30 H) H}\times  (-0.052 - 0.053 \Omega_{m,0}) \Omega_{m,0} +  e^{(5.6 - 7.4 a_D) a_D + (53 - 61 H) H}\times (-3.2\cdot10^{-7} + (-0.000061- 0.0029 \Omega_{m,0}) \Omega_{m,0})\Big)$ & $10^{-6}$& QLattice \\
            \hline
            C4a & unknown & $10^3\expval{\delta z}/{\delta t_0}= 0.17 + (-11 - 14 \expval{z}) \expval{z}$ & 10  & QLattice\\
            \hline
            C4b & unknown & $-0.81 - 0.258 \expval{z} + 1.14 \cos\paa{(-3.22 + \expval{z}) (2.19 + \expval{z})} + \frac{35.9 \expval{z} (0.762 + \expval{z}) f_o }{\sin\paa{3.79 - f_o}} + \paa{-0.785 + \frac{0.785 \expval{z}}{f_o}} \sin(f_o)$ & $10^{-2}$ & GPG\\
            \hline
            C4c & unknown & $0.3 - 1.85 \Omega_{Q,0}^2 + \Omega_{Q,0} (-1.82 +  \expval{z} (53.4 + \frac{55.3 \expval{z}}{\Omega_{R,0}} )) - 1.85 \Omega_{m,0}  \Omega_{R,0}  + (1.82 \expval{z} -  1.83 \Omega_{R,0} ) \sin(5.5 + \Omega_{R,0})$ & $10^{-2}$ & GPG\\
            \hline
            C4d & unknown & $-21\cdot 10 + \frac{15 - \frac{2.4 \expval{z} \Omega_{m,0}^5 \Omega_{R,0}^4}{\Omega_{Q,0}^5}}{H_D} +68 \sqrt{\abs{\frac{\expval{z}^5 \Omega_{m,0}^5 \Omega_{Q,0}^2}{H_D \Omega_{R,0}^3}}} -23\cdot 10^3 \sqrt{ \abs{\expval{z}^2 H_D^3 \Omega_{m,0}^4 \Omega_{Q,0}^4\Omega_{R,0}^5}} + 22 \sin\paa{\frac{\expval{z}^2 \Omega_{m,0}^4\Omega_{R,0}^5}{\Omega_{Q,0}^3}}$ & $10^{-3}$  & ITEA\\
            \hline
            C4e & unknown & $0.00136 - 11.5 \expval{z} \Omega_{m,0}+ 0.345 \expval{z} \cos(2.77 \expval{z})$ &  $10^{-5}$ & Operon\\
            \hline
            C5a & $\rho_{\rm NFW} = \frac{1}{r(1+r)^2}$ & $\frac{1}{r(1+r)^2}$ & $10^{-31}$ & AI-Feynman\\
            \hline
            C5b & $\rho_{\rm NFW} = \frac{1}{r(1+r)^2} + \rm{(1\% \ error)}$ & $\frac{1}{r(1 + r)^2}$  & $10^{-4}$ & AI-Feynman \\
            \hline
            C5c & $\rho_{\rm NFW} = \frac{1}{r(1+r)^2} + \rm{(10\% \ error)}$ & $\frac{x}{r(r+x)^2}$ & $10^{-2}$ & PySR\\
            \hline
            C5d & $\rho_{\rm NFW} = \frac{1}{\frac{r}{R_0}(1+r/R_0)^2}$  & $\frac{R_0^3}{r(r+R_0)^2}$ & $10^{-31}$ & PySR\\
            \hline
            C5e & $\rho_{\rm core} = \frac{R_0^3}{(r+R_0)(r^2+R_0^2)}$ & $\frac{R_0^3}{(r+R_0)(r^2+R_0^2)}$ & $10^{-33}$ & PySR\\
            \hline
            C5f & $  \rho_{\rm NFW/core} = \frac{1}{2}\left(\rho_{\rm NFW} + \rho_{\rm core}\right) + \frac{x}{2}\left(\rho_{\rm NFW} - \rho_{\rm core}\right)$ & $0.0598 + \frac{1}{r (r + 1.76 R_0)} 2.02 R_0^2 x  \cos\paa{0.323 + \frac{0.624 r}{R_0}} \big(-\frac{1.03 \sin(r)}{r + R_0} + (-0.0154 + 0.0743 r) r \sin(2.66 + R_0) + \sin(1.01 + x)\big)$ & $10^{-5}$ & GPG\\
            \hline
            C6a & $ h_{+} = -\frac{2M\eta}{R}\Big[ \big(-4M^2\left(-3t + v_0\right)^{-10/3} + 2\left( -3t+v_0 \right)^{2/3}\big)\cos\left( -3t^2/M + 2v_0t/M \right) + \left( 4M(-3t + v_0) \right)^{-4/3}\sin\left( -3t^2/M + v_0t/M \right)\Big]$  & $0.846 \sin(2.66 (-3.54 + t) t)   \sin\paa{ \sin\paa{0.478 (17.7 + t)}}-0.0502$ & $10^{-2}$ & GPZGD\\
            \hline
            C6b &   $ h_{+} = -\frac{2M\eta}{R}\big[ \left(-\dot r^2 + r^2\dot \Phi^2 + \frac{M}{r}\right)\cos(2\Phi) +2r\dot r\dot\Phi\sin(2\Phi)  \big]$ & $0.000492 + \frac{ 0.0172 \dot{r} \sin(1.59 - 0.762 \dot{r} + 2 \Phi) (-11.9 + \sin(r))}{r\sin(\dot{r})}$ & $10^{-5}$ & GPG\\
            \hline
            C6c & $ h_{+} = -\frac{2M\eta}{R}\Big[ \big(-4M^2\left(-3t + v_0\right)^{-10/3} + 2\left( -3t+v_0 \right)^{2/3}\big)\cos\left( -3t^2/M + 2v_0t/M \right) + \left( 4M(-3t + v_0) \right)^{-4/3}\sin\left( -3t^2/M + v_0t/M \right)\Big]$ & $0.00891 -0.0937 m_1 \cos\paa{\frac{36.7 t}{3.34 m_1 + m_1 t}}  (1.66 - t + \sin(2.25 - m_1)) (6.5553 + 1.44 m_1 + t + \sin(m_1) (5.68 t^2 - 5.68 \sin(t)))$ & $10^{-1}$  & GPG\\
            \hline
        \end{tabular}
        \end{adjustbox}
        \caption{Summary of best (smallest MSE or correct formula) symbolic expression for each dataset. Ground truths are written with the target on the left hand side while any non-numeric on the right hand side represents a feature of the dataset. To increase readability we only include 2-3 significant digits. The MSE is given as order of magnitude. We highlight a single MSE in red because the reported MSE deviates significantly from the MSE of the reported equation.}
        \label{table:best_results}
    \end{table}
    We now move on to discuss the results presented in Tables \ref{fig:cosmo-results} and \ref{table:best_results}. We had expected the dataset $C1a$ to represent a fairly simple regression task. However, only a single algorithm identified the correct expression, namely uDSR. When adding $10\%$ Gaussian error ($C1b$), none of the algorithms identified the correct expression. The datasets $C5a$ (no error), $C5b$ ($1\%$ Gaussian error) and $C5c$ ($10\%$ Gaussian error) show a similar tendency: Seven algorithms identify or nearly identify the correct expression for the dataset with no error, while this is only true for three algorithms for the dataset with $1\%$ error. For the dataset with $10\%$ error, only PySR identifies the correct expression, with the only ``error'' being that it explicitly includes the category label $x$ instead of simply 1 (we comment more on this below). uDSR is also the only algorithm that succeeded in identifying correct ground truths for the remaining datasets of $C1$. We expect this to be due to the high complexity and/or larger numbers of features in these datasets.
    
    \begin{figure}
    \centering
    \includegraphics[scale = 0.5]{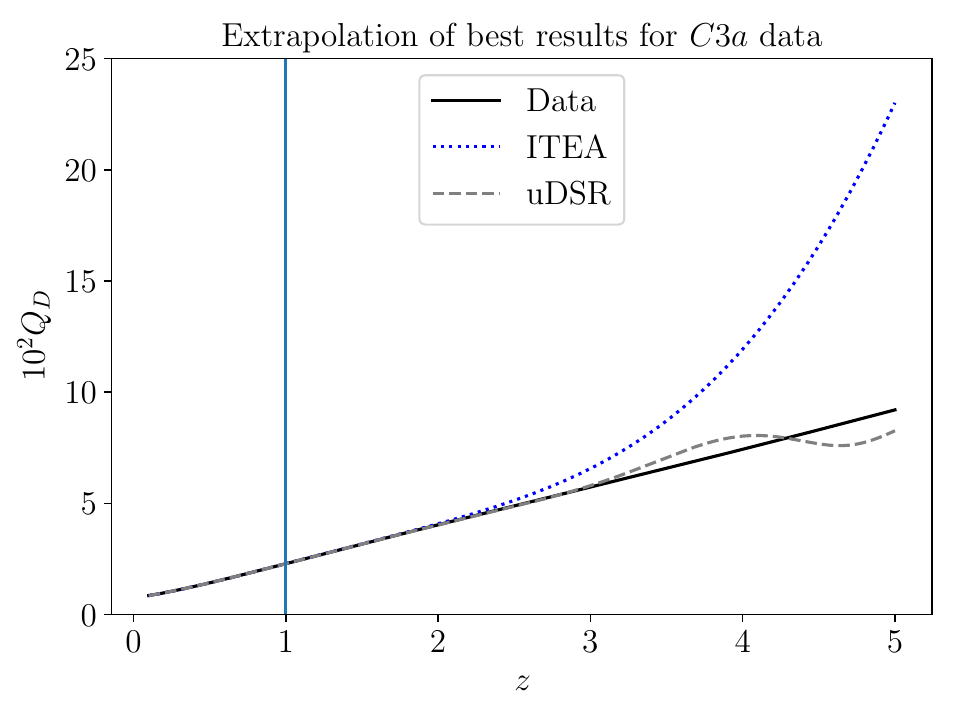}
    \caption{Comparison of correct data corresponding to dataset C3a extrapolated to $z = 5$ compared with best results from ITEA and uDSR. To ease readability of the graph, we only show lines corresponding to the single value of $f_o = 0.2$. The region to the right of the vertical line indicates redshifts above the training region.}
    \label{fig:C7}
    \end{figure}

    For $C2a$ we note that nine algorithms identify the correct formula up to at least 3 significant digits but that Qlattice has a somewhat smaller MSE than most of the other algorithms, presumably due to the numerical uncertainties in the internal models. ITEA also nearly finds the correct formula, the error being one highly suppressed extra term of order $\mathcal{O} (10^{-10})$ which we have dropped it in Table \ref{tab:C2a}. The dropped term is retained in the raw data. AI-Feymnan, FFX, and GPZGD also find the correct form of the data, although with somewhat incorrect coefficients, and for GPZGD the difference is so big that we classify it as only ``almost correct'' and provide the MSE. Although $C2a$ and $C2b$ represent the same datasets, our benchmark shows that their different parameterizations are quite important; only Qlattice succeeds with finding the correct formula for $C2b$. Thus, the simpler symbolic form of the dataset (represented by the dataset $C2a$) is important for the algorithms' ability to identify the correct formula. This is particularly interesting because the parameters in $C2a$ are inter-dependent which we had suspected would make it more difficult for the algorithms to identify the correct formula. Such inter-dependence of parameters turns out to apparently be less important than the complexity of the formula. The datasets $C3h$ and $C3g$ also represent symbolic expressions where the features are interdependent. None of the algorithms find the correct expressions for these datasets. Especially considering the results from $C2a$ and $C2b$ we cannot claim that the low performance of the algorithms on $C3h$ and $C3g$ is due to inter-dependence of features. Instead, we note that the symbolic expressions for $C3h$ and $C3g$ are more complicated than those of the group $C2$ and we suspect this to be the reason the algorithms fail on these datasets. We also note that the algorithms do not do note-worthily better on $C3h$ despite it representing a less complex expression\footnote{Note that there are various ways to define the complexity of a function (see e.g. \cite{Kommenda_2015} for a summary). One of the more traditional methods is a simple count of number of operators, but often also the number of embeddings are evaluated. In both cases, the expression represented by $C3g$ is simpler than that of $C3h$.}.
        \newline\indent
    None of the algorithms find symbolic expressions that represent ground truths for any of the datasets $C3a-C3e$. However, these datasets do highlight an interesting shortcoming of using the MSE as the stand-alone criterion for choosing the final symbolic expression from the pareto front. In particular, we note that ITEA and uDSR have the lowest MSEs for the dataset $C3a$. In Figure \ref{fig:C7} we show how these expressions extrapolate to $z = 5$ for a specific value of $f_o$. It is quite clear from the comparison with the data that neither ITEA nor uDSR found a ground truth relation. We have made similar comparisons between the extended datasets with unknown ground truths with the best expressions found by each algorithm and note that although none of the algorithms find an expression that extrapolates well outside the training region, the expressions with the lowest MSE are often not the ones that extrapolate the best. For instance, AI Feynman and FFX both find symbolic expressions that extrapolate fairly well to $z = 5$ for the $C3a$ dataset. However, for $C3a$, the symbolic expressions that extrapolate the best out to $z = 5$ are those found by Genetic Engine and QLattice which deviate at most 2\% and 0.8\%, respectively, from the data in the redshift interval $z\in[0,5]$. Upon extrapolating up to $z = 10$ we find that the deviations increase up to over 10\% and 8\%, respectively, (and are still growing linearly at this point). We therefore conclude that the expressions do not represent (approximate) ground truths, but that they may nonetheless be on the right track. Overall, we find that the symbolic expressions with the lowest MSE on an extended region is not necessarily the expressions that have the lowest MSE in the training region.
    \newline\indent
    The datasets $C3e$ and $C3f$ are further interesting because they contain a feature, $H_{D_0}$, that is constant throughout the two datasets and which we would thus expect the algorithms to ignore in their identified symbolic expressions. While we find that most algorithms succeed with avoiding $H_{D_0}$ in their final expressions, ITEA does include this feature in its result for $C3e$ while both ITEA and Genetic Engine include it for $C3f$. The datasets $C5a-C5d$ similarly contain a superfluous classification feature (which becomes important in $C5e$). Several algorithms include this categorization feature even though it takes the constant value $1$ throughout these datasets. For many of the algorithms, the resulting expression is in principle correct, but structurally, the expressions become misleading as they include the category $x$ rather than simply using 1. PySR is the only algorithm that succeeds with identifying the ground truth for $C5d$ and $C5e$. Ignoring the use of $x$ rather than 1, PySR identifies the ground truth expressions for the entire group $C5$ except for $C5f$.
    \newline\indent
    \begin{figure}
        \centering
        \includegraphics[width=0.99\linewidth]{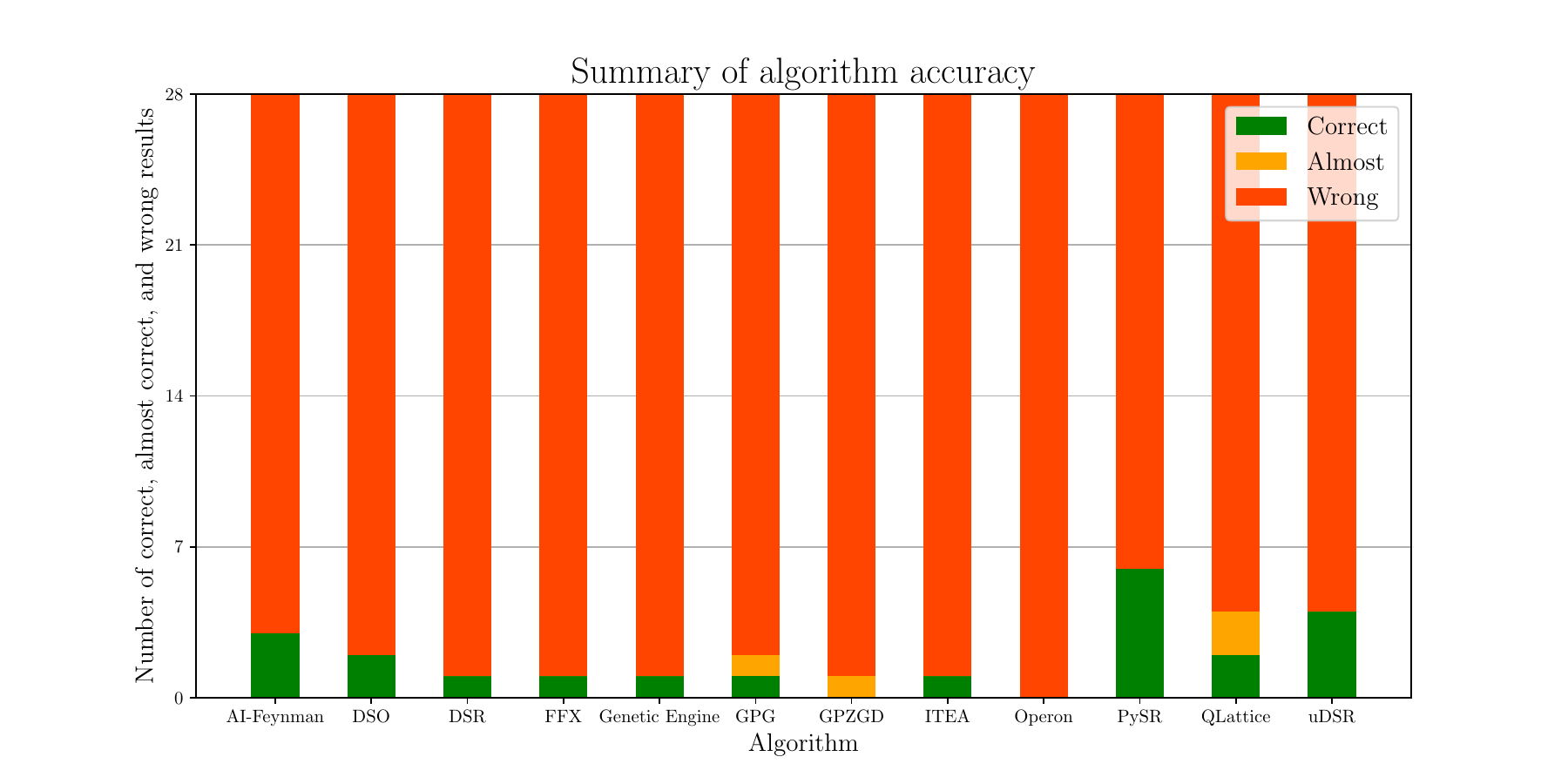}
        \caption{Summery of the results for running the 12 algorithms on 28 datasets. The category ``correct'' refers to finding the correct equation, ``almost'' refers to the result almost being correct as specified in section \ref{sec:procedure}, and ``wrong'' is when the equation is not even close to being correct (according to section \ref{sec:procedure}). Note that we here categorize the cases where the $x$ parameters is used in the dark matter profile dataset ``correct'' for this aggregation, as long as setting $x$ to the respective value yields the correct result.}
        \label{fig:results}
    \end{figure}

    \begin{figure}
        \centering
        \includegraphics[width=0.99\linewidth]{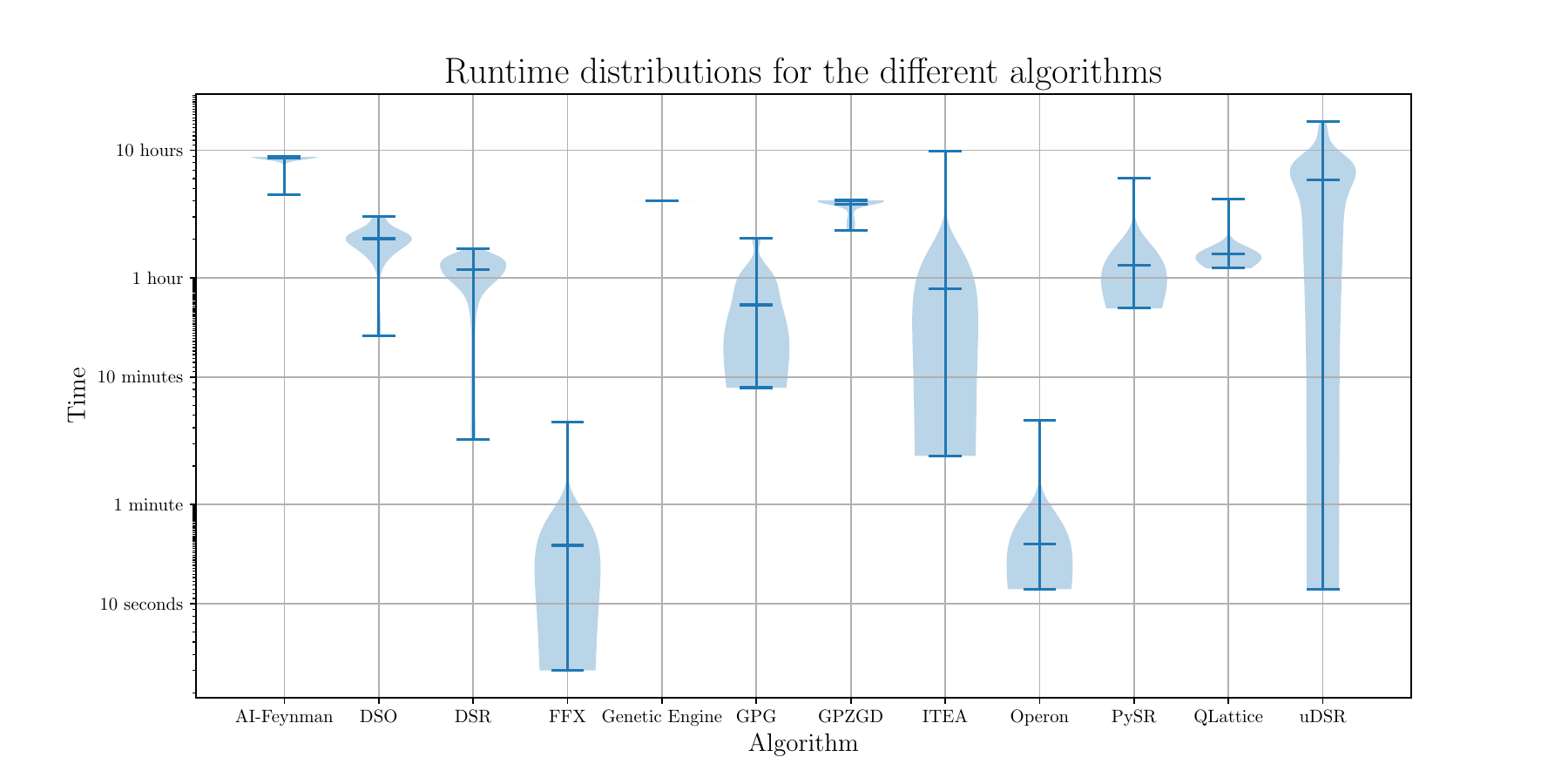}
        \caption{The spread of runtimes for the different algorithms across the benchmark of the 28 datasets.}
        \label{fig:runtime}
    \end{figure}
    The datasets $C6a-C6c$ oscillate in the features which is a quality not exhibited by the other considered datasets. By including this dataset we can therefore assess if some of the considered algorithms do significantly better/worse on oscillating datasets than on overall more monotonous datasets. The results presented in Table \ref{fig:cosmo-results} do not indicate that this should be the case as no algorithms identify the correct expressions for the oscillating datasets and the reported MSEs are generally quite high. Similarly, no algorithms succeeded with identifying expressions with low MSEs for the datasets in the group $C4$. This is perhaps not too surprising since these datasets have a statistical error and in addition we cannot be certain that such a ground truth even exists for these datasets.
    \newline\newline
    Based on the ability to identify correct ground truths for the cosmological datasets, PySR and uDSR overall perform the best of all the benched algorithms, identifying several of the ground truth expressions. We also note that for the cases where uDSR found the ground truth it was incredible fast, in part thanks to its ability to stop early when the precision is deemed good enough. These two algorithms did {\em not} perform noticeably better than the other algorithms on the test dataset (see Appendix \ref{app:StandardDataset}). Indeed, on the test data, ITEA and Operon identified the correct formula for 4 of the datasets while uDSR, QLattice and PySR all identified the correct formula in only 3 instances. The remaining algorithms either identified one or zero correct symbolic expressions. It is specifically noteworthy that Operon performed among the best on the standard datasets, while also being very fast, but performed the worst of all algorithms on the cosmological datasets, not even obtaining a single ``almost'' correct identification as highlighted in Figure \ref{fig:results}. This strengthens our suspicion that it might not be prudent to choose SR algorithms for specific tasks based on how well they perform on standard datasets.
    \newline\newline
    Figure \ref{fig:results} summarizes the performance of the algorithms on the cosmological datasets by showing the number of correct, ``almost correct'' and incorrect expressions, following the definitions of these categories given in Section \ref{sec:procedure}. 
    
    Lastly, we also consider runtime of the algorithms with the spread shown in Figure \ref{fig:runtime}. We see that algorithms which are time-constrained (such as AI-Feynman, Genetic Engine, and GPZGD) have a low spread in runtimes, unlike uDSR which was reasonably fast when it found the correct expression, triggering an early stop, but also had the longest runtime at +16 hours in one case. We note that both FFX and Operon are very fast algorithms yielding results in less than 10 minutes. In general, we see that algorithms using some kind of early stopping have a larger spread, which is to be expected assuming they can get close enough to a solution in some of the scenarios to trigger this early stop. Genetic Engine, and GPZGD were set to stop after 4 hours as no further improvements were seen when more time was permitted. While AI-Feynman does have a max time feature, this does not correlate directly to the effective time where it will time out as this max time refers to a subpart of the algorithm. The effective max time of AI Feynman was therefore around 8-9 hours. For most algorithms the runtime is effectively determined by the maximum number of epochs or generations permitted, but the runtime will in general also depend much on the number of parameters of the input data and the hyperparamters.

\section{Summary, discussion and conclusions}\label{sec:Summary}
    We have presented a tool, cp3-bench, that makes it easy to install 12 symbolic regression (SR) algorithms and compare their performance on given datasets. We demonstrated the use of the tool by introducing a demonstration benchmark of the algorithms on 28 datasets split into six groups representing different physical scenarios. We emphasize that the benchmark is not exhaustive and should not be used as a guide into which SR algorithms to use for a given task at hand. We instead encourage users of SR algorithms to perform their own benchmarks/comparisons, focusing on the type of test datasets most sensible for their needs. In our benchmark we thus focused on six different groups of cosmological datasets, all chosen to emphasize different strengths and weaknesses of datasets and algorithms. The evaluation of the performance of the considered SR algorithms was done considering the scenario where SR algorithms are used for identifying ground truths. SR can also be used for others goals such as simply obtaining ``good fits'' within a specific parameter region.
    \newline\indent
    The results from benchmarking with the two dataset groups $C1$ and $C5$ demonstrate the importance of dataset precision. We found that while one algorithms could identify the correct ground truth for data in group $C1$ without error, none of the algorithms identified the correct ground truth when a $10 \%$ error was added. This is not particularly surprising since SR algorithms usually assess symbolic expressions at least partially based on some type of error function. For the dataset $C5$ we again found that adding $10\%$ error to the datasets significantly impaired the algorithms' ability to identify ground truths. Adding $1\%$ had less of an impact and several algorithms were able to still identify the correct ground truth. This result is particularly encouraging for the use of SR in cosmology where data is reaching percent precision.
    \newline\indent
    Our results from the dataset group $C2$ indicate that it is not a significant obstacle for the tested SR algorithms if features are inter-dependent. It is more important that the analytical expression of the ground truth is simple. The dataset groups $C3$ and $C4$ represent genuine SR mysteries, i.e. the target of these datasets have hitherto unknown symbolic expressions. These datasets e.g. highlight the difficulties one faces when trying to identify a ground truth, including not knowing which parameters to include as features of a dataset. This means that one may want to consider several different combinations of features but also different targets, in case some targets prove to have simpler ground truths. With the specific datasets $C3e$ and $C3f$ we also included a parameter which took the same value for each point in the dataset and which should therefore ideally not be included in the results of the SR algorithms. We found that most algorithms did not include this parameter in their results. This same behavior was tested with datasets in the group $C5$ where a superfluous category feature with value $1$ was added to several datasets. In this case, several algorithms included the category in their final result, yielding expressions that are in principle correct but misleading since the number $1$ could just as well have been used.
    \newline\indent
    Our results for all dataset groups indicate that the algorithms that do the best on standardized datasets do not necessarily perform best on our cosmological datasets. For instance, we found that PySR and uDSR did the best on the cosmological datasets but were less successful than Operon and ITEA on the standardized data which had been used for hyperparameter tuning. For uDSR, this particular difference is likely due to the fact that the Hubble data, which uDSR performed well on, contains polynomial features which this algorithm is particularly optimized for. On the other hand, Operon did the best on the standardized data but performed the worst on the cosmological datasets. We also note that FFX failed to produce any correct equations for the standard datasets but did find one equation for the cosmological dataset. This overall supports our suggestion that researchers using SR should consider performing their own comparisons of SR algorithms rather than choosing an algorithm solely based on how well it does on ``standard'' datasets which we assess are mainly useful for developers of SR algorithms. Our results might also be highlighting the importance of hyperparameter tuning. Specifically, we note that to perform an unbiased benchmark, our hyperparameter tuning was done on standardized datasets and hence ``blind'' towards the cosmological datasets. It is likely that all (or some of) the algorithms would perform better on the cosmological datasets if hyperparameters had been tuned for each specific dataset. A simple example is the basis functions that the algorithms can use in their search. We use the default basis of each algorithm. A simple way to optimize the algorithms is to remove basis functions that are deemed irrelevant for ones data (e.g. removing sine and cosine functions from the basis if oscillatory behavior is not expected in the data). We also note that relying solely on MSE inside the training region for evaluating the goodness of fit for an expression may not be prudent when seeking to identify hitherto unknown ground truths. We for instance notice that for some of our datasets, the expressions with the smallest MSEs quickly become wildly incorrect outside the training regions while several expressions with somewhat larger MSEs are correct to around 1 \% precision far outside the training region.
    \newline\newline
    Although we did not discover any new physical relations in the presented benchmark, we believe the framework of cp3-bench and our presented benchmark could help the community reach such goals in the future, by providing a platform for easy access to (so far) 12 different algorithms that can be tuned and tested on the datasets relevant for a given task.

    Our findings lead us to conclude that having a benchmark suite like cp3-bench, with algorithms using a rich variety of ML methods, can be a prudent part of the strategy, when aiming at discovering new physics with ML methods. Our results overall indicate that there is still much room for enhancing the performance of symbolic regression algorithms. Based on our specific results, we identify possible paths for developers of symbolic regression algorithms. Firstly, one could consider methods for constraining the output using dimensional analysis. In principle, we do not need to consider dimensional analysis when doing symbolic regression since the pre-factors can be considered to implicitly contain any factors of e.g. $c, G$ necessary for the dimensions of ones results to be correct. However, including dimensional analysis in symbolic regression analysis has been discussed by others in e.g. \cite{physics_motivation, choi2011dimensionallyconstrainedsymbolicregression,  villar2022dimensionlessmachinelearningimposing} and might enhance the algorithms' abilities in relation to physics data. To our knowledge, a systematic study of the impact of using dimensional analysis on modern physics data has not been conducted. By leveraging cp3-bench, such systematic studies become much simpler than if one must install all algorithms from scratch. Secondly, algorithms might be improved by carefully considering the choice of complexity allowed. Some algorithms like PySR have very good features for controlling the nesting, while others lack such features entirely, merely focusing on the number of terms. This can lead to rather unrealistic terms such as $e^{e^{\cos(x)}}$ which rarely appears in physics formulae. See e.g. \cite{Kommenda_2015} for further discussions of complexity measures. In addition, the algorithms (not surprisingly) in particular become inadequate when the feature space has dimension above $\sim 2$. Thus, it could be useful if future development of symbolic regression algorithms for astrophysics purposes were focused e.g. on increasing performance of the algorithms on datasets with just a few more dimensions. Lastly, we note that, as also discussed in e.g. \cite{Michaud_2023}, ML algorithms are not generally aimed at the high precision required for identifying ground truth symbolic expressions. Methods aimed at increasing the precision of machine learning should therefore also be a focus when seeking to improve symbolic regression algorithms.
    
    As a future improvement of cp3-bench one could develop an abstraction layer using the \texttt{format\_output} function of the procedure class such that this would return the equation in a format compatible across the package. With this, one could compute the MSE from the output equation of the algorithm and as an option output the MSE from the output equation and not the underlying full model. Alternatively, one could also do a check with this and warn the user if the two methods of computing the MSE are significantly different. Having this abstraction layer could also help control the output format of the equation to and force them to a common and easier to use format. Currently, the output equations are largely shown as the underlying algorithms output them, which is not the same across the different algorithms. A further extension of cp3-bench could be to automate the comparison of output symbolic expressions with ground truths which must currently be done manually. Finally, one could also consider adding support for automatic tuning of the hyperparameters with an algorithm that can scan and optimize the hyperparameters of the different models.
    
\vspace{6pt} 
\begin{acknowledgments}
We thank the authors of \cite{benchmark} and in particular Fabricio Olivetti de França and William La Cava for correspondence, and we thank Héloïse Delaporte for help with a Mathematica script for converting symbolic expressions into easier readable formats and Patricia Sobotkova for help with Mac compatibility. We thank the anonymous referee for suggestions that have significantly improved the manuscript.
\newline\indent
SMK is funded by VILLUM FONDEN, grant VIL53032. Part of the numerical work done for this project was performed using the UCloud interactive HPC system managed by the eScience Center at the University of Southern Denmark.
\newline\newline
{\bf Author contribution statement}: MET wrote the benchmark code cp3-bench and Things-to-bench based on a technical study of the presented algorithms and a literature study conducted by SMK. MET performed hyperparameter tuning with minor contributions from SMK who also contributed to feedback on the installation. The considered datasets were produced by SMK. Both authors have contributed significantly to the writing of the manuscript and the development of the project.

\end{acknowledgments} 
	
\appendix
	
\section{Standard datasets}\label{app:StandardDataset}
    The table below shows the dataset we have used to tune hyperparameters. The dataset is a combination of symbolic expressions found in standard datasets. The dataset and parameter/feature ranges were selected by considering the datasets listed in appendix A of \cite{makke2023interpretable}, although parameter ranges were modified somewhat to improve the hyperparameter tuning.

    \begin{table}[!htb]
        \centering
              \begin{adjustbox}{max width=\textwidth}
        \begin{tabular}{c c c}
            \hline\hline
           Dataset name & Symbolic expression &  Parameter interval \\
            \hline
            \Tstrut
            F1 & $x^5 - 2x^3 +x$ & $x\in[-10,10], N = 1000$ \\
            F2 & $0.3\cdot x\sin(2\pi x)$ & $x\in[-10,10], N = 1000$\\
            F3 & $x^3\cdot e^{-x}\cdot \cos(x)\cdot \sin(x)\left( \sin^2(x)\cos(x)-1 \right)$ & $x\in[0,10], N = 1000$  \\
            \hline
            F4 & $2.5x^4 - 1.3x^3 + 0.5y^2 - 1.7y$& $x,y\in[-3,3], N = 100$\\
            F5 & $1.5e^x + 5\cos(y)$ & $x,y\in[-3,3], N = 100$\\
            F6 & $\frac{\exp\left( -(x-1)^2 \right)}{1.2+(y-2.5)^2}$ & $x,y\in[0.3,4], N = 100$\\
            \hline
            F7 &$0.23 + 14.2\frac{x+y}{3z}$ & $x,y,z\in[-5,5], N = 30$\\
            F8 & $ 6.78 + 11\sqrt{7.23xyz}  $ & $x,y,z\in[0,10], N = 30$\\
            \hline
        \end{tabular}
              \end{adjustbox}
        \caption{Dataset of standard symbolic expressions used for hyperparameter tuning. The dataset is a small selection of the datasets listed in appendix A of \cite{makke2023interpretable}. The parameter intervals are indicated by giving upper and lower limit in the interval and the number ($N$) of equidistant points inside the interval. When expressions contain more than one parameter, the parameter intervals were identical for all parameters, with $N$ indicating the number of data points in each dimension.}
        \label{table:Standard}
    \end{table}

\begin{table}
    \centering
    \begin{adjustbox}{max width=\textwidth}
    \begin{tabular}{ |c||c|c|c|c|c|c|c|c|c|c|c|c| }
    \hline
    \multicolumn{13}{|c|}{Test dataset results} \\
    \hline
    \hline
    &\multicolumn{12}{c|}{Algorithm} \\
    \hline
    Dataset & AI-Feynman & DSO  & DSR & FFX & GE & GPG & GPZGD & ITEA & Operon & PySR & QLattice & uDSR \\
    \hline
    F1 & $10^{1}$  & \checkmark  & \checkmark  & $10^{6}$  & \checkmark  & $10^{8}$ & $10^{5}$  & \checkmark & $10^{0}$ & \checkmark & \checkmark ($10^{-7}$)  & \checkmark \\
    \hline
    F2 & $10^{-1}$ & $10^{0}$ & $10^{0}$ & $10^{0}$ & $10^{0}$ & $10^{0}$  & $10^{-4}$ & $10^{0}$ & \checkmark & $10^{0}$ & $10^{0}$ & $10^{0}$ \\
    \hline
    F3 & $10^{12}$ & $10^{11}$ & $10^{10}$ & $10^{11}$ & $10^{11}$ & $10^{11}$ & $10^{10}$ &  $10^{12}$ & $10^{8}$ & $10^{10}$ & $10^{9}$  & $10^{10}$\\
    \hline
    F4 & $10^{2}$ & $10^{1}$ & $10^{1}$ & $10^{1}$ & $10^{1}$ & $\checkmark(10^{-11})$ & $10^{0}$ & \checkmark & $10^{0}$ & $10^{0}$ & \checkmark  ($10^{-8}$) & \checkmark \\
    \hline
    F5 & $10^{-2}$  & $10^{1}$ & $10^{1}$ & $10^{0}$ & $10^{1}$ & $10^{-2}$ & $10^{-2}$ & \checkmark & \checkmark  & \checkmark  & $10^{-9}$ & $10^{-5}$\\
    \hline
    F6 & $10^{-1}$ & $10^{-2}$ & $10^{-3}$ & $10^{-5}$ & $10^{-2}$ & $10^{-4}$ & $10^{-3}$ & $10^{-3}$ & $10^{-3}$ & $10^{-3}$ & $10^{-9}$ & $10^{-7}$\\
    \hline
    F7 & \checkmark & $10^{2}$ & $10^{2}$ & $10^{4}$ & $10^{2}$ & \checkmark & $10^{4}$ & $10^{4}$ & \checkmark  & \checkmark  & \checkmark ($10^{-6}$) & \checkmark \\
    \hline
    F8 & $10^{2}$ & $10^{3}$ & $10^{3}$ & $10^{2}$ & $10^{1}$ & $10^{2}$ & $10^{3}$ & \checkmark & \checkmark & $10^{3}$ & $10^{2}$ & $10^{2}$\\
    \hline
    \end{tabular}
    \end{adjustbox}
    \caption{Result of the tuning of parameters on the test datasets.}
    \label{fig:test-results}
\end{table}
\newpage

\section{Detailed results}\label{app:full-results}
     This section contains the raw results for the all the cosmological datasets. This means we report what the algorithm outputs, up to algebraic simplification. Numbers of the order $\mathcal{O}(10^{-10})$ or smaller are set to zero, and we round to 5 significant digits if the algorithm reports more than that. To increase readability we reduce the number of significant digits to 5 (in some cases down to 2 significant digits, see the table caption). Note that some algorithms already have a rewriting method that rounds to 5 or 6 significant digits. The runtime is given in the format HH:MM:SS (hours, minutes, seconds).
     
    \begin{table}
        \centering
\begin{tabular}{ |p{2.52cm}||p{2.15cm}|p{7.54cm}|p{1.43cm}| }
         \hline
         \multicolumn{4}{|c|}{Cosmological dataset: C1a} \\
         \hline
         \hline
         Algorithm & MSE & Equation & Runtime\\
         \hline
         AI-Feynman     & $4.6203\cdot 10^{-8}$ & $0.070053 + 0.10276 e^{\sqrt{e^{-z}}} z$&  08:53:25\\
         DSO            & $7.6999\cdot 10^{-5}$ & $\sin \paa{\sqrt{\frac{e^z}{z + e^5}}}$  & 01:26:47\\
         DSR            & $1.7932\cdot 10^{-5}$ & $\frac{z}{2\sqrt{z} + 4z +2 \sin \paa{2z}}$ &  01:01:03\\
         FFX            & $1.7949\cdot 10^{-6}$ & $0.0719 + 0.0394z + 0.015z^2$ & 00:00:08\\
         Genetic Engine & $3.0572\cdot 10^{-6}$ & $\frac{1}{-0.01 + z) z}\log(1 + \abs{0.1 - \frac{0.0001}{z} - 0.0011 z}) \bigg(-0.001 +  z \big(-0.000326228 + z (0.0226228 + z) \big) + (-0.01 + z) z \sqrt{\log \paa{1 + \frac{0.1}{\abs{z}}}}\bigg)$   & 04:00:02\\
         GPG            & $1.6301\cdot 10^{-3}$ & $0.13409$ & 00:08:40\\
         GPZGD          & $3.0792\cdot 10^{-13}$ & $0.13409 - 0.0011566 (211.91 + z (23.471 + z (3.8176 + z))) (\sin\paa{0.018917 + 0.07528 z} - \sin\paa{\sin \paa{0.24493 z}})$ & 04:00:01\\
         ITEA           & $1.5824\cdot 10^{-14}$ & $0.0717424 - 7.29851\cdot 10^{-6} e^{z^2} + 0.030679 z^2 - 0.0042627 \abs{z}^{3/2} - 0.0001696 \cos \paa{z^2} + 0.0332752 \sin \paa{z}$ & 00:07:08\\
         Operon         & $1.6917\cdot 10^{-5}$ & $0.05763 + 0.073092 z$ & 00:00:14\\
         PySR           & $2.1030\cdot 10^{-9}$ & $0.071640 + 0.028106 z^2 + 0.031315 \sin(z)$ & 00:35:36\\
         QLattice       & $1.8361\cdot 10^{-10}$ & $-0.11561 + 0.18718 e^{-0.53161 z} + 0.13165 z$ & 01:32:31\\
         uDSR           & $1.1703\cdot 10^{-30}$ & $0.07159 \sqrt{1 + z (0.9 + (0.9 + 0.3 z) z)}$ & 00:00:13\\
         \hline
        \end{tabular}
        \caption{Results from benchmarking the Hubble dataset C1a.}
        \label{tab:C1a}
    \end{table}
    \begin{table}
        \centering
        \begin{tabular}{ |p{2.52cm}||p{2.15cm}|p{7.54cm}|p{1.43cm}|  }
         \hline
         \multicolumn{4}{|c|}{Cosmological dataset: C1b} \\
         \hline
         \hline
         Algorithm & MSE & Equation & Runtime\\
         \hline
         AI-Feynman     & $2.0738\cdot 10^{-4}$ & $0.53006 + \log\paa{\cos\paa{\cos\paa{\frac{1}{\pi - \sqrt{1 + z}}}}}$ & 08:53:25\\
         DSO            & $2.5167\cdot 10^{-4}$ & $e^{-3+\sqrt{z}}$ & 01:53:07\\
         DSR            & $2.1320\cdot 10^{-4}$ & $\frac{z}{z^{1/4} + 5z +2 \sin \paa{2z}}$ & 01:02:47\\
         FFX            & $2.0062\cdot 10^{-4}$ & $0.083 + 0.0397 z + 0.0143 \max \paa{0, -1.24 + z} + 0.00010654 \max \paa{0, 1.75 - z} \max (0, -0.987 + z) +  0.0149 \max (0, -0.48 + z)$ & 00:00:08\\
         Genetic Engine & $2.0889\cdot 10^{-4}$ & $ \log\paa{1 + \frac{\abs{\frac{-0.0041623 + z (4.0623 + 100z)}{z - 1000 \log\paa{1 + \abs{z}})}}}{\log\paa{1 + 2 \abs{z}}}}$ & 04:00:02\\
         GPG            & $1.9977\cdot 10^{-3}$ & $0.13401$ & 00:09:30\\
         GPZGD          & $5.1140\cdot 10^{-4}$ & $0.13401 +  0.043802 (0.18649 z (1.8686 + z)^2 + 0.153168 (-4.785 + z) z \sin \paa{\sin(z)})$ & 03:42:23\\
         ITEA           & $1.9684\cdot 10^{-4}$ & $0.16115 + 0.00210702 z^3 - 0.0703969 \cos{z} -  0.00078241 \ln \paa{\frac{1}{z^3}} + 0.0039121 \ln(z^5)$ & 00:07:10\\
         Operon         & $2.2276\cdot 10^{-5}$ & $0.05603 + 0.074546 z$ & 00:00:13\\
         PySR           & $3.1870\cdot 10^{-4}$ & $0.077633 z + 0.077633 \cos\paa{\sin\paa{z}}$ & 00:34:39\\
         QLattice       & $1.9738\cdot 10^{-4}$ & $0.19878 - 0.16512 \tanh\paa{1.04296 - 0.56534 z}$ & 01:11:31\\
         uDSR           & $1.9544\cdot 10^{-4}$ & $0.931898 + z(0.278604 + z (-1.30941 + z (2.60802 + (-2.07853 + 0.576118 z) z))) +  \sin\paa{e^{z + \cos\paa{e^{\sin(z)}}}}$ & 02:03:41\\
         \hline
        \end{tabular}
        \caption{Results from benchmarking the Hubble dataset C1b.}
        \label{tab:C1b}
    \end{table}
    
    \begin{table}
        \centering
        \begin{tabular}{ |p{2.52cm}||p{2.15cm}|p{7.54cm}|p{1.43cm}|   }
         \hline
         \multicolumn{4}{|c|}{Cosmological dataset: C1c} \\
         \hline
         \hline
         Algorithm & MSE & Equation & Runtime\\
         \hline
         AI-Feynman     & $6.8164\cdot 10^{-6}$ & $0.39026 - \frac{1}{\pi + e^z \Omega_{m,0} - \Omega_{m,0} \cos(z)}$ & 08:53:26\\
         DSO            & $1.8711\cdot 10^{-4}$ & $e^{-e^{\frac{1}{1 + \frac{1}{\sqrt{\Omega_{m,0}}}}}} \sqrt{z \Omega_{m,0}}$ & 02:03:08\\
         DSR            & $1.4347\cdot 10^{-4}$ & $\Omega_{m,0} \sin \paa{\frac{\sqrt{e^z}}{e^{\sqrt{\Omega_{m,0} + \sqrt{2} \Omega_{m,0}}} + 4 \Omega_{m,0}}}$ & 01:18:44\\
         FFX            & $3.1016\cdot 10^{-6}$  & $0.0758 + 0.00387 z^2 + 0.156 z \Omega_{m,0} -  0.00854 \Omega_{m,0}^2 + 0.01711 z \ln(z)$ & 00:00:33\\
         Genetic Engine & $3.6670\cdot 10^{-5}$ & $\ln\big(1 + \big| -10^{-6} + \Omega_{m,0} (0.00891 + 0.99 \Omega_{m,0} + z (99\cdot 10^{-5}(1 -  z) - 0.99 \Omega_{m,0} +0.99 z \Omega_{m,0})) +(-0.0001 + 0.99 \Omega_{m,0}) \sqrt{\abs{z}}   \Omega_{m,0} + 10 \ln \paa{1 + \abs{\Omega_{m,0}}} \big|\big) -  \ln{ \abs{\Omega_{m,0} + 10 \ln\paa{1 + \abs{\Omega_{m,0}}}}}$ & 04:00:02\\
         GPG            & $3.6054\cdot 10^{-8}$ & $0.088766 + \paa{0.016939 + \frac{ 0.0019861 \Omega_{m,0} \cos(z)}{z+  \cos(5.558 - z)}} \times \bigg(\frac{(2.7357 +       0.02813 \Omega_{m,0}) \Omega_{m,0} z (4.1279 + z)}{1.0522 + \Omega_{m,0}} - 1.113 \sin(2.0505 - z)\bigg)$ & 00:52:44\\
\hline
        \end{tabular}
        \caption{Results from benchmarking the Hubble dataset C1c. We include only 4 significant digits for GPZGD to increase readability.}
        \label{tab:C1c}
    \end{table}
    
    \begin{table}
        \centering
        \begin{tabular}{ |p{2.52cm}||p{2.15cm}|p{7.54cm}|p{1.43cm}|  }
         \hline
         \multicolumn{4}{|c|}{Cosmological dataset: C1d} \\
         \hline
         \hline
         Algorithm & MSE & Equation & Runtime\\
         \hline
         AI-Feynman     & $4.457\cdot 10^{-5}$ & $-0.0005896+ e^{z \sqrt{\frac{\Omega_{m,0}}{\cos\paa{\Omega_{m,0}}}}} H_0$ & 08:53:37\\
         DSO            & $4.142\cdot 10^{-6}$ & $H_0 + \sin \bigg(H_0 z \times \sqrt{\Omega_{m,0} (z + \paa{2^{\frac{1}{4}} H_0^{\frac{1}{4}} + \sqrt{z} + z} \Omega_{m,0})}\bigg)$ & 02:03:23\\
         DSR            & $1.267\cdot 10^{-5}$ & $e^{z \sqrt{\Omega_{m,0}}} H_0$ & 01:09:36\\
         FFX            & $1.396\cdot 10^{-5}$ & \footnotesize $0.0279 + 1.6 H_0^2 - 0.0154 z + 0.00453 z^2 - 0.0673 \Omega_{m,0} + 0.0864 z \Omega_{m,0} - 0.0335 \Omega_{m,0}^2 + H_0 (0.253 + 0.509 z + 1.76 \Omega_{m,0}) + 0.0165 \max(0, -1.75 + z) + 0.0155 \max(0, -1.49 + z) + 0.0174 \max
         (0, -1.24 + z) + 0.0169 \max(0, -0.987 + z) + (0.382 H_0 - 0.011 z + 0.00573 \Omega_{m,0}) \max(0, -0.733 + z) + (0.00319 + 0.118 H_0 - 0.0153 z +0.0282 \Omega_{m,0}) \max(0, -0.48 + z) + (0.000791 + 0.0154 \Omega_{m,0} - 0.0466 \max(0, -0.48 + z)) \max(0, 0.18 - \Omega_{m,0}) +(0.000012 \Omega_{m,0} - 0.0157 \max(0, -0.48 + z)) \max(0, 0.287 - \Omega_{m,0}) + \max(0, 0.0914 - H_0) (-0.0187 + 1.56 H_0 - 0.105 z -0.2 \Omega_{m,0} + 0.471 \max(0, 0.18 - \Omega_{m,0}) + 0.295 \max(0, 0.287 - \Omega_{m,0}) + 0.124 \max(0, 0.34 - \Omega_{m,0}) + (-0.0241 z + 0.0494 \Omega_{m,0}) \max(0, 0.34 - \Omega_{m,0})$ & 00:04:25\\
         Genetic Engine & $8.032\cdot 10^{-6}$ & $\ln\Big(1 + \bigg| 2 H_0^2 (1 + z) + 0.01 z \Omega_{m,0} + H_0 (0.02 + z (1 + z) \Omega_{m,0}) + \log\paa{1 + \abs{\frac{H_0}{1 + H_0}}}\bigg|\Big)$ & 04:00:03\\
\hline
        \end{tabular}
        \caption{Results from benchmarking the Hubble dataset C1d. To increase readability, we only include 4 significant digits.}
        \label{tab:C1d}
    \end{table}
    \begin{table}
        \centering
        \begin{tabular}{ |p{2.52cm}||p{2.15cm}|p{7.54cm}|p{1.43cm}| }
         \hline
         \multicolumn{4}{|c|}{Cosmological dataset: C2a} \\
         \hline
         \hline
         Algorithm & MSE & Equation & Runtime\\
         \hline
         AI-Feynman     & $1.3715\cdot 10^{-10}$ & $0.071568 - H + 0.071611 z$ & 08:53:26\\
         DSO            & $3.6475\cdot 10^{-4}$ & $e^{-e^{\cos{H}} - H + \sqrt{z}} - H$ & 03:02:26\\
         DSR            & $1.5928\cdot 10^{-4}$ & $-2 H + \sin \paa{\frac{(H^2 z)^{1/4}}{e^{1/4}}}$ & 01:20:14\\
         FFX            & $8.4654\cdot 10^{-6}$ & $0.0704 - 0.946 H + 0.066 z$ & 00:00:20\\
         Genetic Engine & $3.9676\cdot 10^{-5}$ & $1/\Bigl(H (-0.01 + z) z \log\paa{1 + \abs{z}}\Bigr)\times \Bigl(H (H^2 (0.002 - 0.2 z) + 0.01 z^2 - 0.1 H z^2 +  H^3 (-0.02 + 2z))+(0.028748 H z^2 +  H^3 (0.0057496 + (-0.55965 - 1.5311 z) z)+(-0.00031623 +  0.031623 z) z^2 \log\paa{1 + 0.1 \abs{z}}) \log\paa{1 + \abs{z}}\Bigr) $ & 04:00:03\\
         GPG            & $4.6841\cdot 10^{-10}$ & $0.071639 - 1.0001 H + 0.071558 z$ & 00:43:55\\
         GPZGD          & $1.6298\cdot 10^{-31}$ & $0.017062 - 0.05722 H + 0.0395799 z$ & 04:00:04\\
         ITEA           & $1.5181\cdot 10^{-32}$ & $0.07159 - H + 0.07159 z$ & 00:37:57\\
         Operon         & $1.1035\cdot 10^{-3}$ & $0.09842 - 0.62533 H$ & 00:00:26\\
         PySR           & $1.0775\cdot 10^{-31}$ & $0.07159 - H + 0.07159 z$ & 01:19:06\\
         QLattice       & $2.1592\cdot 10^{-18}$ & $0.07159 - H + 0.07159 z$ & 01:34:42\\
         uDSR           & $2.1666\cdot 10^{-30}$ & $0.07159 - H + 0.07159 z$ & 00:00:33\\
         \hline
        \end{tabular}
        \caption{Results from benchmarking the redshift dataset C2a.}
        \label{tab:C2a}
    \end{table}    
    \begin{table}
        \centering
        \begin{tabular}{ |p{2.52cm}||p{2.15cm}|p{7.54cm}|p{1.43cm}|  }
         \hline
         \multicolumn{4}{|c|}{Cosmological dataset: C2b} \\
         \hline
         \hline
         Algorithm & MSE & Equation & Runtime\\
         \hline
         AI-Feynman     & $2.5761\cdot 10^{-3}$ & $0.00061452 \sqrt{\cos\paa{z - \omega}}$ & 08:53:26\\
         DSO            & $8.7609\cdot 10^{-4}$ & $e^{-2 z} z^3 \omega \sin\paa{\omega}$ & 02:14:18\\
         DSR            & $8.2501\cdot 10^{-4}$ & $z \omega \cos\paa{\omega} \paa{\omega - \sin\paa{\omega}}$ & 01:02:51\\
         FFX            & $8.3131\cdot 10^{-6}$ & $0.00088 + z (-0.0127 - 0.00114 z - 0.0861 \omega) + (-0.0058 z - 0.152 \omega) \max\paa{0, 0.264 + \omega} + (0.0113 - 0.14 z - 0.0676 \Omega) \max\paa{0,0.396 + \omega} + 0.0596 \omega \max{0, 0.66 + \omega} +0.00984 \omega \max\paa{0, 0.792 + \omega}$ & 00:00:29\\
         Genetic Engine & $3.2259\cdot 10^{-5}$ & $\frac{1}{(0.1 + z) \sqrt{\abs{\omega}} \sqrt{\abs{-\frac{0.01}{\omega} + \ln\paa{2}}}}\times \big( 0.00001 +  z (-0.000911 + (-0.009019 - 0.00009 z) z) + (10^{-7} - 0.0001 z^4 +z^3 (-0.010021 - 0.00101 \omega) + (0.000011 + 0.0001 \omega) \omega +z (9.9\cdot10^{-7} + (-0.0000911 - 0.01001 \omega) \omega)+z^2 (-0.0010012 + (-0.10111 - 0.0001 \omega) \omega)) \sqrt{\abs{\omega}}\big)$ & 04:00:03\\
         GPG            & $6.7220\cdot 10^{-8}$ & $-0.000082 - 0.090442 z \big(1.6273 - \frac{2 z + 2\omega  (z + \omega ) - \cos(z)}{-3.957 +   z (1.333 + z) \omega }\big) \paa{0.32197 + \sin\paa{\sin(\omega)}}$ & 00:57:17\\
\hline
        \end{tabular}
        \caption{Results from benchmarking the redshift dataset C2b.}
        \label{tab:C2b}
    \end{table}
    \begin{table}
        \centering
        \begin{tabular}{ |p{2.52cm}||p{2.15cm}|p{7.54cm}|p{1.43cm}|  }
         \hline
         \multicolumn{4}{|c|}{Cosmological dataset: C3a} \\
         \hline
         \hline
         Algorithm & MSE & Equation & Runtime\\
         \hline
         AI-Feynman     & $2.7984\cdot 10^{-6}$ & $-0.34629 + 2 z + \sin \paa{\ln\paa{1 + \frac{1}{\sqrt{z}}}}$ & 08:53:25\\
         DSO            & $1.1025\cdot 10^{-4}$ & $e^{\sin\paa{\sqrt{z^{9/4}}}} \sin\paa{z}^{1/8}$ & 01:51:17\\
         DSR            & $9.4667\cdot 10^{-5}$ & $\sqrt{z + \paa{\sqrt{\sqrt{z} \sqrt{z^2}} \sqrt{\sqrt{z} + z}}^{1/4}} + \sin\paa{z^2}$ & 01:22:39\\
         FFX            & $1.8697\cdot 10^{-4}$ & $0.854 + 1.07 z - 0.106 \max\paa{0, 0.88 - z} + 0.00322 \max\paa{0, -0.64 + z} +0.0764 \max\paa{0, -0.52 + z} + 0.0869 \max\paa{0, -0.4 + z} + 0.329 \max\paa{0, -0.28 + z}$ & 00:00:09\\
         Genetic Engine & $7.7503\cdot 10^{-6}$ & $-0.0101 + \frac{0.01}{(-0.001 + z)} + z + \sqrt{ \ln\paa{1 + \abs{z}^{1/4}}} + \ln\paa{1 + \abs{-0.1 + z}} (0.01 z + \ln\paa{1 + \abs{z}})$ & 04:00:02\\
         GPG            & $1.7598\cdot 10^{-1}$ & $1.5248$ & 00:08:49\\
\hline
        \end{tabular}
        \caption{Results from benchmarking the two-region backreaction dataset C3a.}
        \label{tab:C3a}
    \end{table}
    \begin{table}
        \centering
                    \begin{adjustbox}{max width=\textwidth}
        \begin{tabular}{ |p{2.55cm}||p{2.2cm}|p{8.4cm}|p{1.4cm}|  }
         \hline
         \multicolumn{4}{|c|}{Cosmological dataset: C3b} \\
         \hline
         \hline
         Algorithm & MSE & Equation & Runtime\\
         \hline
         AI-Feynman     & $1.4727\cdot 10^{-4}$ & $1.74408 \paa{-1 + \cos\paa{\ln\paa{\frac{z}{1 + \cos\paa{\cos\paa{\sqrt{z}}}}}}}$ & 08:53:25\\
         DSO            & $6.0531\cdot 10^{-3}$ & $-e^{\cos\paa{\sqrt{z}}} + \sqrt{z} + \cos\paa{\frac{1}{\sqrt{z}}}$ & 02:01:01\\
         DSR            & $6.3937\cdot 10^{-4}$ & $\frac{z - \sqrt{\cos\paa{z}} - 2 \cos\paa{z}}{3 z + \cos\paa{\sqrt{1 + z}}}$ & 00:54:03\\
         FFX            & $2.0215\cdot 10^{-4}$ & $\frac{-1.3112 + 0.70091 z}{0.30212 + z (0.53112 + z)}$ & 00:00:07\\
         Genetic Engine & $4.6980\cdot 10^{-4}$ & $\frac{1}{(0.01 + z)(-0.0075724 + z (-0.024276 + z))}\times\frac{1}{(0.13206 + z (-1.7439 + z (9.8243 + 1. z)))^2}\times\bigg(0.00001258 + z (-0.00029719 +  z (-0.000095321 +  z (0.081972 + z (-1.2064 + z (7.55163 +  z (-16.908 + z (-28.497 + z (100.011 + z (0.899891 + (-2.93686 - 0.2 z) z)))))))))\bigg)-\frac{ 0.99 \sqrt{\abs{z}}}{0.01 + z}$  & 04:00:02\\
         GPG            & $6.8084\cdot 10^{-1}$ & $-1.31$ & 00:08:15\\
\hline
        \end{tabular}
        \end{adjustbox}
        \caption{Results from benchmarking the two-region backreaction dataset C3b.}
        \label{tab:C3b}
    \end{table}
    
    \begin{table}
        \centering
        \begin{adjustbox}{max width=\textwidth}
        \begin{tabular}{ |p{2.52cm}||p{2.15cm}|p{7.54cm}|p{1.43cm}| }
         \hline
         \multicolumn{4}{|c|}{Cosmological dataset: C3c} \\
         \hline
         \hline
         Algorithm & MSE & Equation & Runtime\\
         \hline
         AI-Feynman     & $1.7755\cdot 10^{-3}$ & $5.723 \sin\paa{\sin\paa{f_o (1 + f_o - \cos\paa{e^z})}}$ & 08:53:25\\
         DSO            & $3.2599\cdot 10^{-3}$ & $4 f_o (1 + z + z^2)$ & 01:46:58\\
         DSR            & $2.3962\cdot 10^{-3}$ & $f_o \paa{4 + (4 + e^z) z + z \sin\paa{z}}$ & 01:30:05\\
         FFX            & $4.7415\cdot 10^{-4}$ & $0.247 + 1.48 f_o^2 + (0.0229 + 0.33 z) z + f_o (3.09 + 5.93 z) +0.11769  \ln\paa{f_o}$ & 00:00:14\\
         Genetic Engine & $4.2219\cdot 10^{-4}$ & $\frac{9.95037}{\sqrt{|(0.1 + z)}}\sqrt{\log\paa{1 + \abs{z}}}\times \sqrt{ |(f_o (-0.01 + f_o + z^2) (-0.101 + \sqrt{ \abs{f_o}}))|} $ & 04:00:03\\
         GPG            & $1.0670\cdot 10^{-5}$ & $-0.04437 + 5.0559 f_o (0.62133 + f_o + z + z^2) \cos(2.7819 f_o z) \cos\big(7.643 (-0.96138 + f_0) f_0 (f_o - 0.45282 \cos(f_o))\big)$ & 00:16:09\\
\hline
        \end{tabular}
        \end{adjustbox}
        \caption{Results from benchmarking the two-region backreaction dataset C3c.}
        \label{tab:C3c}
    \end{table}
    \begin{table}
        \centering
        \begin{tabular}{ |p{2.52cm}||p{2.15cm}|p{7.54cm}|p{1.43cm}| }
         \hline
         \multicolumn{4}{|c|}{Cosmological dataset: C3d} \\
         \hline
         \hline
         Algorithm & MSE & Equation & Runtime\\
         \hline
         AI-Feynman     & $5.5780\cdot 10^{-3}$ & $-3.31857 \paa{1 + \ln\paa{ \frac{\cos\paa{1 + f_o}}{\sqrt{\sin\paa{z}}}}}$ & 08:53:25\\
         DSO            & $4.7046\cdot 10^{-2}$ & $-e^{\cos\paa{e^{f_o + z}} + \cos(z)} + f_o$ & 02:09:10\\
         DSR            & $5.1333\cdot 10^{-2}$ & $-\frac{z + \cos\paa{e^{f_o + f_o (e^z + 2 f_o) z^{3/2}}}}{z}$ & 01:26:32\\
         FFX            & $1.1563\cdot 10^{-3}$ & $\frac{-1.109 + 0.24877 z + f_o (1.6621 + 3.2234 f_o + 1.0082 z)}{0.27248 + (0.21063 +  0.10872 z) z + f_o (-0.73297 + z)}$ & 00:00:14\\
         Genetic Engine & $4.8526\cdot 10^{-3}$ & $\frac{-0.969377 + (0.002 + 2. f_o) z + \sqrt{\abs{0.1 + f_o}} \sqrt{\ln\paa{1 +  \abs{z}}}}{\ln\paa{1 + \sqrt{\abs{0.01 + 0.1 z + f_o (0.1 + z)}}}}$& 04:00:02\\
         GPG            & $1.8124\cdot 10^{-5}$ & $-\frac{10.228 \cos\paa{\frac{6.8966 f_o (f_o -  z)}{3.313 + z}}}{-1.518 + f_o + \frac{0.40741 f_o^2)}{z} + \frac{(-1.0701 + f_o) \cos(5.4249 - f_o)}{ 0.0545889 + z}}-6.12395 $ & 00:15:22\\
\hline
        \end{tabular}
        \caption{Result from benchmarking the two-region backreaction dataset C3d.}
        \label{tab:C3d}
    \end{table}    
    \begin{table}
        \centering
        \begin{tabular}{ |p{2.52cm}||p{2.15cm}|p{7.54cm}|p{1.43cm}| }
         \hline
         \multicolumn{4}{|c|}{Cosmological dataset: C3e} \\
         \hline
         \hline
         Algorithm & MSE & Equation & Runtime\\
         \hline
         AI-Feynman     & $2.1808\cdot 10^{-3}$ & $2.0406 e^{\sec \paa{1 + \Omega_{Q,0} }\sin\paa{z}} \Omega_{m,0}$ & 08:53:25\\
         DSO            & $1.2224\cdot 10^{-3}$ & $2 \Omega_{m,0} + z \paa{\Omega_{m,0} + \sqrt{2 \Omega_{m,0} + 2 z \Omega_{m,0}}}$ & 01:50:32\\
         DSR            & $1.2318\cdot 10^{-3}$ & $2 \Omega_{m,0} +  z \bigg(\sqrt{2} \sqrt{\Omega_{m,0}} +  \sqrt{\Omega_{m,0} \paa{\Omega_{m,0} +z \sqrt{ \sin\paa{2 \Omega_{m,0}}}}}\bigg)$ & 01:41:27\\
         FFX            & $5.6198\cdot 10^{-4}$ & $0.976 + z (0.0709 + 0.341 z - 3.94 \Omega_{Q,0}) - 0.861 \Omega_{Q,0} + \ln\paa{z} (0.12464 z - 0.041117 \ln\paa{\Omega_{m,0}}) + \ln\paa{\Omega_{m,0}} (0.35569 + 0.43758 \ln \paa{\Omega_{R,0}})$ & 00:00:22\\
         Genetic Engine & $1.3613\cdot 10^{-3}$ & $\Big((0.00031623 + 0.031623 z) z^5 + 0.001 \Omega_{m,0} +z^2 \ln \big(1 + \ln \paa{1 + \abs{\Omega_{Q,0}}}\big)\Big)/\Big(z^2 \ln \big( 1 + \frac{0.1}{\sqrt{\abs{z (-0.01 + \Omega_{R,0})}}}\big)\Big)$ & 04:00:02\\
         GPG            & $6.8433\cdot 10^{-6}$ & $1.3725 + 1.0573\Omega_{Q,0} +  ( 0.30976 \Omega_{m,0} (H_0 +\Omega_{Q,0}) ((4.7922 +  H_0)\Omega_{Q,0} +  z (-1 + \Omega_{Q,0})))/(\Omega_{Q,0}^2 (0.60834 +  z (-0.1516 + z) \Omega_{m,0} (H_0 +\Omega_{Q,0}) \cos\paa{6.9275\Omega_{Q,0}})) - 1.0572 \cos \paa{H_0 (-1.2993 + z) + z +   
         \sin(\Omega_{m,0})} + (-1.0218 H_0 - 1.0218\Omega_{Q,0}) \cos(H_0) \sin(z)$ & 00:18:55\\
\hline
        \end{tabular}
        \caption{Result from benchmarking the two-region backreaction dataset C3e.}
        \label{tab:C3e}
    \end{table}    
    \begin{table}
        \centering
        \begin{tabular}{ |p{2.52cm}||p{2.15cm}|p{7.54cm}|p{1.43cm}| }
         \hline
         \multicolumn{4}{|c|}{Cosmological dataset: C3f} \\
         \hline
         \hline
         Algorithm & MSE & Equation & Runtime\\
         \hline
         AI-Feynman     & $1.7502\cdot 10^{-4}$ & $-6.60223 \paa{\Omega_{R,0} - \sin\paa{\sqrt{\sin\paa{\frac{z}{1 + \Omega_{Q,0}}}}}}$ & 08:53:26\\
         DSO            & $2.5639\cdot 10^{-2}$ & $\frac{(z + \Omega_{m,0} -  2 \Omega_{R,0}) \Omega_{R,0}}{\sqrt{z}}$ & 02:16:06\\
         DSR            & $8.3090\cdot 10^{-3}$ & $-e^{\cos\paa{4 z}} + z - \Omega_{Q,0} -\cos\paa{\sqrt{\Omega_{m,0}} + 2 \Omega_{m,0}} - \cos\paa{\Omega_{Q,0}}$ & 01:11:15\\
         FFX            & $1.5102\cdot 10^{-3}$ & $0.985 + 0.19 \Omega_{m,0} - 1.03 \Omega_{Q,0} - 2.42 \Omega_{R,0} + 1.46 \abs{\Omega_{Q,0}} + 1.0684 \ln\paa{z} - 1.1205 \ln\paa{\Omega_{R,0}} -  0.413 \max \paa{0, 0.4 - z} - 0.472 \max \paa{0, 0.52 - z} -  0.329 \max \paa{0, 0.64 - z} - 0.248 \max \paa{0, 0.76 - z}$ & 00:00:20\\
         Genetic Engine & $3.2492\cdot 10^{-2}$ & $\frac{1}{z \Omega_{Q,0} \sqrt{\abs{\frac{(H0 + z) \Omega_{Q,0}}{-1 +  z}}}} (\Omega_{Q,0} (0.00010207 + \Omega_{Q,0} \ (0.0010207 + 0.10207\Omega_{R,0}) +   0.010207 \Omega_{R,0}) +   H_0 (-0.00010207 -  0.010207 \Omega_{R,0} + \Omega_{Q,0} (0.00010207 - 0.10207 z + 0.010207 \Omega_{R,0} -   10.207 z \Omega_{R,0})))$ & 04:00:03\\
         GPG            & $1.4640\cdot 10^{-5}$ & $-2.7969 + \frac{\frac{7.0644 H_0}{z - \Omega_{Q,0}} +  5.7685 z\Omega_{Q,0} + 5.7685 \cos(H_0)}{3.7189 + H_0 + (0.26466 - z) z + \sin(z)} + \frac{5.7685 H_0 - 5.7685\Omega_{R,0} + 5.7685 \sin(z)}{z \paa{1.1555 +\frac{1}{\Omega_{R,0}}} +\Omega_{R,0}} - 5.7684\Omega_{Q,0} \sin\paa{\cos((z -\Omega_{R,0}) \sin(\Omega_{m,0}))}$ & 00:18:28\\
\hline
        \end{tabular}
        \caption{Result from benchmarking the two-region backreaction dataset C3f.}
        \label{tab:C3f}
    \end{table}    
    \begin{table}
        \centering
        \begin{tabular}{ |p{2.52cm}||p{2.15cm}|p{7.54cm}|p{1.43cm}| }
         \hline
         \multicolumn{4}{|c|}{Cosmological dataset: C3g} \\
         \hline
         \hline
         Algorithm & MSE & Equation & Runtime\\
         \hline
         AI-Feynman     & $1.0899\cdot 10^{-3}$ & $\frac{9.52774 v}{v + \sin \paa{e^h + H}}$ & 08:53:26\\
         DSO            & $1.3183\cdot 10^{-2}$  & $\frac{v}{\sqrt{e^{h + H}} \sin(H)}$ & 02:18:15\\
         DSR            & $8.5126\cdot 10^{-3}$ & $3 v + \sqrt{\sin\paa{\sqrt{v}}} + \sin\paa{e^{-e^v (-1 + h + v)} v}$ & 01:28:53\\
         FFX            & $4.1661\cdot 10^{-4}$ & $1.47 + (2.52 + 0.955 v) v - 0.110311 \ln \paa{H} + 0.212804 \ln \paa{v} + 154 v \max \paa{0, 0.0952 - H} + \max \paa{0, 0.0875 - H}(266. v + 1447. \max \paa{0, 0.0952 - H}) +  64.9 v \max \paa{0, 0.103 - H} +28.4 v \max \paa{0, 0.111 - H} + (16.5 v + 28.2 \max \paa{0, 0.111 - H}) \max \paa{0,   0.119 - H} - 0.877 \max \paa{0, 0.242 - v} - 0.492 \max \paa{0, 0.294 - v} -  1.04 \max \paa{0, 0.346 - v}$ & 00:00:20\\
         Genetic Engine & $4.9531\cdot 10^{-4}$ & $\ln\paa{1 + \abs{0.001 + H + 2 v + 10 h^2 v}}+ \frac{\ln\paa{1 + \ln\paa{1 + \abs{-0.01 + v}}}}{\ln\paa{1 + \abs{-0.001 + H + 0.1 v}}}$ & 04:00:02\\
         GPG            & $5.0753\cdot 10^{-5}$ & $-0.012285 +  2.7879 v (1.19046 + h (h + H - 2 v) + 3.2305 v - 4.6911 H v + \frac{ h - 5.786 H + v}{-1.293 + \cos(H)})$ & 00:15:36\\
\hline
        \end{tabular}
        \caption{Result from benchmarking the two-region backreaction dataset C3g.}
        \label{tab:C3g}
    \end{table}
    \begin{table}
        \centering
        \begin{tabular}{ |p{2.52cm}||p{2.15cm}|p{7.54cm}|p{1.43cm}| }
         \hline
         \multicolumn{4}{|c|}{Cosmological dataset: C3h} \\
         \hline
         \hline
         Algorithm & MSE & Equation & Runtime\\
         \hline
         AI-Feynman     & $4.8209\cdot 10^{-4}$ & $\frac{3.503 \Omega_{m,0}}{\ln \paa{\Omega_{m,0} - \frac{1}{\ln \paa{\sin\paa{a_D}}}}}$ & 08:53:26\\
         DSO            & $2.0520\cdot 10^{-3}$ & $\Omega_{m,0} + \frac{\Omega_{m,0}}{a_D\sin(a_D) \sin( a_D + \Omega_{m,0})}$ & 01:37:21\\
         DSR            & $1.6317\cdot 10^{-3}$ & $\Omega_{m,0} + \frac{\Omega_{m,0}}{a_D^2 (\Omega_{m,0} (a_D + \Omega_{m,0}))^{1/4}}$ & 00:52:37\\
         FFX            & $4.7861\cdot 10^{-4}$ & $-0.515 - 68100 \ddot{a}_D^2 + 49.8 H^2 + 119 \ddot{a}_D \Omega_{m,0} + 45.4 H \Omega_{m,0} - 0.768 \Omega_{m,0}^2$ & 00:00:18\\
         Genetic Engine & $3.4552\cdot 10^{-4}$ & \small $\frac{10 \ddot{a}_D}{a_D^2}+  \frac{0.001 a_D}{\Omega_{m,0}^2} + (0.001 + a_D^2 H (0.01 - 0.1 \Omega_{m,0}) + (-0.02 +  0.1\Omega_{m,0}) \Omega_{m,0} +  H (0.0099 + \Omega_{m,0} (-0.199 +\Omega_{m,0})) + a_D (-0.0001 - 0.001 H^2 +  \ddot{a}_D (0.01 + \Omega_{m,0} (-0.2 + \Omega_{m,0})) + \Omega_{m,0} (0.022 + \Omega_{m,0} (-0.41 + 2 \Omega_{m,0}))))/((0.001 + 0.01 a_D H + a_D^2 (-0.1 + \Omega_{m,0}) -0.01 \Omega_{m,0}) (-0.1 + \Omega_{m,0})) - \frac{0.011} {\ln\paa{1 + \abs{H}}}$ & 04:00:02\\
         GPG            & $5.6788\cdot 10^{-6}$ & $1.5029 +\frac{1370.8 (0.0076011 + \ddot{a}_D)}{-6.82703 + \frac{2.14236 +\Omega_{m,0}}{\ddot{a}_D + H)}}   \Omega_{m,0} \times \cos\paa{ \sin\paa{\cos(\ddot{a}_D) + \sin(a_D)}}^{-1} \times \bigg(\frac{2 (0.89088 + \ddot{a}_D)}{ 0.04397 + a_D - \ddot{a}_D} + (-0.533228 + a_D) a_D \Omega_{m,0} \sin(a_D) + \frac{\cos(1.677 +\Omega_{m,0}) \sin(a_D) (1 + \sin(a_D))}{(-\ddot{a}_D +\Omega_{m,0}) (\ddot{a}_D +  H +\Omega_{m,0})}\bigg)$ & 00:18:43\\
\hline
        \end{tabular}
        \caption{Result from benchmarking the two-region backreaction dataset C3h.}
        \label{tab:C3h}
    \end{table}
    \begin{table}
        \centering
        \begin{tabular}{ |p{2.52cm}||p{2.15cm}|p{7.54cm}|p{1.43cm}| }
         \hline
         \multicolumn{4}{|c|}{Cosmological dataset: C4a} \\
         \hline
         \hline
         Algorithm & MSE & Equation & Runtime\\
         \hline
         AI-Feynman     & $4.5639\cdot 10^{-2}$ & $-18.44 \expval{z} \paa{\expval{z} + \sin\paa{\sqrt{e^{-\sqrt{\expval{z}}}}}}$ & 08:53:25\\
         DSO            & $8.8410\cdot 10^{-2}$ & $-e^{e^{\sin\paa{\sqrt{\expval{z}}}} + \expval{z}} + e^{\cos \paa{e^{\expval{z}} \expval{z}}} - 2 \expval{z}$ & 02:06:51\\
         DSR            & $6.1976\cdot 10^{-2}$ & $-\expval{z} \paa{e^{e^{\paa{3 \expval{z} + \sin\paa{\sqrt{\expval{z}}}}^{1/8}}} + e^{\expval{z}} \expval{z}}$ & 01:05:39\\
         FFX            & $1.1793\cdot 10^{-1}$ & $-0.358 - 15.1 \expval{z} - 2.94 \max(0, -0.733 + \expval{z}) - 5.06 \max(0, -0.6 + \expval{z}) - 3.51 \max(0, -0.467 + \expval{z}) -  3.89 \max(0, -0.333 + \expval{z}) - 6.41 \max(0, -0.2 + \expval{z})$ & 00:00:03\\
         Genetic Engine & $3.9346\cdot 10^{-2}$ & $\frac{\expval{z} (-1010 + \expval{z})}{\ln\paa{1 + \abs{\expval{z}}^2}}  (6.9315\cdot 10^{-10} + \expval{z} (-1.3863 \cdot 10^{-6} +   \expval{z} (0.00069317 +  \expval{z} (-0.00001 +  \expval{z} (-8\cdot10^{-10} + (1.6\cdot10^{-6} -0.0008 \expval{z}) \expval{z})))) -  0.0008 (-0.001 + \expval{z}) -0.85393 +   \expval{z} (-12.488 + \expval{z}^3) -  12.5 \sqrt{\abs{\expval{z}}}) \sqrt{\abs{\expval{z}}})$ & 04:00:03\\
         GPG            & $7.0630\cdot 10^{1}$ & $-10.585$ & 00:11:45\\
\hline
        \end{tabular}
        \caption{Result from benchmarking the two-region redshift drift dataset C4a.}
        \label{tab:C4a}
    \end{table}
    \begin{table}
        \centering
        \begin{tabular}{ |p{2.52cm}||p{2.15cm}|p{7.54cm}|p{1.43cm}| }
         \hline
         \multicolumn{4}{|c|}{Cosmological dataset: C4b} \\
         \hline
         \hline
         Algorithm & MSE & Equation & Runtime\\
         \hline
         AI-Feynman     & $3.0829$ & $18.16 + \frac{\pi}{\ln \paa{\sin\paa{1 +\paa{-1 + e^{f_o}} \expval{z}}}}$ & 08:53:25\\
         DSO            & $8.0794\cdot 10^{-1}$ & $-e^{2 \expval{z} + 7 f_o} + e^{-\expval{z} + \cos \paa{\sqrt{\expval{z}}}}$ & 02:02:58\\
         DSR            & $6.5251\cdot 10^{-1}$ & $-e^{e^{2 f_o} + \expval{z}^{1/4} + 4 f_o} \expval{z} + f_o$ & 01:24:34\\
         FFX            & $1.6590\cdot 10^{-1}$ & \small $-0.189 - 1.21 \max \paa{0, -0.6 + \expval{z}} - 70.3 \expval{z} f_o -1.9 \max \paa{0, -0.467 + \expval{z}} + \max \paa{0, -0.2 + \expval{z}} (-36.6 f_o - 7.3 \max (0, -0.21 + f_o) - 96 \max (0, -0.15 + f_o)) + \max (0, -0.733 + \expval{z}) (-38.6 f_o - 56.2 \max (0, -0.15 + f_o)) -  0.992 \expval{z} \max (0, -0.15 + f_o)$ & 00:00:24\\
         Genetic Engine & $5.8377\cdot 10^{-2}$ &\small  $\Big(f_o (f_o ((-0.001 - 0.01 f_o) f_o + \expval{z}^5 (-0.02 + 2 f_o) +  \expval{z} (-0.001 + f_o (1.09 + f_o)) + \expval{z}^4 (3 + f_o (-100 + 1.98 f_o)) +  \expval{z}^3 (-100 +f_o (3.07+ (-99.1 - 0.02 f_o) f_o)) +  \expval{z}^2 (1.1 +f_o (-99.1 + f_o 0.07 + f_o)))) + (-0.00001 - 0.0001 f_o + \expval{z} (0.01 +2\cdot 10^{-6} f_o +  \expval{z} (0.0998 + (1 + 0.01 f_o) f_o + \expval{z}^3 (-0.02 + 2 f_o) + \expval{z}^2 (3 + (-100 - 0.02 f_o) f_o) +  \expval{z} (-100 + f_o (-0.93 + f_o))))) \sqrt{\abs{\expval{z}}})\Big)\times \frac{1}{\expval{z} (\expval{z} + 0.01 f_o)}$ & 04:00:02\\
         GPG            & $3.5472\cdot 10^{-2}$ & $-0.81024 - 0.25827 \expval{z} + 1.137 \cos\paa{(-3.2231 + \expval{z}) (2.1866 + \expval{z})} + \frac{35.852 \expval{z} (0.76216 + \expval{z}) f_o }{\sin\paa{3.7911 - f_o}} + \paa{-0.78483 + \frac{0.78478 \expval{z}}{f_o}} \sin(f_o)$ & 00:52:12\\
\hline
        \end{tabular}
        \caption{Result from benchmarking the two-region redshift drift dataset C4b. To increase readability, we only include 4 significant digits and for Genetic Engine 3 significant digits.}
        \label{tab:C4b}
    \end{table}
    \begin{table}
        \centering
        \begin{tabular}{ |p{2.52cm}||p{2.15cm}|p{7.54cm}|p{1.43cm}| }
         \hline
         \multicolumn{4}{|c|}{Cosmological dataset: C4c} \\
         \hline
         \hline
         Algorithm & MSE & Equation & Runtime\\
         \hline
         AI-Feynman     & $6.4086\cdot 10^{-1}$ & $-0.1432 + \paa{-62.9 + e^{1 + \expval{z} + \pi}} \Omega_{Q,0}$ & 08:53:27\\
         DSO            & $1.5190$ & $\Omega_{Q,0} - 2 \paa{e^{e^{\expval{z}}} +  8 \expval{z}} \Omega_{m,0} (\expval{z} + \Omega_{R,0})$ & 02:19:20\\
         DSR            & $5.7490\cdot 10^{-1}$ & \small $-\expval{z} \big(e^{e^{\Omega_{m,0}} +\sqrt{\Omega_{m,0}} + \sin \paa{\sqrt{\expval{z}}} + \sin \paa{\Omega_{m,0}}} +  3 \Omega_{m,0}\big)$ & 01:21:24\\
         FFX            & $1.6484\cdot 10^{-1}$ & \small $- 12 \Omega_{m,0} \max \paa{0, -0.73 + \expval{z}} -0.37  + \max (0, -0.47 + \expval{z})  (-20 \Omega_{m,0} - 15 \max (0, -0.22 + \Omega_{m,0})) - 41 \expval{z} \Omega_{m,0} + \max (0, -0.2 +\expval{z}) (-21 \Omega_{m,0} - 0.34 \max (0, -0.22 + \Omega_{m,0}))  $ & 00:00:29\\
\hline
        \end{tabular}
        \caption{Result from benchmarking the two-region redshift drift dataset C4c. To increase readability, we only include 4 significant digits and for Genetic Engine and FFX 2 significant digits.}
        \label{tab:C4c}
    \end{table}
    \begin{table}
        \centering
        \begin{adjustbox}{angle=90}
        \begin{tabular}{ |p{2.52cm}||p{2.15cm}|p{13.35cm}|p{1.43cm}| }
         \hline
         \multicolumn{4}{|c|}{Cosmological dataset: C4d} \\
         \hline
         \hline
         Algorithm & MSE & Equation & Runtime\\
         \hline
         AI-Feynman     & $7.1095$ & $15.06 - e^{e^{1 + \expval{z} \sin\paa{\Omega_{m,0}}}}$ & 08:53:28\\
         DSO            & $7.6305\cdot 10^{-1}$ & $-\expval{z} \paa{1 + e^{\expval{z} + 4 \Omega_{m,0} + \Omega_{R,0}} + \Omega_{m,0}} + \Omega_{Q,0}$ & 01:57:03\\
         DSR            & $5.6043\cdot 10^{-1}$ & $ - e^{e^{\cos\paa{e^{H_D}} + \expval{z} + \Omega_{m,0} + \sqrt{\sin\paa{\Omega_{m,0}}} + \sin\paa{\Omega_{m,0}}}} + \expval{z}+ e^{2 e^{\Omega_{m,0}}}$ & 01:13:09\\
         FFX            & $1.6539\cdot 10^{-1}$ & $-1.7 - 13 \expval{z}^2 + 0.55 \Omega_{m,0}^2 - 0.35 \Omega_{R,0} + 13 \expval{z} \Omega_{R,0} + (-18 +     220 \expval{z}) \ln \paa{\Omega_{R,0}}$ & 00:00:46\\
\hline
        \end{tabular}
        \end{adjustbox}
        \caption{Result from benchmarking the two-region redshift drift dataset C4d. To increase readability, we only include 4 significant digits and for Genetic Engine and FFX 2 significant digits.}
        \label{tab:C4d}
    \end{table}
    \begin{table}
        \centering
        \begin{tabular}{ |p{2.52cm}||p{2.15cm}|p{7.54cm}|p{1.43cm}| }
         \hline
         \multicolumn{4}{|c|}{Cosmological dataset: C4e} \\
         \hline
         \hline
         Algorithm & MSE & Equation & Runtime\\
         \hline
         AI-Feynman     & $1.5504\cdot 10^{-3}$ & $-23.05 + 23.14 \cos\paa{\sqrt{\expval{z}}}^{\Omega_{m,0}}$ & 08:53:26\\
         DSO            & $7.7612\cdot 10^{-4}$ & $\expval{z} \Omega_{R,0} (-\expval{z} - 14 \Omega_{m,0} + \Omega_{R,0})$ & 01:34:04\\
         DSR            & $9.0970\cdot 10^{-3}$ & $\sqrt{\Omega_{m,0}} \Big(-2 \expval{z} - 2 \Omega_{m,0} -  \sqrt{\expval{z}^2} (e^{e^{\Omega_{m,0}}} +  \expval{z} + \Omega_{m,0}) + \Omega_{R,0}\Big)$ & 01:25:18\\
         FFX            & $3.4198\cdot 10^{-3}$ & $0.00905 - 0.406 \expval{z}^2 - 11 \expval{z} \Omega_{m,0}$ & 00:00:28\\
\hline
        \end{tabular}
        \caption{Result from benchmarking the two-region redshift drift dataset C4e. To increase readability, we only include 4 significant digits and for Genetic Engine and GPG 2 significant digits.}
        \label{tab:C4e}
    \end{table}

    \begin{table}
        \centering
        \begin{tabular}{ |p{2.52cm}||p{2.15cm}|p{7.54cm}|p{1.43cm}| }
         \hline
         \multicolumn{4}{|c|}{Cosmological dataset: C5a} \\
         \hline
         \hline
         Algorithm & MSE & Equation & Runtime\\
         \hline
         AI-Feynman     & $ 1.2193\cdot 10^{-31}$ & $\frac{1}{r (1 + r)^2}$ & 04:28:48\\
         DSO            & $ 1.0894\cdot 10^{-31}$ & $\frac{x}{(r + r^2) (r + x)}$ & 00:20:59\\
         DSR            & $ 1.2010\cdot 10^{-31}$ & $\frac{1}{r x (r + x) (r + x (x^2)^{1/4})}$ & 00:03:14\\
         FFX            & $ 1.3937\cdot 10^{-2}$ & $\frac{-0.0094 + 0.00427 r}{1 - 0.0189 r - 0.999 x}$ & 00:00:06\\
         Genetic Engine & $ 7.0480\cdot 10^{-32}$ & $\frac{1}{r (r^2 + x (-1. + 2. x) + r (-1. + 3. x))}$ & 04:00:02\\
         GPG            & $ 3.2492\cdot 10^{-13}$ & $6.6928 \cdot 10^{-7} + \frac{x}{r (r + x)^2}$ & 00:09:18\\
         GPZGD          & $ 4.1697\cdot 10^{-1}$ & $1.2842 + 1.3495 \sin\paa{3.98 + 0.87933 r}$ & 02:27:06\\
         ITEA           & $ 3.1813\cdot 10^{-9}$ & $-1.5459 + \frac{0.92039}{r x^2} - 0.10831 \abs{r}^2 \abs{x} + 1.2138 \sqrt{\abs{\frac{x^5}{r}}} - 0.69571 \ln\paa{\frac{1}{r^2 x^2}} - 0.2733 \sin(r x)$ & 00:08:30\\
         Operon         & $ 1.7487\cdot 10^{-4}$ & $0.0599 + \frac{0.96555 e^{-1.7006 r}}{r}$ & 00:00:13\\
         PySR           & $ 1.2193\cdot 10^{-31}$ & $\frac{x}{r (r + x)^2}$ & 00:41:57\\
         QLattice       & $ 3.8100\cdot 10^{-16}$ & $3.4125\cdot 10^{-8} + \frac{1}{2.0652\cdot 10^{-8} + r (1 + r)^2}$ & 01:29:46\\
         uDSR           & $ 6.9462\cdot 10^{-4}$ & $\frac{1}{e^r r x + r^2 x}$ & 07:07:42\\
         \hline
        \end{tabular}
        \caption{Result from benchmarking the dark matter dataset C5a.}
        \label{tab:C5a}
    \end{table}
    
    \begin{table}
        \centering
        \begin{tabular}{ |p{2.52cm}||p{2.15cm}|p{7.54cm}|p{1.43cm}| }
         \hline
         \multicolumn{4}{|c|}{Cosmological dataset: C5b} \\
         \hline
         \hline
         Algorithm & MSE & Equation & Runtime\\
         \hline
         AI-Feynman     & $ 1.0798\cdot 10^{-4}$ & $\frac{0.99982}{r (1 + r)^2}$ & 08:53:25\\
         DSO            & $ 1.0883\cdot 10^{-4}$ & $\frac{1}{-1 + x + r (1 + r (1 + x (r + x)))}$ & 01:56:18\\
         DSR            & $ 1.1504\cdot 10^{-4}$ & $\frac{1}{2 r^2 + r^3 + r (r + \cos\paa{r^{1/4}})^{1/32}}$ & 01:26:27\\
         FFX            & $ 1.5045\cdot 10^{-2}$ & $\frac{-0.00935 + 0.00424 r}{1 - 0.0189 r - 0.999 x}$ & 00:00:06\\
         Genetic Engine & $ 1.8027\cdot 10^{-4}$ & $(-100 - 1000 r^3 + 90000 x)/(2 r^4 + r^3 (199.91 - 2199 x) + r^2 (-9.0005 + x (-19801 + 198900 x)) + x (5 + x (-98 + x (460 + 200 x))) +  r (-0.05 + x (901.03 + x (-19045 + 100400 x))) + ( r^4 +  r^3 (100.01 - 1100 x) + x (-0.05 + (0.48 + 0.2 x) x) +  r^2 (0.60002 + x (-10006 + 99998 x)) + r (0.0005 + x (-60.005 + x (580 + 200 x)))) \sqrt{\abs{r}})$ & 04:00:03\\
         GPG            & $ 1.1944\cdot 10^{-4}$ & $-0.071443 + \frac{0.023833 r ( \frac{39.888}{0.9652 + r} + \cos(r)}{\sin(r)( r^2 + \sin(r))} + (-0.0064821 r + 0.0064821 x) (x - \cos(x))  \frac{r - x + \sin(x)}{\cos(5.5776 + r - x)}$ & 00:30:57\\
\hline
        \end{tabular}
        \caption{Result from benchmarking the dark matter dataset C5b.}
        \label{tab:C5b}
    \end{table}

    \begin{table}
        \centering
        \begin{tabular}{ |p{2.52cm}||p{2.15cm}|p{7.54cm}|p{1.43cm}| }
         \hline
         \multicolumn{4}{|c|}{Cosmological dataset: C5c} \\
         \hline
         \hline
         Algorithm & MSE & Equation & Runtime\\
         \hline
         AI-Feynman     & $ 1.2824\cdot 10^{-2}$ & $-0.195346 + \frac{1}{1 + r - \cos\paa{r + \sin\paa{r}}}$ & 08:53:25\\
         DSO            & $ 1.4519\cdot 10^{-2}$ & $\frac{e^x}{r + e^r r (1 + x) (r + x)}$ & 02:01:41\\
         DSR            & $ 1.3460\cdot 10^{-2}$ & $\frac{1}{\sqrt{e^{r \sqrt{\frac{1}{r^{3/4}} + x + r \sin(r)}}} r + r^2}$ & 01:22:02\\
         FFX            & $ 3.0528\cdot 10^{-2}$ & $0.459   + 13.7 \max\paa{0, 0.48 - r} -  50.8 r \max \paa{0, 0.48 - r} + 5.01 \max(0, 0.733 - r) - 0.176 r -  11.4 r \max(0, 0.733 - r)+ 0.0173 r^2- 11.6 \max(0, 0.48 - r) \max(0, 0.733 - r) +  0.322 \max(0, 0.987 - r) + 0.328 r \max(0, 0.987 - r) -  0.186 \max(0, 0.48 - r) \max(0, 0.987 - r) +  5.83 \max(0, 0.733 - r) \max(0, 0.987 - r) - 0.096 \max(0, -1.75 + r) +  0.0188 \max(0, -1.49 + r) - 0.118 \max(0, -0.987 + r) + 0.00255 r \max(0, -0.987 + r) - 0.0375 \max(0, -0.733 + r) -  0.166 r \max(0, -0.733 + r) -  2.64 \max\paa{0, 0.987 - r} \max\paa{0, -0.733 + r} +  0.374 \max\paa{0, -0.987 + r} \max\paa{0, -0.733 + r}$ & 00:00:08\\
         Genetic Engine & $ 1.3271\cdot 10^{-2}$ & $\frac{\paa{-0.00589 + (2.97 + 0.13 r) r - 0.6477 r^2 \ln{1 + \abs{r}}}r^{-1}}{2.94\cdot10^{-6} + 0.00294 x +  r (0.00882 + r (5.88 + r + 0.5 x) + 2.94 x)}$ & 04:00:03\\
         GPG            & $ 2.8787\cdot 10^{-1}$ & $-1.5578 + 0.2565 r - 0.24779 x + 0.42686 \cos(r) + (2.8206 + x (-1.5899 + \frac{1.5899 x}{r^2}) \sin\paa{\frac{2.3556 r}{2.9873 + x}} - \frac{0.17026 x (0.47155 - \sin(r)) \sin\paa{\frac{5.11373 \sin(x)}{r}}}{\frac{r}{r + 0.53201 x} - \frac{1.8184 r}{\cos\paa{\frac{3.5952}{r}}}} $ & 00:27:13\\
\hline
        \end{tabular}
        \caption{Result from benchmarking the dark matter dataset C5c. To increase readability, we only include 3 significant digits for Genetic Engine.}
        \label{tab:C5c}
    \end{table}

    \begin{table}
        \centering
        \begin{tabular}{ |p{2.52cm}||p{2.15cm}|p{7.54cm}|p{1.43cm}| }
         \hline
         \multicolumn{4}{|c|}{Cosmological dataset: C5d} \\
         \hline
         \hline
         Algorithm & MSE & Equation & Runtime\\
         \hline
         AI-Feynman     & $ 2.1582$ & $0.985447 - \cos\paa{\frac{R_0 \sin\paa{\sin(x)}}{r}}$ & 08:53:26\\
         DSO            & $ 3.2736\cdot 10^{-3}$ & $ \frac{R_0^2}{r (2 r + r^4 + R_0)}$ & 01:52:15\\
         DSR            & $ 3.1622\cdot 10^{-3}$ & $\frac{R_0}{\paa{e^r + \frac{r}{R0^2}}\sin(r)}$ & 00:49:03\\
         FFX            & $ 3.7398\cdot 10^{-2}$ & $0.0112 + (0.165 + 0.161 R_0) R_0 +(0.010423 - 0.56458 R_0) \ln(r) + \max(0, 0.48 - r) (-20.8 + 13.1 R_0 - 10.163 \ln(r) - 1.3 \max(0, -0.967 + R_0) - 0.119 \max(0, -0.7 + R_0)) + \max(0, -0.967 + R_0) (0.036 - 0.1494 \ln(r) - 0.017 \max(0, -0.7 + R_0)) + \paa{0.045 - 0.15157 \ln(r)} \max(0, -0.7 + R_0)$ & 00:00:29\\
         Genetic Engine & $ 5.5226\cdot 10^{-4}$ & $(3.3183\cdot10^{-6} +  0.00033183 x - 0.036501 r^2 + R_0 (0.0033183 + 0.33183 x) ) \frac{r\ln\paa{1+ \frac{\abs{\frac{R_0}{r}}}{\ln\paa{1 + \abs{x}}}}}{0.001 R_0^2 + r (-0.00001 + r (-0.01 + x) + 0.001 x)}$ & 04:00:03\\
         GPG            & $ 1.5333\cdot 10^{-6}$ & $0.68028 + \frac{1.0022 R_0^2 x}{1.6219 r^2 + r R_0} - 1.2699 \cos \paa{\cos\paa{\frac{R_0}{r + R_0}}}$ & 01:06:36\\
\hline
        \end{tabular}
        \caption{Result from benchmarking the dark matter dataset C5d.}
        \label{tab:C5d}
    \end{table}

    \begin{table}
        \centering
        \begin{tabular}{ |p{2.52cm}||p{2.15cm}|p{7.54cm}|p{1.43cm}| }
         \hline
         \multicolumn{4}{|c|}{Cosmological dataset: C5e} \\
         \hline
         \hline
         Algorithm & MSE & Equation & Runtime\\
         \hline
         AI-Feynman     & $ 6.1925\cdot 10^{-2}$ & $0.985249 - \cos \paa{\sin\paa{\frac{R_0 \sin(x)}{r}}}$ & 08:53:26\\
         DSO            & $ 3.4661\cdot 10^{-4}$ & $e^{\frac{r \paa{-x + \sin\paa{e^x}}}{R_0 x}}$ & 01:27:12\\
         DSR            & $ 3.7350\cdot 10^{-4}$ & $e^{\frac{(r (-e^x + x))}{R_0}}$ & 00:32:03\\
         FFX            & $ 1.0211\cdot 10^{-4}$ & $0.0137 - 0.0412 r + 0.222 R_0 + \ln(r) (-0.10597 - 0.25834 \ln(R_0)) + 0.078173 \ln(R_0) + (0.00332 + 0.309 r - 1.37237 \ln(R_0)) \max(0, 0.48 - r) +  \max(0, 0.733 - r) (0.405 - 0.258 R_0 + 0.00767 \max(0, 0.987 - r)) + (0.144 - 0.0389 R_0) \max(0, 0.987 - r) +\max(0, 1.24 - r) (0.0645 + 0.0278 \max(0, 1.49 - r)) + 0.0505 \max(0, 1.49 - r) + 0.0299 \max(0, 1.75 - r) -  0.00176 \max(0, 0.833 - R_0) - 0.0142 \max(0, 0.967 - R_0) -  0.00973 \max(0, 1.1 - R_0) - 0.0182 \max(0, 1.23 - R_0) -  0.0117 \max(0, 1.37 - R_0)$ & 00:00:44\\
         Genetic Engine & $ 1.9818\cdot 10^{-4}$ & $\frac{r (-0.11 + (-0.299 + 0.003 r) r + R_0 (-0.01 - 0.99 x))}{-0.01 + 1.2 r^2 + r (1 - 0.02 R_0) R_0 + r (r + x) \ln\paa{1 + \abs{r}}}$ & 04:00:03\\
         GPG            & $ 3.8320\cdot 10^{-7}$ & $-0.029222 - ( 1.0941 R_0 x)/((R_0 (-1 + x) + r (0.3951 + 2.3459 x)) (-2.4352 + \cos\paa{\frac{r}{R_0}} + \cos(0.44184 x)))$ & 00:48:56\\
\hline
        \end{tabular}
        \caption{Result from benchmarking the dark matter dataset C5e.}
        \label{tab:C5e}
    \end{table}

    \begin{table}
        \centering
        \begin{tabular}{ |p{2.52cm}||p{2.15cm}|p{7.54cm}|p{1.43cm}| }
         \hline
         \multicolumn{4}{|c|}{Cosmological dataset: C5f} \\
         \hline
         \hline
         Algorithm & MSE & Equation & Runtime\\
         \hline
         AI-Feynman     & $ 9.7738\cdot 10^{-1}$ & $-0.26435 + \sin\paa{\cos\paa{\ln\paa{1 + e^{\cos\paa{\frac{R_0}{r}}}}}}$ & 08:53:29\\
         DSO            & $ 3.8792\cdot 10^{-2}$ & $\frac{r + e^{\sin(R_0 x)} R_0}{3 r + r^2 + 4 r^3}$ & 02:21:13\\
         DSR            & $ 1.2143\cdot 10^{-2}$ & $\frac{R_0}{e^r - x \cos\paa{2 r - R_0^{1/4} + x}}$ & 00:57:04\\
         FFX            & $ 1.1135\cdot 10^{-1}$ &\small $-0.193 + (0.449 + 0.0184 R_0) R_0 + (-0.46904 R_0 - 0.051681 x) \ln(r) + (0.287 x - 0.018544 \ln(r)) \max(0, -0.7 + R_0) + 1.39\cdot10^{-8} \max(0, -0.6 - x) - 0.0342 x \max(0, -0.6 - x) + \max(0, 1.37 - R_0) (-0.0848 - 0.0461 R_0 + 0.839 \max(0, -0.6 - x) +     0.0262 \max(0, -0.333 - x)) - 0.000231 x \max(0, -0.333 - x) + \max(0, 0.48 - r) (1.91 R_0 + 0.776 x - 5.1681 \ln(r) -  3.29 \max(0, 1.37 - R_0) + 0.876 \max(0, -0.7 + R_0) -  22.9 \max(0, -0.6 - x) - 0.982 \max(0, -0.333 - x) -    0.169 \max(0, -0.0667 - x)) - 0.0264 x \max(0, -0.0667 - x)$ & 00:00:49\\
         Genetic Engine & $ 6.5033\cdot 10^{-3}$ & $\frac{R_0 (0.001 - 0.1 r^3 + R_0 - 0.01 x) \ln\paa{1 + \ln\paa{1 + \abs{1 + R_0}}}}{-0.003 R_0 + r (-0.1 + r + R_0)}\times \ln\paa{ 1 + \abs{1 - r + x}} $ & 04:00:02\\
         GPG            & $ 3.3156\cdot 10^{-5}$ & $0.059812 + \frac{1}{r (r + 1.7573 R_0)} 2.0227 R_0^2 x \times \cos\paa{0.32347 + \frac{0.6238 r}{R_0}} \big(-\frac{1.0267 \sin(r)}{r + R_0} + (-0.015434 + 0.074322 r) r \sin(2.6631 + R_0) + \sin(1.0097 + x)\big)$ & 02:01:52\\
\hline
        \end{tabular}
        \caption{Result from benchmarking the dark matter dataset C5f.}
        \label{tab:C5f}
    \end{table}

    \begin{table}
        \centering
        \begin{tabular}{ |p{2.52cm}||p{2.15cm}|p{7.54cm}|p{1.43cm}| }
         \hline
         \multicolumn{4}{|c|}{Cosmological dataset: C6a} \\
         \hline
         \hline
         Algorithm & MSE & Equation & Runtime\\
         \hline
         AI-Feynman     & $ 1.8923\cdot 10^{-1}$ & $0.33998 \sin\paa{1 + \pi \sin\paa{t^{3/2}}}$ & 08:53:25\\
         DSO            & $ 1.7713\cdot 10^{-1}$ & $e^{-\sqrt{t}} \sin\paa{e^{e^{t^{1/4}}} \sqrt{t}}$ & 02:31:34\\
         DSR            & $ 1.8647\cdot 10^{-1}$ & $\sin\paa{e^{e^{-t (-2 + \sqrt{\sqrt{t} + t})} + t} - t} \sin(t)$ & 01:06:29\\
         FFX            & $ 1.5930\cdot 10^{-1}$ & $12.88 + 3.77 \max(0, 0.6 - t) + 4.84 \max(0, 1. - t) -  0.186 \max(0, 0.6 - t) \max(0, 1. - t) - 5.23 \max(0, 1.4 - t) -  3.66 \max(0, 0.6 - t) \max(0, 1.4 - t) +  3.97 \max(0, 1. - t) \max(0, 1.4 - t) + 0.266 \max(0, 1.8 - t) -  10.7 \max(0, 1. - t) \max(0, 1.8 - t) +  7.88 \max(0, 1.4 - t) \max(0, 1.8 - t) - 2.92 \max(0, 2.2 - t) +  14. \max(0, -2.6 + t) - 9.19 \max(0, -2.2 + t) +  15.8 \max(0, -2.6 + t) \max(0, -2.2 + t) + 3.85 \max(0, -1.8 + t) -  26.8 \max(0, -2.6 + t) \max(0, -1.8 + t) +  9.47 \max(0, -2.2 + t) \max(0, -1.8 + t) - 3.73 \max(0, -1.4 + t) -  0.723 \max(0, -1. + t)$ & 00:00:11\\
         Genetic Engine & $ 1.8040\cdot 10^{-1}$ & $(t (-0.0076609 +  t (0.11682 +  t (-7.4921 +  t (107.33 +   t (281.28 +  t (-1547.4 +  t (1777.9 + (-616.76 -   9.2727 t) t))))))))/(0.000048045 +  t (0.0008662 +  t (0.099046 +  t (1.8199 +  t (55.79 +  t (952.91 + t (4832 + t (138.83 + t))))))))$ & 04:00:02\\
         GPG            & $ 2.0735\cdot 10^{-1}$ & $-0.050185$ & 00:09:46\\
\hline
        \end{tabular}
        \caption{Result from benchmarking the gravitational wave dataset C6a.}
        \label{tab:C6a}
    \end{table}
\begin{table}
        \centering
        \begin{adjustbox}{max width=\textwidth}
        \begin{tabular}{ |p{2.52cm}||p{2.15cm}|p{10.8cm}|p{1.43cm}| }
         \hline
         \multicolumn{4}{|c|}{Cosmological dataset: C6b} \\
         \hline
         \hline
         Algorithm & MSE & Equation & Runtime\\
         \hline
         AI-Feynman     & $ 1.2049\cdot 10^{-3}$ & $-\frac{0.20161 \cos(\dot{r} - 2 \Phi)}{r}$ & 08:53:25\\
         DSO            & $ 9.4951\cdot 10^{-2}$ & $-\cos(2 \Phi)$ & 02:49:01\\
         DSR            & $ 5.6447\cdot 10^{-2}$ & $-r + \sin(\Phi)^2$ & 01:03:59\\
\hline
        \end{tabular}
        \end{adjustbox}
        \caption{Result from benchmarking the gravitational wave dataset C6b. To increase readability, we only include 2 significant digits for FFX, Genetic Engine and QLattice.}
        \label{tab:C6b}
    \end{table}

    \begin{table}
        \centering
        \begin{tabular}{ |p{2.52cm}||p{2.15cm}|p{7.54cm}|p{1.43cm}| }
         \hline
         \multicolumn{4}{|c|}{Cosmological dataset: C6c} \\
         \hline
         \hline
         Algorithm & MSE & Equation & Runtime\\
         \hline
         AI-Feynman     & $ 9.6059\cdot 10^{-1}$ & $0.000580941 \sqrt{-\sin\paa{1 - e^{e^m1}}}$ & 08:53:26\\
         DSO            & $ 8.2165\cdot 10^{-1}$ & $e^{-e^{\sqrt{\frac{e^{\frac{\sin\paa{\sqrt{e^{m_1}}}}{m_1}}}{m_1}}} + m_1 t}$ & 02:52:50\\
         DSR            & $ 8.2547\cdot 10^{-1}$ & $\frac{m_1}{-\frac{\sqrt{m_1} + t}{m_1} + \frac{e^{e^{m_1^2 - t}}}{m_1 - \sin(m_1)}}$ & 01:10:29\\
         FFX            & $ 7.5087\cdot 10^{-1}$ & $-0.0211 +\max(0, -1.13 + m_1) (1.16 - 7.3\max(0, 0.6 - t) + 33.1 \max(0, -2.6 + t))$ & 00:00:12\\
         Genetic Engine & $ 7.8982\cdot 10^{-1}$ & $t^2 \ln\paa{1 + \abs{m_1}} ((0.001 + m_1 (-0.1 - t) + 0.01 t) t + (-0.001 -   0.01 t + m_1 (0.1 +  t)) (\ln\paa{1 + \abs{m_1}} +  m_1 t \times \ln\paa{1 + \abs{t}})) \ln\paa{1 + \abs{(-0.011 + 0.01 m_1) t^2}}$ & 04:00:03\\
         GPG            & $ 1.1328\cdot 10^{-1}$ & $0.0089118 -0.093731 m_1 \cos\paa{\frac{36.714 t}{3.3411 m_1 + m_1 t}} \times (1.6574 - t + \sin(2.2466 - m_1)) (6.5553 + 1.4407 m_1 + t + \sin(m_1) (5.6793 t^2 - 5.6793 \sin(t)))$ & 01:06:03\\
\hline
        \end{tabular}
        \caption{Result from benchmarking the gravitational wave dataset C6c.}
        \label{tab:C6c}
    \end{table}
\clearpage
\bibliographystyle{ieeetr} 
\bibliography{main} 

\end{document}